\documentclass[10pt,preprint]{aastex}

\slugcomment{Accepted for publication in PASP}

\shorttitle{ACS Photometric Calibration}
\shortauthors{Sirianni M. et al.}

\begin{document}

\title{The Photometric Performance and Calibration of the HST Advanced Camera For Surveys}

\author{M. Sirianni\altaffilmark{1,2,3}, 
        M.J. Jee\altaffilmark{3},
        N. Ben\'itez\altaffilmark{3,4},
        J.P. Blakeslee\altaffilmark{3},
        A.R. Martel\altaffilmark{3},
        G. Meurer\altaffilmark{3} 
M. Clampin\altaffilmark{6},\\ 
        G. De Marchi\altaffilmark{1,2,7}, 
        H.C. Ford\altaffilmark{3},
        R. Gilliland\altaffilmark{2}, 
        G.F. Hartig\altaffilmark{2}, 
        G.D. Illingworth\altaffilmark{5},  
        J. Mack\altaffilmark{2}\\ \& 
        W.J. McCann\altaffilmark{3}
        }

\altaffiltext{1}{European   Space  Agency,  Research   and  Scientific
  Support   Department} 
 \altaffiltext{2}{Space   Telescope   Science
  Institute,   3700   San   Martin   Drive,  Baltimore,   MD   21218.}
\altaffiltext{3}{Department  of Physics  and Astronomy,  Johns Hopkins
  University,  3400  North   Charles  Street,  Baltimore,  MD  21218.}
\altaffiltext{4}{Istituto de Astrof\'isica de Andaluc\'ia, Camino bajo de Hu\'etor 24, Granada 18008, Spain}
\altaffiltext{5}{UCO/Lick Observatory, University of California, Santa
  Cruz, CA 95064.}  
\altaffiltext{6}{NASA Goddard Space Flight Center,
  Code  681,  Greenbelt, MD  20771.}  
\altaffiltext{7}{Presently:  ESA,
  Astrophysics Division,  Keplerlaan  1, 2200  AG  Noordwijk,
  Netherlands}

\begin{abstract}
We  present the  photometric calibration  of the  Advanced  Camera for
Surveys (ACS).   The ACS was  installed in the Hubble  Space Telescope
(HST)  in March  2002.  It  comprises  three cameras:  the Wide  Field
Channel (WFC), optimized for deep near-IR survey imaging programs; the
High  Resolution Channel (HRC),  a high  resolution imager  that fully
samples the  HST point spread function  (PSF) in the  visible; and the
Solar Blind Channel  (SBC), a far-UV imager.  A  significant amount of
data has  been collected to  characterize the on-orbit  performance of
the three  channels.  We  give here an overview  of the  performance and
calibration of the two CCD cameras   (WFC and  HRC), and a
description of  the best techniques for  reducing   ACS CCD
data.  The overall performance is as expected from pre-launch testing
of the camera. Surprises were a better than predicted sensitivity in 
the visible and near-IR for both WFC and HRC and an unpredicted dip
in the HRC UV response at $\sim$3200 \AA.

On-orbit observations  of spectrophotometric standard  stars have been
used  to revise  the pre-launch  estimate of  the  instrument response
curves to  best match predicted  and observed count  rates.  Synthetic
photometry has  been used to  determine zeropoints for all  filters in
three magnitude  systems and to derive  interstellar extinction values
for the ACS photometric systems.

Due to  the CCD  internal scattering of  long wavelength  photons, the
width  of the  PSF  increases  significantly in  the  near-IR and  the
aperture correction for photometry with near-IR filters depends on the
spectral  energy distribution  of the  source. We  provide  a detailed
recipe to correct for the latter effect.

Transformations between the ACS  photometric systems and the UBVRI and
WFPC2 systems  are presented. In general, two  sets of transformations
are available: one based on  the observation of two star clusters; the
other  on  synthetic photometry.  We  discuss  the  accuracy of  these
transformations and their sensitivity  to details of the spectra being
transformed.

Initial  signs  of  detector  degradation  due to  the  HST  radiative
environment are already visible.  We discuss the impact on the data in
terms  of dark rate  increase, charge  transfer inefficiency,  and hot
pixel population.

\end{abstract}


\maketitle


\tableofcontents

\newpage

\section{INTRODUCTION}\label{intro}

The  Advanced Camera  for Surveys  (ACS) was  installed in  the Hubble
Space Telescope (HST)  on March 7, 2002, during  Servicing Mission 3B,
in the axial bay previously  occupied by the Faint Object Camera.  ACS
was  built  in  a  collaborative  effort  between  the  Johns  Hopkins
University,  Ball  Aerospace  \&  Technologies Corporation,  the  NASA
Goddard Space  Flight Center (GSFC),  and the Space  Telescope Science
Institute (STScI).   The description of the camera  and the philosophy
underlying its design  are presented in Ford et  al.  (1998, 2002).  The
status and  the on-orbit  performance of the  camera are  described in
Clampin et al.  (2002).   An overview of the capabilities of ACS
is provided in the ACS  Instrument Handbook (Pavlovsky, et al.  2004a,
hereafter referred to as AIHv5).
 
\begin{figure}
\centering
\includegraphics[angle=270,width=4.5in,clip=true]{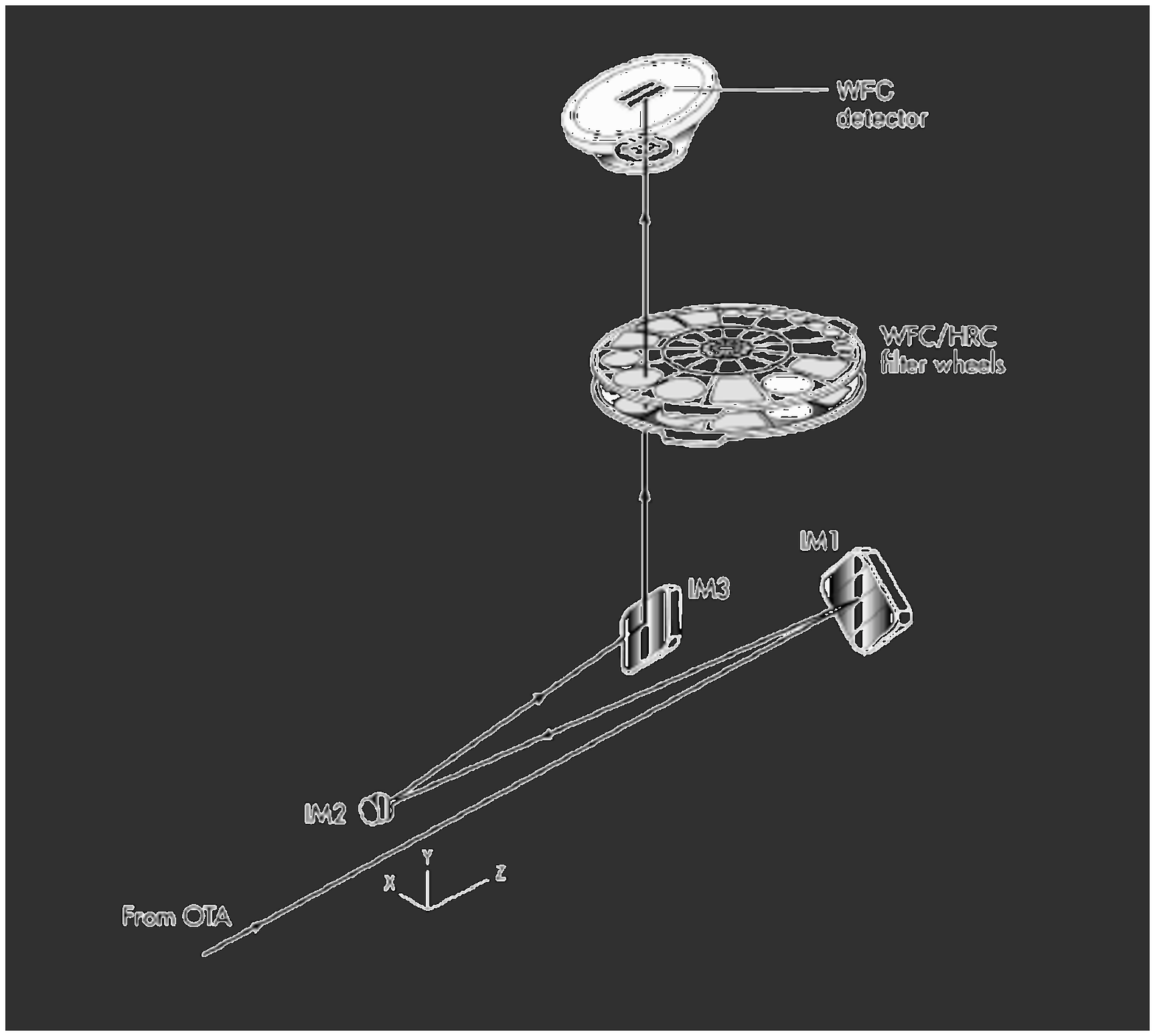}
\includegraphics[angle=270,width=4.5in,clip=true]{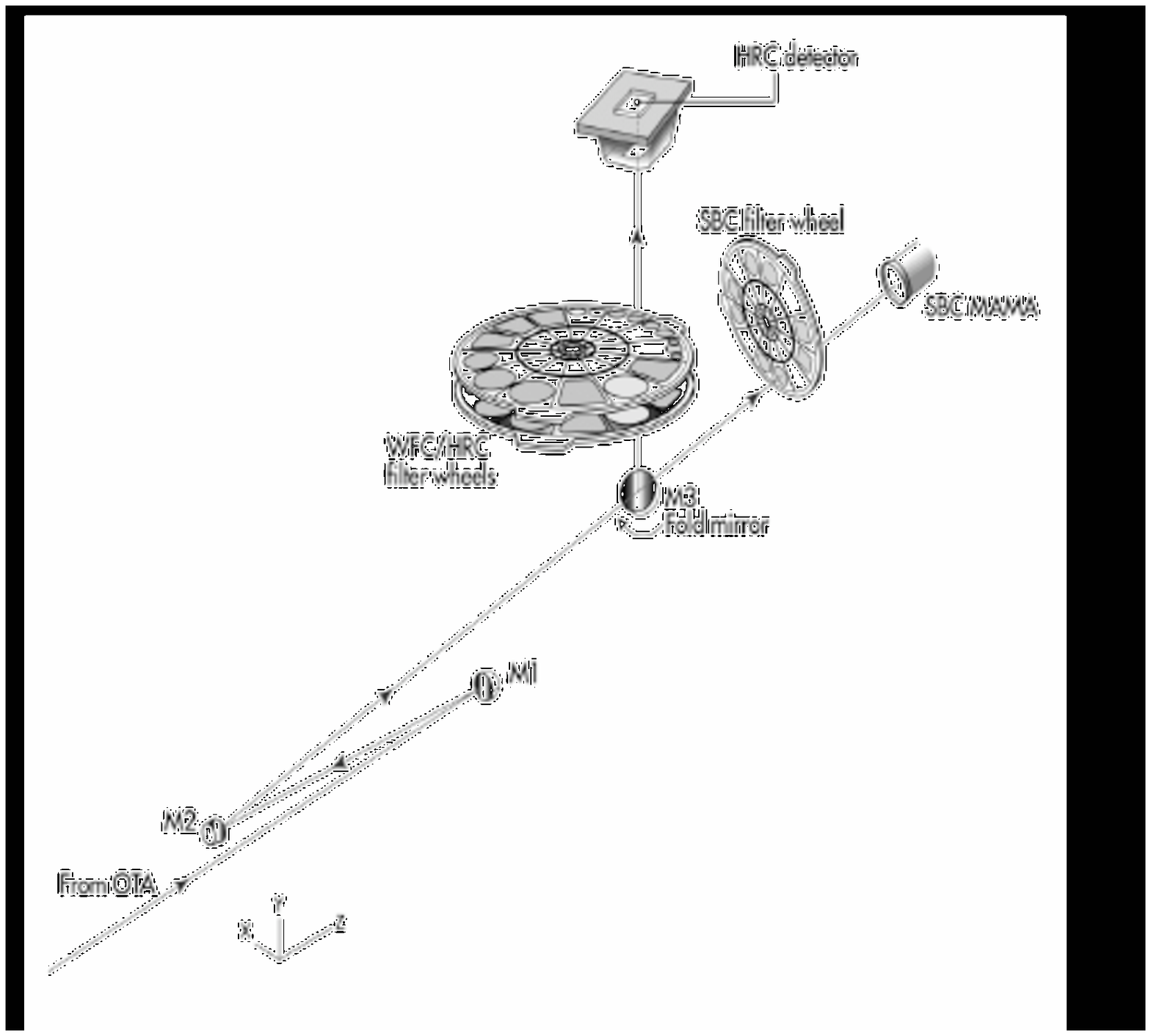}
\caption{Schematic showing the optical design for the WFC (top) and
HRC/SBC (bottom).}
\label{schematic}
\end{figure}

The ACS consists of three  channels, the Wide Field Channel (WFC), the
High Resolution Channel (HRC) and  the Solar Blind Channel (SBC).  The
WFC   and    HRC   have    separate   optical   paths    and   mirrors
(Fig.~\ref{schematic}).   The  WFC  is  a high-throughput,  wide  field
imager  (202\arcsec  $\times$ 202\arcsec)  designed  for deep  imaging
surveys in the optical and near-IR.   WFC provides a factor 10 gain in
discovery  efficiency  (throughput x  field  of  view) at  8000\,\AA\,
compared to  the Wide Field  Planetary Camera-2 (WFPC2).   The primary
goal of  the WFC design  is to maximize  the camera throughput  in the
near-IR, and  this has  been achieved by  using only  three reflective
optics and by coating the  mirrors with Denton protected silver.  This
configuration provides a reflectivity  50\% higher than three aluminum
coated  mirrors  would provide  at  800  nm.   The three  mirrors  are
designated  IM1, IM2,  and IM3.   The  light from  the Optical  Assembly
Telescope (OTA)  first encounters IM1, a spherical  mirror which images
the HST pupil onto the IM2.  This mirror is an anamorphic asphere figured
with the inverse  spherical aberration of the HST  primary mirror, and
thus corrects the  spherical aberration in the HST  primary mirror and
the field  dependent astigmatism of the  HST at the center  of the WFC
field-of-view  (FOV). The light from the IM2 mirror  is reflected by a
Schmidt-like  plate (IM3)  through the  two filter  wheels to  the WFC
CCDs. The Schmidt plate corrects astigmatism over the WFC FOV.

In the near-UV ( $<$ 3700\,\AA) the reflectivity of the silver coating
falls  rapidly  and  sets  the  lower  limit  of  the  spectral  range
observable with the WFC at  about 3000\,\AA.  The f/25 focal length of
the WFC  gives a  plate scale of  $0\farcs05$ pixel$^{-1}$ and  a near
critical sampling in the I-band.   The focal plane detector array is a
mosaic of  two Scientific Imaging  Technologies (SITe) 2048 x  4096 15
$\mu$m Charge  Coupled Devices (CCDs) (Clampin et  al.  1998, Sirianni
et al.  2000, 2003).  The  two CCDs, although physically  separated in
the focal  plane, have been cut  from the same silicon  wafer and have
gone through similar post-manufacturing processes.  There is therefore
a significant continuity in responsivity across the focal plane array.

The HRC  is a  near-UV to near-IR  high resolution f/70  imager, which
provides  critically sampled images  in the  visible over  a 29\arcsec
$\times$  26\arcsec  \, field  of  view    with a  resolution  of
$0\farcs025$ pixel$^{-1}$.
The  HRC optical  design employs  three aluminum  coated  mirrors with an
MgF$_2$ overcoating.   Two mirrors,  M1 and M2,  shared with  the SBC,
have a coating optimized for  maximum reflectivity at 1216\,\AA. M2 is an
anamorphic asphere  figured with  the inverse spherical  aberration on
the HST primary mirror, to correct the spherical aberration in the HST
primary  mirror.   The  third  mirror   M3,  is  a  fold  mirror  with
reflectivity optimized  at wavelengths longer than  2000\,\AA\, and it
is inserted  at 45\arcdeg\, into the  light path to  redirect the beam
through the two  filter wheels to the HRC focal  plane array.  The HRC
detector is  a SITe 1024 x  1024 21$\mu$m CCD (Clampin  et al.  1998),
based on  the Space Telescope Imaging Spectrograph  (STIS) CCD (Kimble
et al. 1998).

The SBC is a far-UV imager optimized for high throughput at 1216\,\AA,
with  a 31\arcsec  $\times$  35\arcsec \,  FOV  and a  plate scale  of
$0\farcs032$  pixel$^{-1}$.  In order  to maximize  far-UV throughput,
the  SBC  optical  design  is  a  two  mirror  system,  with  its  own
independent filter wheel. The two mirrors are shared with the HRC. The
SBC is selected  when the fold mirror is moved out  of the light beam.
The SBC uses a STIS-based  photon-counting detector with an opaque CsI
photocathode Multi  Anode Microchannel plate Array  (MAMA) (Quijano et
al. 2003)

More detailed description of the instruments are available in a series
of      papers      that      can      be      found      at      \url
{http://acs.pha.jhu.edu/instrument/papers/}.

In addition  to the  key design  goal of each  camera, the  other main
element  which determines  the quality  of  the science  that ACS  can
produce is its  filter complement.  The complement of  ACS filters and
dispersers is described  by Ford et al. (1996) and  is listed in Tables
\ref{wfilter} and \ref{hfilters}.  The  ACS optical design allows both
the WFC  and HRC to share  two filter wheels. These  filter wheels are
populated with 17 bandpass filters, a set of five linear ramp filters,
a grism,  a prism  for the  HRC, and three  visible and  three near-UV
polarizers.   The SBC  has its  own filter  wheel with  five long-pass
filters,  two prisms,  a narrow-band  Lyman-$\alpha$ filter,  and four
opaque positions which act as  shutters.  The CCD camera filter wheels
contain filters with two  different sizes. Some filters are full-sized
and can be used with both the HRC and WFC.  Others are smaller, giving
a full unvignetted FOV with the HRC, but a vignetted area of 72\arcsec
$\times$  72\arcsec \,  when used  with the  WFC.  Eleven  filters are
allocated to broad-band imaging from the near-UV to the near-IR.  This
includes the Sloan g (F475W),\,  r (F625W),\, i (F775W) and z (F850LP)
filters. The two near-UV high throughput filters F220W and F250W are similar to
the  F218W and  F255W currently  in the  WFPC2 and  provide continuity
between the  two instruments.   These are complemented  by one  set of
broad band  U (F330W),  B (F435W),  V (F555W), broad  V (F606W)  and I
(F814W)  filters, also similar  to the  WFPC2 filters.   To complement
these filters, there is a medium-band continuum filter F550M, which is
centered   at  $\sim$5500   \AA\,  with   a  band   width   of  $\sim$
10\%\footnote{  When  expressed in  percent  the  full  width at  half
  maximum  transmittance is divided  by the  center wavelength  of the
  band and multiplied by 100 percent}.

Narrow-band  filters for  the CCD  cameras  in ACS  take three  forms:
standard narrow-band filters  which cover the entire field  of view of
both cameras,  linear ramp  filters similar to  those in WFPC2  and an
HRC-size methane band filter designed for the observation of the giant
planets.  The  standard narrow-band filter complement  is NeV (F344N),
[OIII] (F502N),  H$\alpha$ (F658N) and  [NII] (F606N).  The  five ramp
filters  (four narrow-band  $\sim$2\% and  one broad-  band $\sim$9\%)
allow  narrow-  or  medium-band   imaging  centered  at  an  arbitrary
wavelength.  The ramp filters  cover the wavelength region from $\sim$
3700 \AA\, to  $\sim$ 1.05 $\mu$m.  Each is  divided into three strips
covering  $\sim$500 \AA\,  per  segment.   The  desired bandpass  is
selected  by  a  combination  of  target  position  and  filter  wheel
rotation.  The central  strip of each of the  five linear ramp filters
can be used with the HRC.  The transmittance curve for each filter can
be                             obtained                             at
\url{http://acs.pha.jhu.edu/instrument/filters/general/Master\_Table.html}.
The  total throughput  curve, including  the OTA  and the  ACS optical
elements   and   detector   quantum   efficiency  are   available   at\hfil\break 
\url{http://acs.pha.jhu.edu/instrument/photometry/}

This paper  presents data relevant to the  photometric performance and
calibration of  the two  CCD cameras of  ACS.  Information on  the SBC
photometric performance  are available in  AIHv5, Tran et  al. (2002a,
2002b),  and Sirianni  et al.   (2002a).  Section  2 presents  a brief
description  of the  WFC and  HRC data  and FITS  file  structure.  In
Section 3 we  describe the main steps of  the ACS calibration pipeline
at  STScI, including a  discussion of  the accuracy  of bias  and dark
subtraction, flat field normalization, geometric distortion and charge
transfer efficiency.  Section 4 describes the spatial variation of the
point  spread function,  the  encircled energy  profile  and the  long
wavelength scattered light problem.   The analysis of the variation of
the encircled energy profile with  the effective wavelength is used to
estimate the aperture correction  in all filters.  Section 5 discusses
the  gain   settings,  and  the   linearity  and  saturation   of  the
CCDs. Section  6 presents the  absolute flux calibration of  ACS.  The
definition of the  photometric system of ACS and  the determination of
the zeropoints  are presented in Section  7.  In Section  8 we present
observed  and synthetic  photometric transformations  between  WFC and
HRC.  We also discuss  transformations from the ACS photometric system
to the UBVRI  and WFPC2 systems. Sections 9 and 10  discuss the HRC UV
red  leak and  the  reddening  in the  ACS  system, respectively.   We
inserted several sections in the appendix to include material that can
be useful  to some  readers which  are also ACS  users.  We  provide a
cookbook  for the  calibration  of point  source photometry,  resolved
object photometry and surface  photometry, and additional material for
transformations  to non-ACS  photometric systems  of stars  and galaxy
templates.
 
This document represents  our understanding of the CCD  cameras of ACS
as of Spring  2005.  Any of the discussions  or results presented here
could be subject to revision as our understanding of ACS evolves.

\section {ACS DATA FORMAT}
The ACS Data Handbook (Pavlovsky et al. 2005, hereafter referred to as
ADHv4) contains all the information needed  to manipulate,
process and analyze  data from ACS.  In this  section we summarize the
key elements of the ACS data structure.

\subsection{WFC Data}
\begin{figure}
\centering
\epsscale{.8}
\plotone{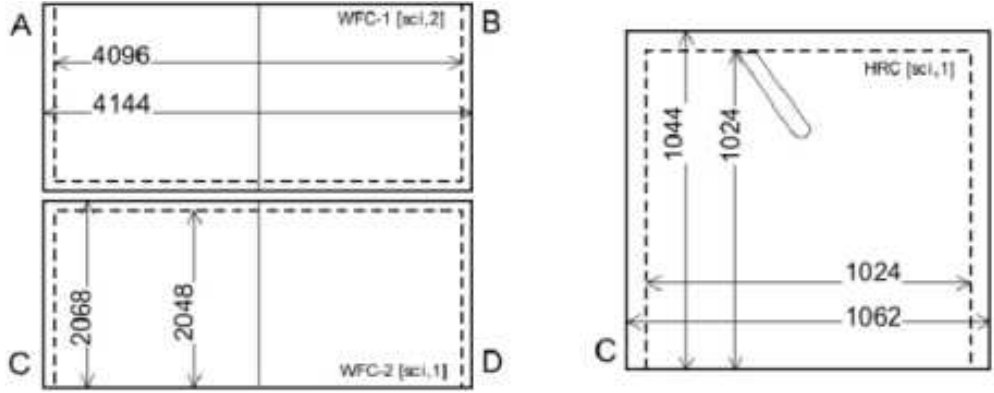}
\caption{Structure  of the  WFC and  HRC chips  (not  scaled).  Dashed
  lines  show the  overscan regions,  dotted lines  show  the quadrant
  division. The dimension in pixels are shown. The gap between the two
  WFC  chips is  about 50  pixel wide.   The position  of  the readout
  amplifiers  used in  the default  configuration and  the approximate
  size and position of the HRC Fastie finger are also shown.}
\label{wfc_hrc}
\end{figure}

  Fig.~\ref{wfc_hrc}  shows a schematic  of the WFC focal  plane array
  and readout amplifiers.   The WFC focal plane array  consists of two
  mosaiced CCDs.  The  imaging area of each of  the two chips consists
  of  4096 columns  each containing  2048  pixels.  Each  pixel is  15
  $\times$ 15 $\mu$m  in size.  The serial register,  along one of the
  4096 pixel edges  of the device, is split in the  middle so that the
  entire device  can be  read out  through either one  or both  of the
  amplifiers that  terminate each end  of the serial  register.  There
  are 24  additional extended pixels (physical overscan)  on each side
  of the serial readout register, just before each amplifier.  The WFC
  raw data  also contain  a virtual overscan  of 20 rows,  obtained by
  over-clocking after  the readout of the  last CCD row.   In order to
  minimize the readout time and  reduce the number of serial transfers
  the default configuration is a 2-amp readout.  Hence, each amplifier
  is used  to read a single  2048 $\times$ 2048  quadrant.  WFC-1 uses
  amplifiers A and B, and WFC-2  uses the amplifiers C and D.  The WFC
  detectors are  operated in  Multi Pinned Phase  (MPP) mode.   MPP is
  used  to hold  the signal  charge away  from the  surface Si-Si0$_2$
  interface,  thus  reducing  dark  current rate,  minimizing  surface
  residual  image   effects  and  providing   better  charge  transfer
  efficiency.  The WFC detectors are  operated in MPP mode only during
  the  integration  and  not   during  the  readout.   Ground  testing
  demonstrated  that  this configuration  allows  a  higher full  well
  capacity and provides better cosmetics.

 The  physical separation  between the  two CCDs  is equivalent  to 50
 pixels, which  corresponds to $2.5''$.   The structure of the  data and
 data  products  of the  ACS  calibration  pipeline  (CALACS; Hack  \&
 Greenfield  2000) is  based  on the  previous  HST instruments'  file
 format and consists of multi-extension FITS files which store science
 (SCI), data quality (DQ) and error (ERR) arrays.  Since WFC data come
 from two CCD chips, they  are treated as separate observations with a
 specific SCI,  DQ and ERR  array for each  chip, and with  both chips
 being stored in the same FITS  file. Thus the array SCI-1 in the FITS
 extension [1] stores data from WFC-2 and the SCI-2 array, in the FITS
 extension [4] store  WFC-1 data.  A more detailed  description of the
 ACS file  structure is available in  ADHv4. At the end  of the CALACS
 steps, when  WFC data is drizzled to  combine dithered\footnote{For a
   detailed description  of the  advantage of using  a camera  in {\em
     dithering}  mode and  of {\em  drizzling} as  image reconstruction
   technique  see Fruchter  \& Hook,  2002.}  exposures  and/or remove
 geometric  distortion,  both chips  are  included  in  a single  FITS
 extension.

\subsection{HRC Data} 
 The imaging area of HRC consists  of 1024 columns each of 1024 pixels
 21 $\times$ 21  $\mu$m in size. This CCD  has four output amplifiers,
 one at each end of the two serial registers.  The array can therefore
 be divided  into 512 $\times$  512 quadrants to minimize  the readout
 time.   The  default configuration  is  a  single amplifier  readout,
 namely amplifier C,  which was selected for its  lower read noise and
 better  cosmetics.   The HRC  CCD  is  operated  in MPP  mode  during
 integration  time  and readout.   There  are  19 additional  extended
 pixels  (physical  overscan)  on  each  side of  the  serial  readout
 register.  The HRC data also  contains a virtual overscan, 20 rows on
 the top of the chip, obtained by overclocking the readout of the last
 CCD  row. The  structure of  the  HRC data  is simple;  a single  HRC
 exposure comes from a single chip and has only three data extensions.
 One particularity of all HRC  images is the presence of a finger-like
 shadow departing  from one edge  of the image  (Fig.~\ref{wfc_hrc}).
 This  is the shadow  of a  fixed occulting  mask called  the ``Fastie
 finger'',  a  0.8\arcsec\, wide,  5\arcsec\,  long reflecting  finger
 permanently located at the  CCD entrance window.  This mask, together
 with the two coronographic circular  stops that can be flipped in and
 out of  the light path,  constitutes the coronographic  capability of
 HRC (Krist et al. 2003a).

\section{ACS DATA TREATMENT} \label{pipe}

ACS raw data is processed  by the ACS calibration pipeline (CALACS) at
the STScI.  A detailed description of the individual calibration tasks
in CALACS is available in ADHv4.  Briefly the CALACS pipeline consists
of the following steps performed on the raw CCD scientific data: 

\begin{itemize}
\item{} Subtraction  of the  bias level  
\item{} Subtraction of a superbias frame
\item{} Subtraction of a superdark frame
\item{} Application of a flat field
\item{} Cosmic ray rejection of combined  CR-split observations
\end{itemize}

The SITe  CCDs are very stable;  ground testing have  shown that their
response  does not  change over  time  or after  thermal cycling.   As
consequence the flat fielding strategy for ACS is the same adopted for
WFPC2 (Holtzman  et al.1995a) which relies on  well characterized high
signal-to-noise flat fields acquired through all filters during ground
testing.   Second  order adjustment  of  the  ground  flat fields  are
performed  with  limited  on-orbit  observations  (see  \S  \ref{ff}).
Additional   calibration  steps   are  performed   using  MultiDrizzle
(Koekemoer et al.  2002)\footnote{Until September 2004 PyDrizzle (Hack
  et al.  2003) was used  instead of MultiDrizzle.}  which applies the
correction  of the  geometric distortion  of non-dithered  images, and
converts  the  data  units  from electrons  to  electron-rates.   When
multiple  dithered observations are  available, MultiDrizzle  runs the
Drizzle  algorithm (Fruchter  \& Hook  2002) to  register them  into a
single product.  In the following, we briefly describe the main steps,
as  well   as  relevant  instrumental  effects  not   treated  by  the
processing.

\subsection{Subtraction of the Bias Level}
{\bf WFC} - Normal WFC images  are read out using all four amplifiers,
therefore  each   2K  $\times$  2K   quadrant  needs  to   be  treated
independently in the calibration process.
All  four  amplifiers  produce  a  horizontal  ramp  in  the  physical
overscan, the region that serves to correct the bias level; it extends
up to 18 columns toward the active area (Sirianni et al.  2001).
 Consequently,  for  accurate  bias  subtraction, only  the  last  six
 columns  of the  leading  physical overscan,  those  adjacent to  the
 active area,  are used by CALACS  to estimate the bias  level of each
 amplifier.    In  CALACS,   the  {\em   doblev}  routine   creates  a
 sigma-clipped  average column  of the  selected area  of  the leading
 physical  overscan,  performs a  linear  fit  along  this column  and
 subtracts the bias  level row-by-row using a single  fitted value for
 each.   The  virtual  overscan  is  not used  to  estimate  the  bias
 level. This region shows large  scale structure which is quadrant and
 gain dependent. This structure, due  to the bias fixed pattern noise,
 makes  the  use of  the  virtual  overscan  in an  automatic  routine
 problematic.  Moreover,  this region can be  contaminated by deferred
 charge due to degradation in parallel charge transfer efficiency (see
 \S \ref{CTE}).

 Analysis of the active area in bias frames shows small differences in
 bias level between the leading  physical overscan and the active area
 (Sirianni  et al.  2002b). Ideally,  no  offset would  be seen.   This
 offset varies from  amplifier to amplifier and it can  be as large as
 3.5 analog-to-digital converter units (hereafter, ADU).

 In principle, if the residual offset between the imaging area and the
 overscan  region were  always  of the  same  amplitude, a  full-frame
 superbias subtraction  (see next  section) would remove  any residual
 difference.   Unfortunately, the  offset  is not  constant but  shows
 random variations of the order of  a few tenths of an ADU.  The cause
 of the residual offsets and their variation with time are still under
 investigation.   They occur  at a  level below  what could  have been
 tested on  the ground.   They are likely  the result  of interference
 between  the  WFC integrated  electronics  module  and the  telescope
 and/or scientific  instruments.  The non-MPP readout  mode could also
 play a role.  The accuracy of  the bias level subtraction in a single
 quadrant is limited by this random effect.

The difference between  these two bias levels, i.e.,  the amplitude of
the residual  offset level, varies between quadrants.   Since the bias
level is measured in the leading physical overscan and subtracted from
the active  area of each quadrant,  the center of  the resulting image
shows a ``jump'' between two adjacent quadrants,
This  offset  is  present  in  all  calibration  and  science  frames.
Therefore science images, after bias  and dark subtraction, may show a
quadrant-to-quadrant  jump  as large  as  a  few  ADUs (see  Fig.~1  in
Sirianni  et al.   2002b).   For point  source  photometry, the  local
background  is  typically subtracted  in  an  annular  region, so  the
residual  offset   has  essentially   no  impact  on   the  integrated
magnitudes.  However, if  the objects fall right across  the jump, the
photometry  should  be  considered  as suspect.   Similarly,  extended
objects  may spread  over  two  or more  quadrants,  so their  surface
brightness profile will suffer from the residual offset.  More studies
are  in  progress to  better  characterize  this  problem and  find  a
solution  or   develop  a  correction  to  apply   directly  into  the
calibration pipeline.  At  the moment we suggest to  fit the sky level
in each quadrant separately.  A  stand alone code, based on the STSDAS
{\em pedsky} task has been written  for the first order removal of the
bias offset in  observations of sparse fields (Sirianni  et al. 2005).
This script  first reverses the multiplicative  flat field correction,
then solves iteratively for the bias differences between the quadrants
and removes them, before re-applying  the flat field.  This script has
been used to produce sky-flats and for the data reduction of the Ultra
Deep Field.  It will be made  available in the future on the ACS STScI
Web site.

Finally, WFC  images may  show apparent amplifier  cross-talks between
the  four amplifiers, which  produces very  low level  dark  ``ghosts''
evident as  mirror images  of bright objects  in the  other quadrants.
The  amplitude of  this  effect is  usually  below any  scientifically
significant level,  and shows dependency  on the background  level and
gain setting.  A more detailed  characterization of this problem is in
presented in Giavalisco (2004a, 2004b).

{\bf HRC}  - HRC images are  read out using only  one amplifier. 
As in the case of WFC the readout amplifier produces a horizontal ramp
in  the leading overscan  which extends  up to  10 columns  toward the
active area.   The bias  level is therefore  measured in the  last six
columns  of the  leading  physical overscan.   The  task {\em  doblev}
removes the bias level row-by-row  as for the WFC quadrants.  There is
not a  significant difference between  the bias level in  the overscan
area and the active area.

\subsection{Superbias Subtraction} 
 Even  after  the  bias  level  subtraction from  the  overscan,  some
 structure remains  in the bias frames.  This  structure includes real
 bias pattern and  also some dark current.  In  every image, including
 the  zero-exposure time bias  frames, dark  counts accumulate  for an
 additional time  beyond the integration  time, primarily in  the time
 required to read out the  detector.  Since WFC is operated in non-MPP
 readout,  the dark  current rate  during the  readout is  higher than
 during  integration  time.  In  order  to  remove  this structure,  a
 superbias is subtracted from each image.  Bias frames are being taken
 as part of  the ACS daily monitoring program  for each supported gain
 state.   Superbiases  are created  weekly  by  combining seven  daily
 frames  (Mutchler   et  al.  2004).    There  is  some   evidence  for
 intermittent  variations  in  the   WFC  bias  structure  at  sub-ADU
 levels. On  a few occasions,  some frames show  bias jumps in  one or
 more  quadrants that  can run  for several  hundreds of  rows.  These
 effects occur at few tenths of ADU level and they might appear in one
 or   more   quadrants  simultaneously.    The   probable  source   is
 interference  with  the  electronics   by  other  activities  on  the
 spacecraft.  At  the moment there  is no automatic detection  of bias
 jumps  within the  calibration pipeline.   Though bias  jumps  at the
 sub-ADU  level are not  important for  most applications,  they might
 exist in processed data and  we recommend a careful inspection of the
 data for their presence.
 
\subsection{Dark Subtraction}

CALACS  removes  the  dark  current  from  each  scientific  frame  by
subtracting a superdark frame.  Every day
four  1000  sec  long  dark  frames  are acquired  for  both  WFC  and
HRC. Every  two weeks  all the  daily darks are  combined in  order to
derive an  accurate average  dark rate for  pixels which do  not vary.
These include normal pixels and hot pixels, i.e., pixels that generate
a dark current higher than  the average.  Permanent hot pixels are due
to  native  silicon   lattice  imperfections  or  impurities.   During
on-orbit  operations, CCDs are  exposed to  radiation sources  such as
protons and neutrons that induce lattice damage and therefore generate
hot pixel. As a consequence the population of hot pixels changes on a
daily  basis.   Therefore,  the  STScI calibration  pipeline  uses  as
baseline a  bi-weekly superdark  which is updated  with the  hot pixel
population  of a  specific  day  (Mutchler et  al.   2004).  For  each
science image,  CALACS subtracts the  appropriate superdark multiplied
by  the exposure  time  and divided  by  the gain.   Any dark  current
accumulated during  the readout time  is automatically removed  by the
superbias subtraction.
 
Sirianni  et  al. (2004)  analyzed  the  evolution  of the  hot  pixel
population in  two years of on-orbit  operation from WFC  and HRC dark
frames.   The  number of  hot  pixels  increases  linearly with  time.
Pixels with  a dark current  greater than 0.08  electron/pixel/sec are
flagged has ``hot''  in the data quality array in  the FITS file.  For
this threshold Sirianni et al. (2004)  report a daily rate of about 47
and 800 new hot pixels  for HRC and WFC, respectively.  Most radiation
induced hot pixels can be annealed by elevating the temperature of the
CCD,  thus   allowing  the  reconstruction  of   the  damaged  lattice
structure.   Analysis of  the effectiveness  of the  monthly annealing
process  in  the first  two  years of  ACS  operation  shows that  the
annealing  rate  of  new  hot  pixels with  this  threshold  has  been
$\sim$82\%  and   $\sim$86\%  for  the  WFC   and  HRC,  respectively.
Therefore the population of permanent hot pixels is growing at a daily
rate of  about 6 new pixels  and 140 new  pixels for the HRC  and WFC,
respectively.  Currently (Apr 2005) about 0.9\% of the  pixels of the
WFC  and the  HRC CCDs  are  permanently classified  as ``hot''.   For
comparison,  a fraction of  1.5\% to  3\% of  the pixels  is typically
impacted by cosmic rays in a 1000 second exposure.
Hot  pixels could for  the most  part be  eliminated by  the superdark
subtraction.   However  since  the  dark  current  in  hot  pixels  is
dependent on  the signal  level (Janesick 2001),  this process  is not
perfect.  Moreover, there are indications that the noise in hot pixels
is  much  higher  than  the  normal  shot noise  (Riess  2002).  As  a
consequence,  since the  location of  hot  pixels is  known from  dark
frames,  they are  flagged  and discarded  during image  combinations.
Careful planning  of the  observations including a  dithering strategy
will remove the impact of hot  pixels on the science frame.  See AIHv5
and ADHv4 for the best strategy for hot-pixel removal.

\subsection{Flat Field} \label{ff}
Appropriate flat  fielding normalizes the  system illumination pattern
together  with pixel-to-pixel  non-uniformities  to generate  accurate
count rates which are independent of field position.  Since in any WFC
image each quadrant is read  out independently, the ability to combine
photometry from  different chips/quadrants depends on  the accuracy of
the cross-quadrant and cross-chip normalization.

Intensive  ground testing was  devoted to  the creation  of laboratory
flat fields  to use for reduction  of on-orbit data  by simulating sky
illumination of  the CCDs.  The  Refractive Aberrated Simulator/Hubble
Opto-Mechanical-Simulator  (RAS/HOMS)  was used  to  produce high  S/N
ratio flat  fields which include  both the low frequency  (L-flat) and
the  high  frequency  pixel-to-pixel  (P-flat)  structure  (Martel  \&
Hartig, 2001  and Bohlin et al.   2001).  The low  order structure of
the flats  over the field of view  is mostly due to  the variations in
the   sensitivity  of  the   CCD  detectors.    Thickness  variations,
non-uniform  doping  or  anti-reflection   coating  can  result  in  a
different  response  to  incoming  photons  across  the  device  area.
Moreover,  since the  conversion  of photons  to  electrons occurs  at
different  depths within  the CCD  photo-sensitive  silicon substrate,
depending on  the energy of the  photons, the ``cosmetics''  of a flat
field strongly  depends on  the wavelength (see  AIHv5).  Furthermore,
since  the filters  are close  to the  focal plane,  variation  in the
filter transmission also contribute  to the L-flat structure.  The HRC
CCD  shows a  typical sensitivity  variation  from center  to edge  of
$\sim$ 10\%, while the large format WFC minimum to maximum sensitivity
variation  is  $\sim$15\%.  To  normalize  to  unity,  the flats  were
normalized by the average number of  counts in the central 1\% in area
of the  image.  In  the case of  the WFC  frames, the WFC-2  image was
divided by the  WFC-1 central value, in order  to preserve the overall
sensitivity  difference  between the  two  CCDs  across  the gap  that
separates the two chips.

These ``ground'' flats contain  information about the variation in CCD
response  and  spatial  structure  in filter  transmission,  and  were
expected to differ from on-orbit flats principally by the ratio of the
real  HST+ACS  illumination  pattern  and  the  RAS/HOMS  illumination
pattern.  Comparison of the observed  count rates for the same star in
different positions across the  chip, after flat fielding, provides an
estimate of the residual errors of  the flat field and of the accuracy
of  the chip-to-chip normalization.   Early Servicing  Mission Orbital
Verification  (SMOV)  data,  both  of  individual  spectro-photometric
standards  and stellar fields,  suggested the  presence of  a residual
low-frequency flat field structure.   There are a variety of plausible
approaches to  calibrate and remove this effect.   However, the easiest
methods  to  implement  are  either  improper or  not  practical.   The
internal ACS calibration lamps  illuminate the back of the calibration
door, which  is placed in  the light path  when a calibration  lamp is
turned  on.  Hence,  the  photons  arrive at  the  detector through  a
different  light path  from science  observations, bypassing  the OTA.
This  results in  a significantly  different illumination  pattern for
internal  flats  relative to  a  uniform  ``sky".   So, while  internal
flatfield  lamps  exist,  they  are  only  used  to  monitor  temporal
variations.   Bright earth flats  prove to  be impractical  since they
saturate in even the shortest exposures except when the HRC or the two
narrow band filters are employed.  Even then, the resulting images are
highly  structured and  streaky in  appearance due  to  terrestrial and
atmospheric  features passing  through the  field of  view.   The most
obvious approach to correct for  a residual L-flat structure is to use
deep  images of  ``empty'' sky  fields.  However  the creation  of sky
flats with  an appropriate S/N ratio  requires a very  large number of
images in each  filter, and careful masking to  remove galaxies, stars
and other sources.  Such an effort is currently in progress.

In the  interim, a  faster approach  has been taken  by the  STScI ACS
group.  As  described by Mack et  al. (2002a), a  moderately dense star
field (NGC 104,  in most cases) was observed  at typically nine offset
positions  separated   by  hundreds  of  pixels,   resulting  in  flux
measurements of each star  in nine largely separated positions.  Since
the entire  field of view is well  sampled by stars, the  flux data is
fitted to yield  a low order (fourth order)  polynomial L-flat as well
as  the  flat-field  corrected  flux  of  each  star.   The  resultant
polynomial is  combined with  the high-spatial frequency  structure of
the ground flats  to make new 'P+L' flats as described  by Mack et al.
(2002b)  and van der  Marel (2003).   While the  original ground-based
P-flat  fields produced photometric  errors of  the order  of $\pm$5\%
from  corner to  corner.  The  new  'P+L'-flats yield  a more  uniform
photometric  response (within  $\sim$1\%)  for a  given  star for  any
position  in  the field  of  view.   As  more data  becomes  available
improvements will be possible with  the adoption of sky flats. Updates
will be posted in the STScI Web page.

Finally, for  the first  two years of  operation the ACS  filter wheel
movements were  accurate to  one motor step,  which leads to  an error
that can  exceed one  percent in the  flat fields over  small regions.
For  seven filters  on  the WFC  and six  on  the HRC  with the  worst
blemishes, flat fields are available as a function of the filter wheel
offset step  (Bohlin, et al.  2003).  CALACS automatically  selects the
appropriate  flat  field corresponding  to  the  offset  step of  each
observation.  On  March 2004 the  flight software has been  updated to
make the filter wheel movements completely accurate.

\subsection{Cosmic Ray Rejection}\label{crsplit}
As  any other  HST  instruments, ACS  is  subject to  cosmic rays  and
protons  from  the  Earth's  radiation  belts.   Cosmic  rays  usually
generate significant quantities of electrons in more than one pixel.  A
first characterization of the cosmic rays in ACS images (Riess, 2002a)
shows  that the  fraction of  pixels  affected by  cosmic rays  varies
between 1.5\% and 3\% during a 1000 sec exposure for both HRC and WFC.
As consequence great care must  be taken in planning and analyzing ACS
observation to minimize the impact of cosmic rays on science images.

Cosmic ray rejection is ACS frames can be performed in CALACS with the
task ACSREJ or with MULTIDRIZZLE.  ADHv4 describes in great detail the
two different techniques.   Small offsets, even of the  order of a few
tenths  of a  pixel  between pairs  of  images, can  lead to  apparent
changes near a bright  target that swamp expected Poisson differences.
If large  enough, these  offsets can cause  an incorrect  flagging and
rejection of the  center of stars, mistaking the peak of  a star for a
cosmic  ray.  This  problem  is particularly  severe for  undersampled
images.  Optimization  of the  cosmic-ray removal process  may require
fine tuning of the image alignment process (Mack et al.  2002b).

\subsection{Geometric Distortion} \label{geometric}
By  design, the  ACS  focal plane  arrays  are tilted  with respect  to
incoming  rays (20$^{\circ}$ for  the WFC  and 31$^{\circ}$  for HRC).
The result of this design is an image of the sky which is distorted in
two different ways:  1) the pixels are elongated  with a scale smaller
along the  radial direction of  the OTA field  of view than  along the
tangential  direction,  and  2)  the  pixel  area  varies  across  the
detector.

The  distortion  affects  astrometry  and  therefore  the  ability  to
register  images taken  at different  pointings.  Moreover,  since the
distortion  causes  pixels to  have  different  effective  areas as  a
function of their position, photometry  and the flat fielding are also
affected.  The geometrical area of each pixel is imprinted in the flat
field as well as  the photometric sensitivity.  PyDrizzle accounts for
these   effects  and   the   final  geometric   corrected  output   is
photometrically and astrometrically corrected.  {\em If, however, flat
  fielded images  are analyzed prior to the  geometric correction with
  PyDrizzle, the  effective area of  each pixel must be  accounted for
  when doing  integrated photometry}.  The  correct total flux  can be
recovered by  multiplying the  flat-fielded images by  the appropriate
pixel area  map. Once this correction  has been applied  and the image
transformed in electrons/sec the same zeropoint as applied to drizzled
data products may be used to transform to absolute flux units.

The calibration of  the geometric distortion in the  ACS detectors has
been discussed  in detail  by Meurer et  al. (2002,2003)  and Anderson
(2002). Dithered  observations of the  rich NGC 104 star  cluster have
been  used to  calibrate the  distortion.  A  fourth  order polynomial
solution was  found adequate for  characterizing the distortion  to an
average rms accuracy of $\simeq$  0.05 pixels for the WFC and $\simeq$
0.03  pixels for the  HRC over  the entire  field of  view of  the two
cameras.  Although the  main distortion characterization was performed
using  F475W data,  wavelength  dependency of  the  solution has  been
checked  with complementary observations  in F775W  (WFC and  HRC) and
F220W (HRC).  A marginal increase in  the rms in the redder filter was
observed for the  WFC, and little or no increase for  the HRC.  A more
significant increase in  the rms fit (up to  0.11 pixels) was observed
in the  UV HRC  filter F220W,  quite likely due  to the  elongation of
$0.1''$ of the UV PSF.  Similar elongation has been observed also in SBC
PSFs, and  can be attributed to  the optics of either  the ACS HRC/SBC
mirrors or the HST OTA (Hartig et al. 2003).

Until September 2004 the geometric distortion was removed by PyDrizzle
in CALACS using the correction table provided by Meurer et al.  (2002)
which provides an adequate solution for most of the applications which
require an accuracy of the  order of 0.1 pixels.  A more sophisticated
solution, which includes a wavelength dependency based on the Anderson
(2002) and Anderson  \& King (2004)  results, has been  implemented in
CALACS at the end of September 2004 as part of Multidrizzle (Koekemoer
et al. 2002).  The coefficients of the fit of the geometric distortion
and the scripts to transform  these coefficients into a pixel area map
are available on the STScI/ACS web page.

\subsection{Charge Transfer Efficiency} \label{CTE}

Charge transfer is one of  the basic steps of CCD operation. Electrons
are transferred from the pixel where they are collected to the readout
amplifier.  During the transfer of charge from one pixel to the next,
defects in the  silicon can result in traps  that remove small amounts
of  charge from  the charge  packet.  The  Charge  Transfer Efficiency
(CTE) measures the  effectiveness of the charge transfer  in a CCD and
it is  typically measured  as the fraction  of charge  transferred per
pixel transfer.

CTE degradation is well known to affect data from previous HST cameras
such as WFPC2 and STIS.   Several detailed characterization of the CTE
degradation and formulation of empirical correction formulae for WFPC2
and STIS have  been published (Whitmore et al.  1999, Stetson 1998,
Saha 2000, Dolphin 2000b, Goudfrooij and Kimble 2002).

The total  amount of  charge lost increases  with the number  of pixel
transfers.   Therefore even if  ACS has  a better  CTE per  pixel than
WFPC2, the impact  of CTE degradation is expected to  be larger in ACS
data given the greater number of transfers (2048 for WFC, 1024 for HRC
versus 800 for WFPC2).

During on-orbit operations, CCDs  are subject to radiation damage that
degrades their ability to  transfer charges.  CTE degradation can lead
to photometric inaccuracy  (the magnitude of an object  will depend on
the position on the chip), astrometric shifts (the shape of the PSF is
altered) and  decrease in  discovery (the brightness  of an  object is
reduced  and   deferred  charges  can   increase  the  noise   in  the
background).   A comprehensive  discussion  of the  type of  radiation
damage is beyond the scope  of this document.  Excellent review papers
on  the subject  can  be found  in  the literature  (e.g. Janesick  et
al. 1991;  Hopkinson 1991).  The principal factor  that influences the
charge  transfer is  the presence  of spurious  potential  pockets and
defects  in  the  silicon  lattice  structure.   These  defects,  more
generically  called traps,  act as  trapping sites  for  the transient
electrons: part of the electrons contained in the signal packet can be
trapped and if they are  released quickly, they rejoin the main packet
of charge,  but if they fall  out of the  traps at a slower  rate than
they  go into  the traps,  they  are left  behind.  The  net CTE  loss
depends  upon the trap  constants, their  relevant variables,  and the
number  density of  traps  (see Cawley  et  al.  2001  for a  detailed
description of the trap characteristics).

In order  to reduce the  impact of CTE  degradation, the ACS  CCDs, by
design,  incorporate  a  mini-channel,  a narrow  extra-doped  channel
inside  the  buried  channel.    The  higher  potential,  due  to  the
extra-doping, is  designed to store and transfer  small charge packets
($<$7-10  $\times 10^3$ electrons) entirely  within the mini-channel.
The possibility of interaction between the charge packet and the traps
in  the   buried  channel  is  therefore  minimized   resulting  in  a
significant improvement of the CTE at these low signal levels.

A poor CTE  has a strong impact on high  precision photometry.  One of
the major manifestations  of severely degraded CTE is  the presence of
tails of charge apparently streaming  out of stellar point sources, in
a direction away from the serial register.  These tails are due to the
release  of trapped  charges on  timescales slightly  longer  than the
readout clocking rate.  Not all  trapped charges are released within a
few milliseconds; some  traps release the charges in  much longer time
scales and contribute to the background noise of that same exposure or
a  subsequent one.   This complex  mechanism of  capture  and release,
which depends  on numerous factors (such as  temperature, signal size,
readout rate and background level), can result in non-linearity across
the device, reduced image quality,  reduced sensitivity, and a loss in
spatial resolution.   The effect of  poor CTE on extended  sources has
been  examined by  Riess  (2000)  by subtracting  pairs  of images  of
galaxies observed near and far from the output amplifier. As one would
expect,  the charge is  lost primarily  from the  leading edge  of the
galaxy image (the  one closer to the readout  amplifier). The trailing
edge does  not suffer significant loss  because the traps  it sees are
filled. It  actually might see an  increase in signal  due to deferred
charges.   The resulting asymmetry  of extended  objects may  not have
negligible  consequences   for  statistical  measurements   of  galaxy
morphology,  position angle, axial  ratio and  asymmetry concentration
(e.g. such as required for weak gravitational lensing studies of field
galaxies).   The net  loss to  a measured  galaxy magnitude  has small
dependence on  the size of  the aperture used;  it is usually  in good
agreement  with  the  photometric   loss  expected  for  point  source
photometry.

Several programs monitor the CTE  performance of the CCDs in ACS. They
are  based either  on  calibration images  taken  with internal  lamps
(Sirianni et  al. 2003; Mutchler et  al.  2005) or on  the analysis of
the  cosmic ray tails  in dark  images (Riess  2002). These  tests can
track variations in CTE on a timescale of a few weeks.  Monitoring CTE
degradation  is  fairly easy,  but  the  calculation  of a  correction
formula based on stellar photometry  and on the number of parallel and
serial transfers  in the  CCDs is more  difficult.  There  are several
aspects to take into account: background level, signal strength of the
source,  photometric  technique,  etc.   A  large amount  of  data  is
necessary to  constrain all dependencies.  Riess (2003)  and Riess and
Mack (2004) have presented  correction formulae to correct photometric
losses as a  function of a source's position,  flux, background, time,
and aperture size in ACS CCDs.

Riess (2003) found  that the observed photometric losses  in WFC (HRC)
range   from  0.0\%   $\pm$   0.2\%  ($\sim$1\%)   for  bright   stars
($\sim$$10^5$ electrons)  on a significant background  ($\sim$ tens of
electrons) to as much as 7\%-10\% (5-6\%) for faint stars (few hundred
electrons)  and  faint background  ($<$  5  electrons).  The  observed
dependence  of the parallel  transfer CTE  loss (YCTE)  versus stellar
flux  suggests  a  power-law   relation  which  was  utilized  in  the
correction  formula.   Such  a  formula  includes  dependence  on  the
position, stellar flux, background level and time of the observations.
At the moment the correction  formula does not include a dependence on
the number  of serial transfers.   In general the contribution  of CTE
degradation  in the  serial  direction  to the  total  charge loss  is
minimal  due  to  the  faster  readout  clocking  as  demonstrated  by
pre-launch testing.   While internal  calibration data already  show a
degradation even in  the serial direction, it is still  too weak to be
detected with  stellar photometry.  Future observations  will allow 
better constraints on the correction for CTE loss in both directions.

In the future,  the effects of CTE degradation  can be mitigated using
the  post-flash capability  included  in the  ACS.   A relatively  low
amount of charge is added to the general background and hence mitigate
the effects of  CTE degradation.  However the addition  of this charge
will  of course  elevate  the background  contribution  of the  noise.
Also, HRC can be operated in 4-amps readout, thus reducing the maximum
number of parallel transfer from 1024 to 512.

\section{THE ACS POINT SPREAD FUNCTION}\label{PSF}
  
A detailed analysis  of the image quality of the  ACS cameras has been
presented by  Clampin et  al. (2002) and  Hartig et al.   (2003).  The
latter  also  discusses the  stray  light characteristics,  describing
several  ghost  images   and  scattered  light  features.   Naturally,
observed  Point Spread  Functions (PSFs)  vary with  wavelength, field
position and time.  Variations within the FOV arise from a combination
of defocus,  coma, astigmatism and charge  diffusion.  Time variations
occur  from  focus  changes  and  from spacecraft  jitter  during  the
exposures.

\subsection{Spatial Variations}

In general, PSF field-dependent variations in the ACS cameras are less
severe  than in  other HST  cameras.  Krist  (2003b) performed  a very
detailed study of  the WFC and HRC field  dependent PSF variations due
to optical and detector  effects.  Residual optical aberrations reduce
the  flux within  the core  of the  PSF and  redistribute it  into the
wings, degrading the Encircled Energy (EE). ACS/WFC PSF core width and
ellipticity  variations across  the field  are  large enough  to be  a
concern in the case of very small aperture photometry. Therefore, when
performing PSF-fitting  photometry it is more appropriate  to assume a
spatially variable PSF when computing  a PSF model or fitting stars to
the PSF model itself.   In the smaller FOV of the HRC,  the PSF can be
essentially regarded as constant.

The  major cause  of alteration  of  the FWHM  across the  FOV is  the
blurring caused  by the detector  charge diffusion.  All ACS  CCDs are
thinned backside-illuminated devices,  therefore the electrodes are on
the  rear face  of the  chip. Since  the depth  at which  a  photon is
absorbed within the epitaxial layer is proportional to the wavelength,
blue  photons generate  electrons  at the  top  of the  photosensitive
layer.   If the  electric field  is  weak at  the depth  at which  the
electrons are generated, they may  move and be collected into adjacent
pixels.  Charge  diffusion is  expected to be  greater in  the thicker
regions  of the  CCD and  the  largest variation  in charge  diffusion
across the FOV are seen  at the short wavelengths. Thickness variation
in the  ACS CCDs  have been derived  from fringe flat  fields obtained
during  ground tests  of the  camera (Walsh  et al.  2002,2003). These
results show that  the CCD thickness varies between  12.6-17 $\mu$m in
the WFC and 12.5-16.0$\mu$m in the HRC.

Analytical modeling of the PSF in  any position of the FOV of both ACS
CCD  cameras is possible  with TinyTim  (Krist 2003c).   Krist (2003b)
developed a specific blurring  kernel to reproduce the observed amount
of  charge   diffusion  observed  and  these   corrections  have  been
implemented in the latest release of his software.

Dithering is a well established technique for HST imaging and has many
advantages such  as improving the  sampling of the PSF  and mitigating
the  impact of  hot pixels  and detector  blemishes (see  ADHv4).  The
choice of the kernel for drizzling can change the shape of the core of
the  PSF.   All  our  data  were processed  with  the  default  linear
kernel, but a different  kernel is possible depending on the
data and scientific objectives.  For  example, if the square kernel is
applied  to a  single  image, with  the  parameters set  to perform  a
bilinear interpolation, the output image will have strongly correlated
noise  and also degraded  resolution.  In  all cases  where correlated
noise  is a  major concern  more  different drizzle  kernel should  be
considered.  For example, the  ACS Science Team pipeline (Blakeslee et
al.  2003a)  processes all Guaranteed-Time-Observation  (GTO) data with
the  Lanczos3  kernel  which  produces   a  sharper  PSF  core  and  a
significant  decrease in  the apparent  noise correlation  in adjacent
pixels.

\subsection{Encircled Energy Curves} \label{ee}
In order to  measure the encircled energy profile,  i.e.  the fraction
of total  source counts as a  function of the aperture  radius we used
the very high S/N observation of the spectrophotometric standard stars
used for  the zeropoint determinations  (see \S \ref{zp}).   All stars
are near the center  of the chip. In the case of  the WFC the star was
first centered  in WFC-1 and then in  WFC-2. In all cases  a couple of
images have been taken for cosmic ray rejection.  The images have been
processed with  CALACS and the  geometric distortion has  been removed
with PyDrizzle.

The  PSF  of any  instrument  extends  essentially  to infinity.   The
far-field PSF can be measured  only with very high S/N observations of
isolated stars. Still,  it is impossible to measure  the flux beyond a
certain radius  without incurring systematic errors  due to background
determination, large scale flat  field residual errors, nearby objects
in  the field  or detector  edges.  We  have chosen  to  normalize the
encircled energy curves  to the light at a  $5\farcs5$ circular radius
aperture.  Analysis of PSFs of  saturated and unsaturated stars in all
filters shows that  this aperture is a safe  assumption to measure the
total flux  within an ``infinite''  aperture.  It is also  the maximum
radius allowed for  most of the images with stars  centered in the HRC
FOV due to the presence of the Fastie finger.

One  of the  most difficult  task for  the proper  measurement  of the
encircled energy  at large apertures  is the determination of  the sky
background.   In order to  minimize the  impact of  errors in  the sky
determination, in all cases where multiple observations are available,
we used the following technique.   First, using the information in the
Data Quality  (DQ) array,  we masked out  the defective pixels  in the
SCI.   This allowed  us  to exclude  bad  pixels or  pixels masked  by
aperture features (such  as the Fastie finger in  HRC) when performing
statistics on the sky and  the photometry.  Then, for each observation
and for  each filter,  we determined  the mode sky  level in  the sky
annulus (from 6\arcsec\, to 8\arcsec \, for WFC and from $5\farcs6$ to
$6\farcs5$ for  HRC) and  used this value  to derive the  total counts
inside the $5\farcs5$ radius  aperture.  Under the assumption that the
flux of a  standard is constant and that for  large aperture the major
source of  error should be only  the Poisson noise,  we calculated the
median value  of all determinations  of total flux and  calculated the
appropriate sky level in each  image necessary to reproduce the median
flux level.   We then used the  new sky level to  perform the aperture
photometry with DAOPHOT in a wide range of aperture radii.

The mean  EE curves for different  broadband filters are  plotted as a
function of the aperture radius  in Fig.~\ref{halo_wd_wfc} for the WFC
and Fig.~\ref{halo_wd_hrc} for  the HRC.  
All curves are normalized to the mean value
at $5\farcs5$, however  in order to show in  more detail the structure
of  the  curves  at smaller  radii,  only  the  values for  the  inner
3\arcsec\  are plotted.   The  value  of the  EE  for the  broad-band
filters  with  errors  are  tabulated  in  Tables  \ref{eeWFC_WD}  and
\ref{eeHRC_WD}.   In order  to better  sample the  impact of  the long
wavelength halo in the near IR  (see next section) we also include the
narrow-band  filter F892N,  whose effective  wavelength  falls between
F814W  and  F850LP.  The  errors  in  these  tables are  the  standard
deviations from the different observations.

There are some variations between different observations. The main one
is a commanded  focus change operated with a  movement (3.6 $\mu$m) of
the HST secondary  mirror on December 2, 2002.   This focus change has
slightly modified the PSF in all wavelengths, but the most significant
variation is seen  in HRC-UV PSFs which show  a slightly sharper core.
Normal thermally-induced  variations of the HST  focus (breathing) may
typically  produce,  within a  single  orbit,  a  displacement of  the
secondary mirror comparable to the  one commanded in December 2002. We
therefore chose to average the  EE profiles before and after the focus
change  to create  the mean  EE curves.   The larger  errors  at small
aperture radii  in the  EE curves of  the HRC/UV filters  reflect this
variation in  the PSF.   On the  other hand, the  average PSFs  at the
center  of WFC-1 and  WFC-2 are  remarkably similar  and we  decide to
combine the two datasets to improve the statistics.

Running PyDrizzle  with different kernels produces  PSF profiles which
are significantly different only in the core of the PSF (0-10 pixels).
The aperture corrections calculated with  the EE can be applied to the
photometric  data, regardless of  the choice  of kernel,  for aperture
radii larger than 10 pixels.

\subsection{Long Wavelength Scattered Light}\label{halo}
All ACS CCD detectors suffer from scattered light at long wavelengths.
They  are  thinned  backside-illuminated  CCDs  and  consequently  are
relatively  transparent  to  the  near-IR wavelength  photons.   A  16
$\mu$m-thinned  CCD  transmits $\sim$5\%  of  8000\,\AA\, photons  and
$\sim$85\%  of  1$\mu$m  photons.  In   the  case  of  SITe  CCDs  the
transmitted long wavelength light  illuminates and scatters in the CCD
header, a  soda glass  substrate. It is  then reflected back  from the
header's   metalized  rear  surface   and  re-illuminates   the  CCD's
frontside photosensitive surface (Sirianni et al. 1998).  The fraction
of the  integrated light  in the scattered  light halo increases  as a
function of wavelength.  The onset of the halos occurs at a wavelength
of $\sim$  7500 \AA.  The  HRC CCD scatters  the near-IR light  into a
broad halo centered on the target; the intensity of the halo increases
with the  wavelength and reaches about  20\% of the total  energy at 1
$\mu$m (Fig.~\ref{HRC_image}).

In order to rectify the long wavelength halo, the WFC CCDs incorporate
a special  anti-halation aluminum layer  between the frontside  of the
CCD and its  glass substrate (Sirianni et al.  1998). While this layer
is effective at suppressing the IR  halo, it appears to give rise to a
relatively  strong  spike  in   the  row  direction  toward  the  edge
containing amplifiers AC as illustrated in Fig.~\ref{HRC_image}.  This
feature, probably due to scatter  from the CCD channel stop structure,
contains  $\sim$20\%  of the  PSF  energy  at  1$\mu$m but  is  almost
insignificant at 8000\,\AA\ (Hartig et al. 2003).

\begin{figure}
\epsscale{.7}
\plotone{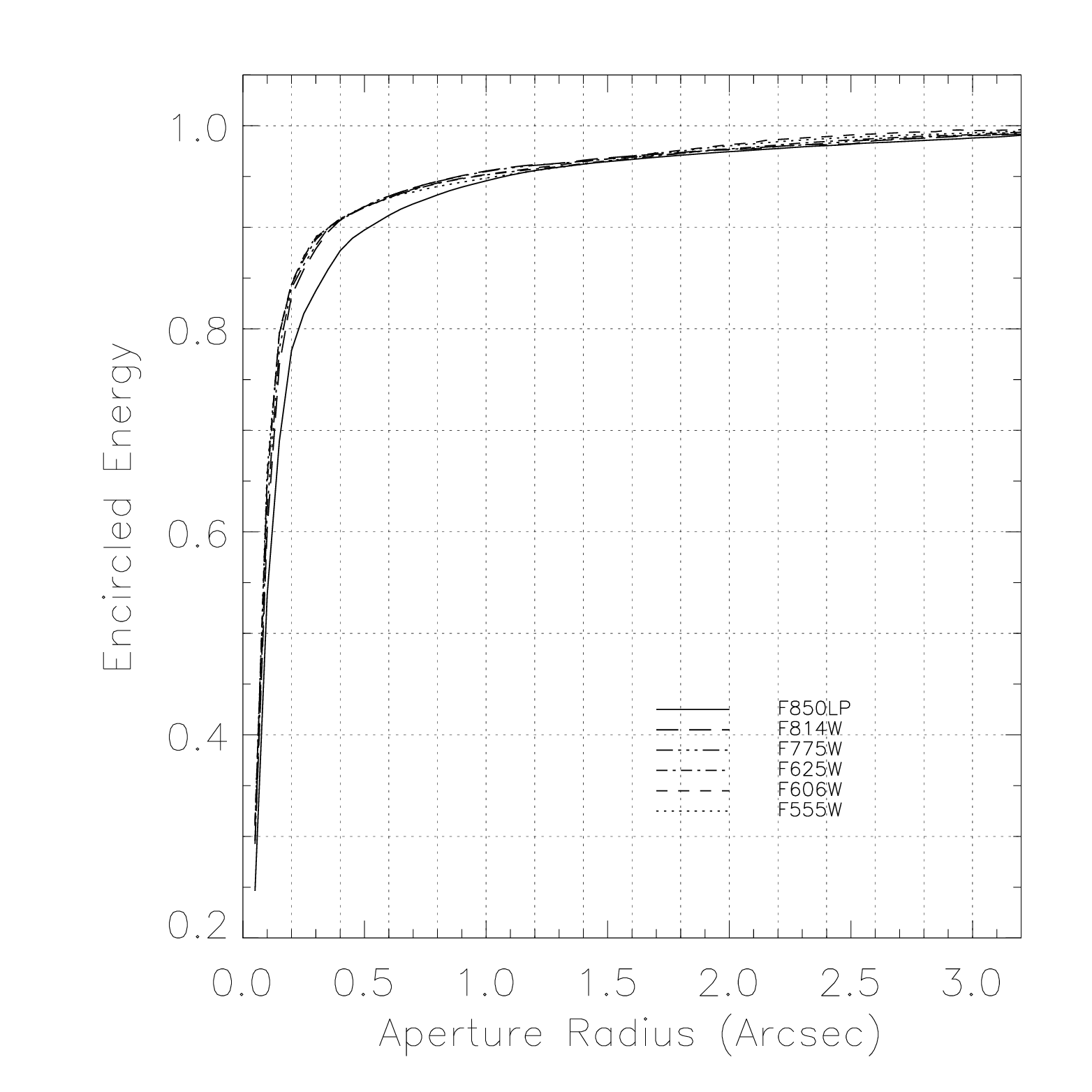}
\caption{Comparison of EE  profiles derived from multiple observations
  of two white dwarves in visual and near-IR filters of the WFC.}
\label{halo_wd_wfc}
\end{figure}

\begin{figure}
\plotone{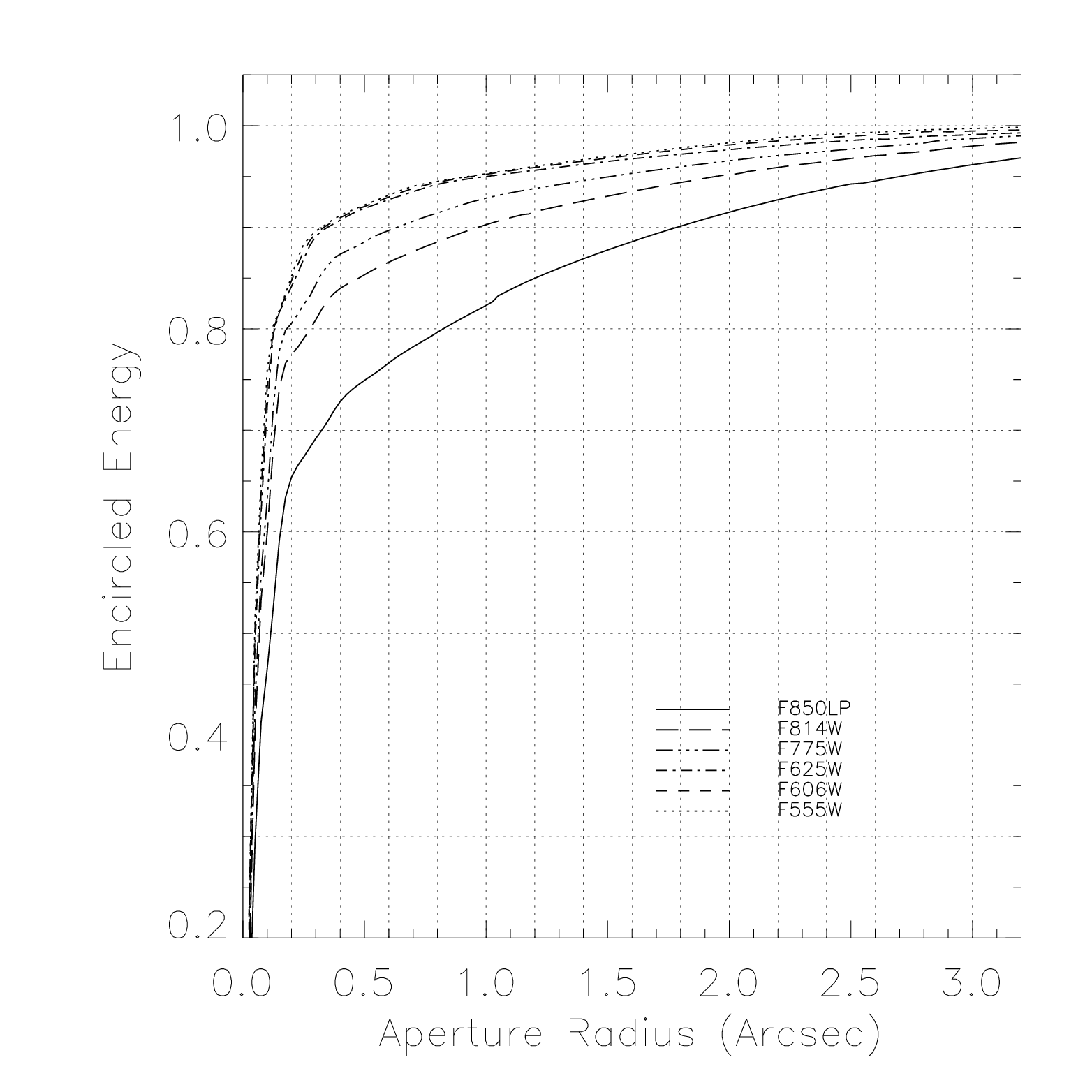}
\caption{Comparison of EE  profiles derived from multiple observations
  of two white dwarves in visual and near-IR filters of the HRC.}
\label{halo_wd_hrc}
\end{figure}

\begin{figure}
\plotone{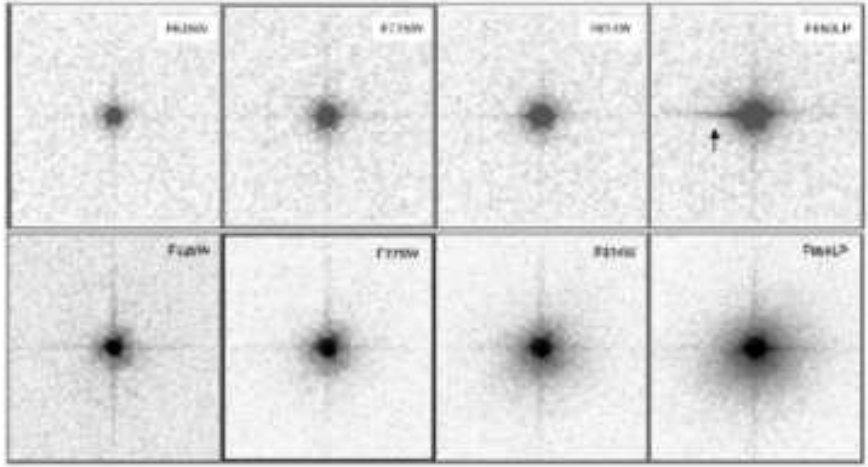}
\caption{WFC (top) and  HRC (bottom) images ($7.5''\times7.5''$) taken
  through the  near-IR broad-band  filters showing the  long wavelength
  scattered halo. The halo grows  considerably in the HRC from left to
  right.  The  anti-halation aluminum  layer  of  the WFC  effectively
  reduces the halo in the WFC but  it gives rise to a spike in the row
direction toward the edge containing amplifiers AC as indicated by the arrow.}
\label{HRC_image}
\end{figure}

The  EE   profiles  change  naturally  with  the   wavelength  in  any
diffraction limited optical  system, however at the onset  of the long
wavelength  halo,  the  PSF  becomes  broader  than  expected.   
Figs.~\ref{halo_wd_wfc} and  \ref{halo_wd_hrc} show the variation  of the EE
profile for  the standard  stars as  a function of  the filter  in the
visible and near-IR for the WFC  and HRC, respectively. In the case of
the WFC,  the long  wavelength halo is  effectively suppressed  by the
anti-halation  aluminum layer;  the EE  profiles of  all  filters from
F555W to  F814W overlap nicely  and only the  F850LP EE shows  sign of
broadening with $\sim$ 2.3\% less flux in a $0\farcs5$ radius aperture
than in the same aperture in F814W.  The effect of the scattered light
is more  severe for HRC (Fig.~\ref{halo_wd_hrc})  where the broadening
of  the  PSF is  clearly  visible at  F775W  and  increases at  longer
wavelength.  With respect  to the flux contained in  a 0\farcs5 radius
aperture in  the F625W  filter, the HRC  F775W, F814W and  F850LP PSFs
contain respectively about 3.3\%, 8.5\%, and 16.9\% less flux.

\begin{figure}
\plotone{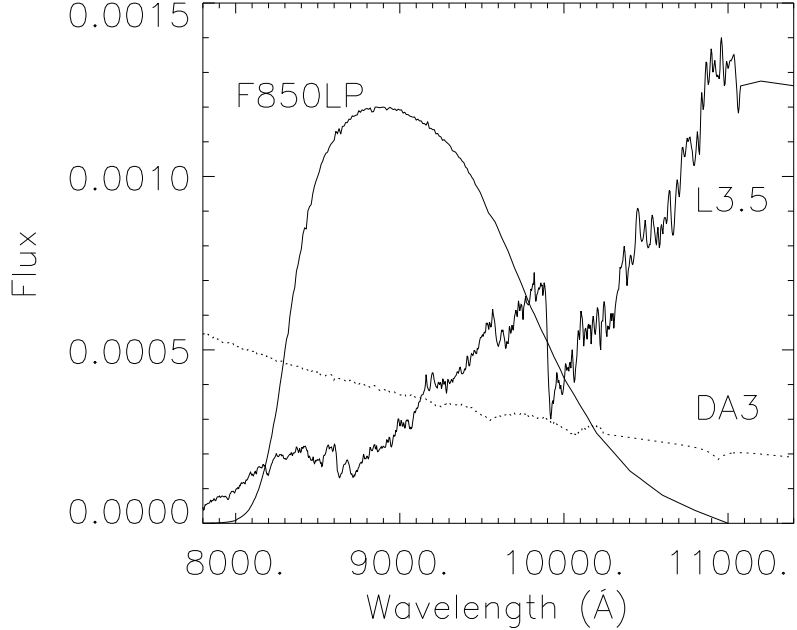}
\caption{SED of the  hot GRW+70 (DA3) star (dotted  line) and the cold
  2M0036+18 (L3.5)  dwarf star (solid line) normalized  to produce the
  same arbitrary  magnitude within the HRC F850LP  filter, also shown.
  The  wavelength-dependent  contribution to  counts  within the  same
  passband  changes  dramatically  for  these  stars.   The  effective
  wavelength is 9107\AA\ for GRD+70 and 9445\AA\ for 2M0036+18. }
\label{photons850}
\end{figure}

The same mechanism responsible for  the variation of the intensity and
extension  of  the  halo as  a  function  of  the wavelength  is  also
responsible for the variation of the shape of the PSF as a function of
the  color of  the source.   The spectral  energy distribution  of two
stars with  different colors is  different within any  given bandpass.
If we take as  an example the F850LP filter and we  observe a hot star
and a  cold star which produce  the same total counts  in this filter,
more long wavelength photons are collected from the red star than from
the blue star (see Fig.~\ref{photons850}).  As a result, the number of
photons transmitted by the F850LP filter that are scattered in the CCD
substrate is larger  for the red star than for the  blue star. The PSF
for the  red star will  therefore be broader  than the one  of the
blue star.  The effective wavelength is  a
source-dependent   passband  parameter   which  represents   the  mean
wavelength  of  the detected  photons  and  can  be used  to  estimate
\textit{shifts}   of   the    average   wavelength   due   to   source
characteristics   (e.g.,   temperature,   reddening   and   redshift)
and is defined as\footnote{This definition include the factor $\lambda$/hc to
convert the energy flux to photon flux as appropriate for photon-counting detectors.}

\begin{displaymath}
\label{efflam}
\lambda_{eff}\,\equiv\,\frac{\int{f_{\lambda}(\lambda)
P(\lambda)\lambda^2
d\lambda}}{\int{f_{\lambda}(\lambda)P(\lambda)\lambda d\lambda}}
\end{displaymath}
where $P(\lambda)$ is the dimensionless passband transmission curve, and 
\( f_{\lambda }(\lambda  ) \) is  the flux distribution  of the
object,  usually  in  \(  ergs \,\,  cm^{-2}s^{-1}$\AA$^{-1}\). 
$\lambda_{eff}$   can  be used  to  estimate  the  impact of  the  light
scattering on the aperture correction for any observation.
 
\begin{figure}
\epsscale{.7}
\plotone{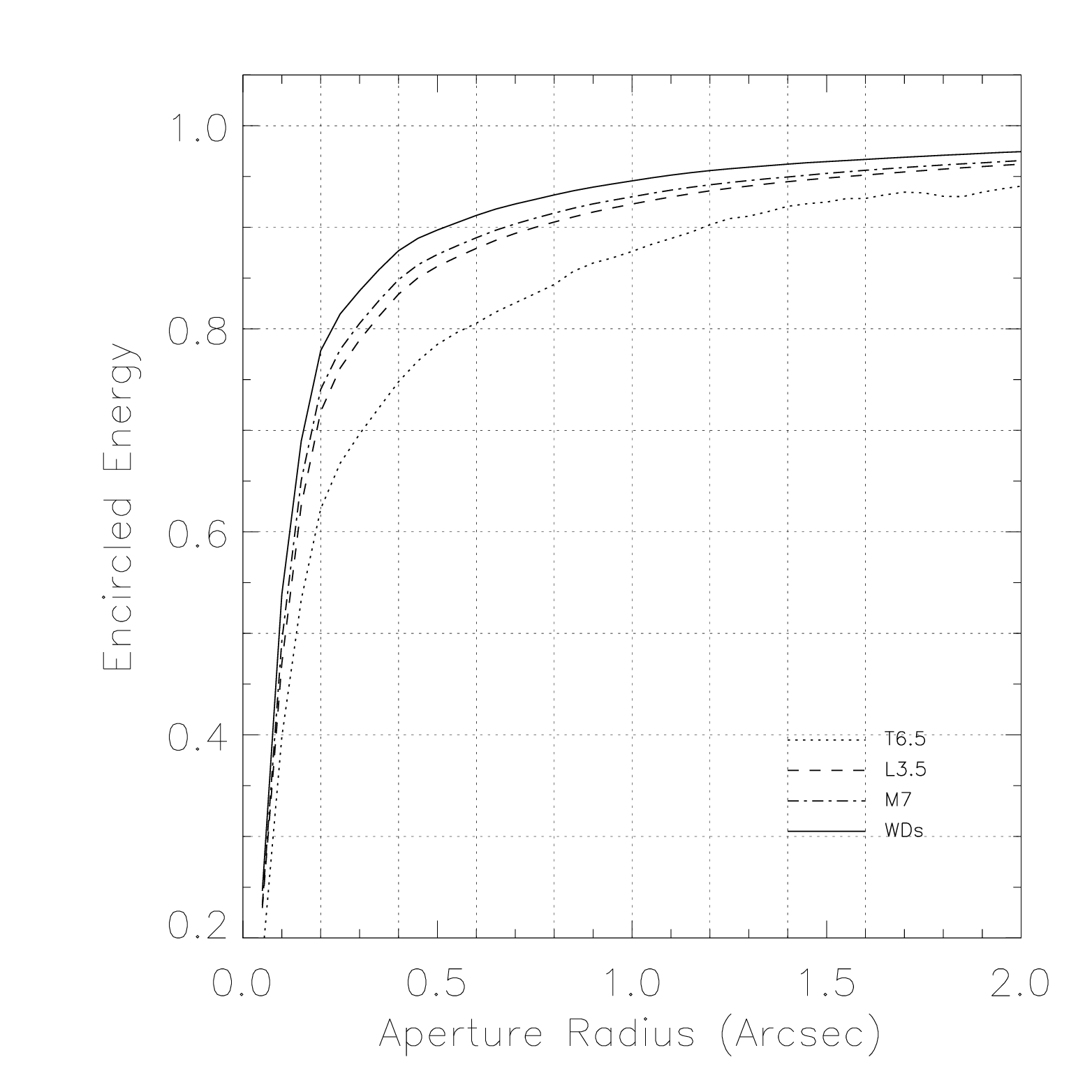}
\caption{Variation of  the EE in  the WFC F850LP filter  for different
  types of stars.}
\label{halo_all_wfc}
\end{figure}

\begin{figure}
\epsscale{.7}
\plotone{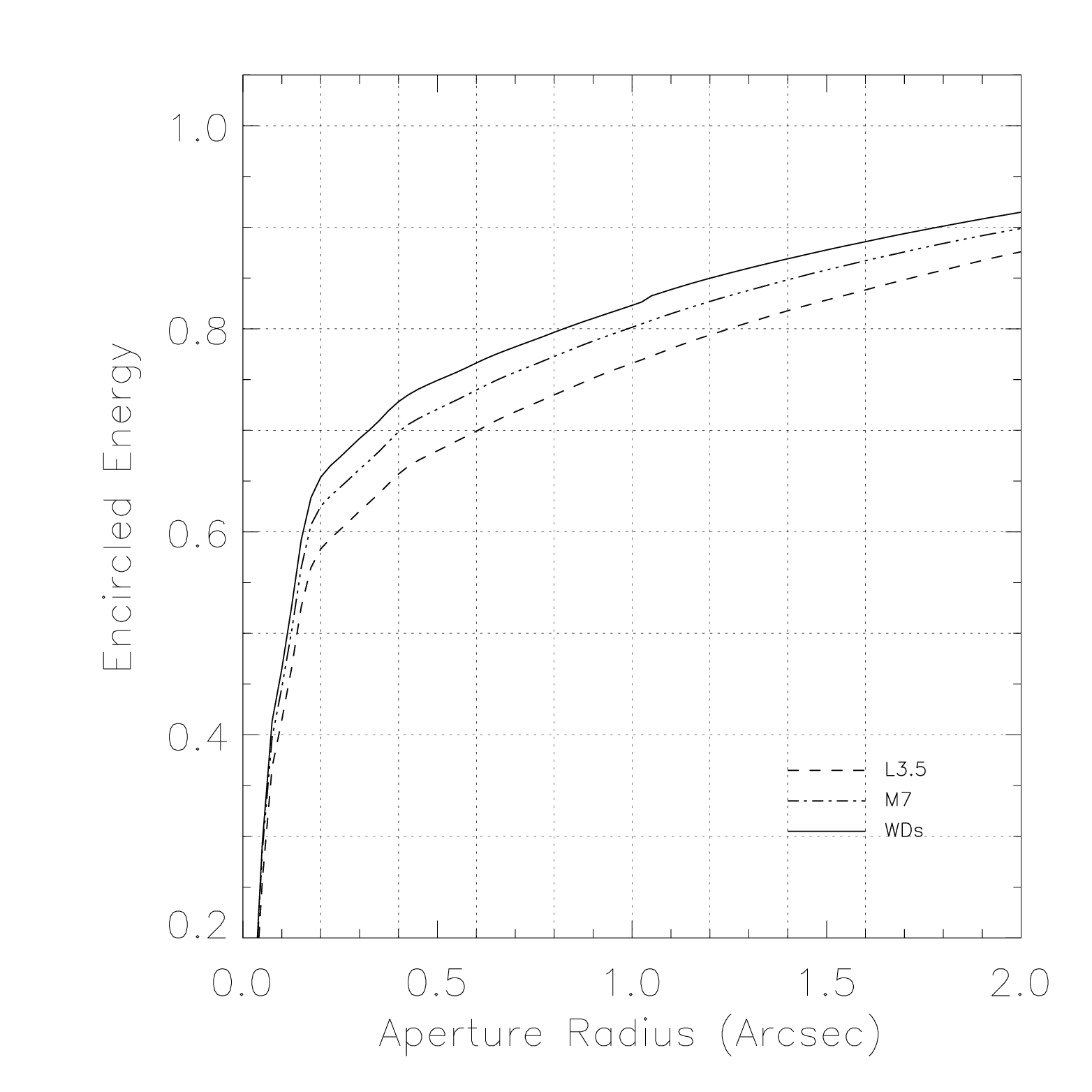}
\caption{Variation of  the EE in  the HRC F850LP filter  for different
  types of stars.}
\label{halo_all_hrc}
\end{figure}

In order to characterize the ACS throughput in the near-IR and the PSF
in  the  filters  affected by  the  long  wavelength  halo, a  set  of
intrinsically  red stars  have been  observed  with both  HRC and  WFC
(Gilliland \& Riess 2002).  We used these three stars with spectral types
M, L,  and T  to calculate  the near-IR EE  curves.  For  the M  and T
stars,  flux-calibrated  spectra are  available  from the  literature,
whereas  for   the  L-dwarf  a   STIS  spectrum  has   been  acquired.
Spectrophotometric data are used  to get information on the wavelength
distribution of the flux in  the near-IR filters, i.e. to calculate the
effective wavelength.   Gilliland \& Riess (2002)  compared the observed
flux of the three standard stars within an aperture of $2\farcs8$ with
the  prediction of  the ACS  exposure time  calculator and  provided a
provisional quantum efficiency correction  in the near-IR.  We derived
the EE  profile from these  stars using the  same data but, as  in the
case of the EE of the  standard stars, the curves have been normalized
to the counts in a $5\farcs5$ radius aperture. Figs.~\ref{halo_all_wfc} 
and  \ref{halo_all_hrc} show  how  the EE  profile changes  with  the  
spectral  type  in  F850LP for  the  WFC  and  HRC respectively.  We 
did not include the  curve for the T6.5 star in Fig.~\ref{halo_all_hrc}  
because it  was impossible  to obtain  an accurate profile  up to  
$5\farcs5$.  It  should  be noted  that although  Fig.~\ref{halo_wd_wfc} 
shows a broadening only  for the WFC F850LP EE, this
does not mean that only this filter can be affected by scattered light
at  long  wavelengths. Those  curves  have  been  determined from  the
observations of two spectrophotometric  standard stars, both hot white
dwarfs.  Although  its effect is greatly reduced  by the anti-halation
correction, light  scattering occurs  also when a  very red  object is
imaged  through   the  F775W  and  F814W  WFC   filters.   From  
Figs.~\ref{halo_all_wfc} and \ref{halo_all_hrc} it  is evident that for all
filters affected by the  long wavelength light scattering the aperture
correction will  be different for  objects with different  colors (see
\S~\ref{acorred}).

The first assessment of the  scientific impact of this PSF artifact in
the  near-IR was made  by Sirianni  et al. (1998)  and Gilliland  et al
(2002).  The presence of an extended  halo centered on the core of the
target has the obvious effects of reducing the signal-to-noise and the
limiting magnitude  of the camera in  the near-IR.  The  halo can also
have an impact on the  photometry on very crowded fields.  The effects
of the  long wavelength  halo should also  be taken into  account when
performing  morphological studies.  The  apparent size  of a  resolved
object in any near-IR filter  can increase with the intrinsic color of
the source.

In general, the effects of the  PSF red halo should also be considered
in doing surface photometry of extended objects, as significant biases
may occur.   For example,  Tonry et al.  (1997, 2001), found  that the
$(V-I)$  colors of  their ground-based  survey galaxies  measured with
thinned  SITe CCDs  (such as  used  in ACS)  were too  red by  several
hundredths of  a magnitude or more.   The bias was  attributed to this
fraction of the  I-band light of the standard stars  being lost to the
red halo.  Since the galaxy photometry was measured over larger areas,
they appeared too red with respect to the standards.  Reobservation of
the  galaxies  with  thick  CCDs  was undertaken  to  ensure  accurate
photometry.
 
In the case of ACS, we have calibrated the standard star photometry to
infinite radius,  so the integrated colors of  extended objects within
very large  apertures should  be correct.  However,  a bias  may still
occur  for  measurements  of   small  aperture  colors  or  gradients,
especially  for   the  HRC.   Michard  (2002)   has  done  experiments
convolving model early-type galaxies with empirical PSFs that included
the CCD red halo.  He found  that the bias was large enough to reverse
the sense  of some $(V-I)$  gradients; the effect on  colors involving
the $z$  band would be  even greater.  Michard  recommended convolving
each image with  the PSF from the other  bandpass before attempting to
measure galaxy color maps.  Blakeslee et al. (2003b) took the opposite
approach  in  measuring ACS/WFC  $(i-z)$  colors  of faint  early-type
galaxies  within the  galaxy  effective radii.   They deconvolved  the
images in each bandpass using empirical WFC PSFs and an implementation
of the  CLEAN algorithm (H{\"o}gbom  1974) that preserved  total flux;
the mean color correction for  their sample was $+$0.05 mag, and twice
this  for  the  most   compact  galaxies.   Thus,  there  are  various
approaches  to the  problem, but  proper care  is required  to achieve
accurate galaxy colors.
 
\subsection{Aperture Correction} \label{acorr}
In order to reduce errors due  to the residual flat fielding error and
background variations, and to  increase the signal-to-noise ratio, the
two  most  popular  photometric  techniques, aperture  photometry  and
PSF-fitting photometry,  are usually  performed by measuring  the flux
within  a  small  radius  around  the  center  of  the  source.   This
measurement  must be  tied to  the total  count rates  by  applying an
aperture  correction.  This correction  could be  the major  source of
systematic errors in the calibration of PSF fluxes.

For all the photometric calibration measurements, we decided to use an
aperture of $5\farcs5$  radius. This allowed us to  estimate the total
flux of spectrophotometric standard  stars and apply correction to the
response curve  of the instrument (see  \S\,\ref{DQE}).  Our zeropoints
are  based on  such  measurements  and therefore  they  refer to  this
fiducial aperture.  However  such a big aperture can  be used only for
bright  and  isolated sources  and  is  impractical  for most  science
observations.

A definition of zeropoints linked  to a smaller aperture is often more
convenient.  For  example, WFPC2 zeropoints are historically defined in
relation to  the flux in an  aperture of radius $0\farcs5$  (Holtzman et al.
1995a).   Using   a  non-infinite  zeropoint   implies  an  additional
correction to the zeropoint  for surface photometry.  This corresponds
to  the  aperture correction  from  $0\farcs5$  to  nominal infinity  which
amounts to 0.1 mag for  WFPC2, irrespective of filters (Holtzman et al.
1995a).  But  even a  $0\farcs5$ aperture is  impractical for  point source
photometry in  a number of  situations (photometry of  crowded stellar
fields,  faint  objects or  targets  with  an  uneven background)  and
smaller  apertures  need to  be  used,  with  the consequent  need  for
aperture  corrections.    In  all  these  circumstances,   it  is  also
impossible  to  measure  the  aperture  correction  from  any  smaller
aperture radius to ``infinity''.

A WFPC2-like  approach for the zeropoint definition  is less practical
for ACS  because, in  both HRC and  WFC, the aperture  correction from
$0\farcs5$ to ``infinity'' fluctuates from filter to filter with variations
up to  15\%.  In \S \ref{zp}  we therefore provide  zeropoints for the
fiducial aperture ($5\farcs5$), which can be used directly for surface
photometry.   This  choice is  also  dictated  by  the fact  that  the
photometric  calibration  defined by  STScI  and  returned by  SYNPHOT
traditionally refers  to an  ``infinite'' aperture, allowing  a better
conversion between point sources and extended sources.

As part of the ACS calibration we have determined the count rate conversion
from an intermediate aperture to ``infinity'.
   In  Table  \ref{ee05}  we  provide  the  aperture  correction  from
   $0\farcs5$  to ``infinity''  (AC05)  derived from  the EE  profiles
   listed  in  Tables  \ref{eeWFC_WD} and  \ref{eeHRC_WD}.  
  The offset between photometry within a smaller aperture and that for 
 photometry  with   a  $0\farcs5$  radius  aperture   can  be  easily
  determined and added to the value listed in Table \ref{ee05}.  This
  is  usually done by  measuring a  few bright  stars in  an uncrowded
  region  of the  FOV  and  applying such  offset  to all  photometric
  measurements.  If we define the observed magnitude as
\begin{equation}
OBMAG = -2.5 \times \log(total \,count \,rate \,\,[e^-/s])
\label{obmag}
\end{equation}
 then for point source photometry Equation \ref{obmag} becomes:
\begin{equation}
OBMAG = -2.5 \times \log(count\, rate \,at \, r\leq0\farcs5\,\, [e^-/s]) - AC05
\label{obmag2}
\end{equation}
where AC05 is listed in  Table \ref{ee05}. The AC05 values are derived
from the photometry of the two spectrophotometric standards stars used
in this paper. They are  both hydrogen white dwarfs and therefore are
not suitable for deriving the  aperture correction for stars with much
redder colors (see next section).

One of  the problems that must  be addressed is that  the variation of
the  PSF as  a  function of  position  on the  image  may require  the
application of different aperture corrections for different regions of
the FOV.  Of course, how  the photometry is carried out determines how
seriously  the  results  are   affected  by  the  space  and  temporal
variations of the  PSF.  If the aperture used is  large enough (about
$0\farcs5$  arcsec in radius),  then the  aperture correction  will be
both small and fairly constant.  When performing photometry in a small
aperture,  instead, the  assumption  of a  single aperture  correction
applied  to  all photometry  in  the entire  chip  may  lead to  large
systematic errors.   To minimize this effect, an  empirical mapping of
the aperture correction should be applied. An appropriate dataset will
eventually be available but meanwhile  it may be possible to compute a
suite of PSF models  using TinyTim. Unfortunately, the current version
of TinyTim does not model the residual long wavelength halo in the WFC
and  the HRC  PSF models  include only  a rough  estimate of  the long
wavelength scattered light.

Our determination  of the EE  profiles, and therefore of  the aperture
correction, is based on aperture  photometry and on the measurement of
the sky level  within an annulus far from  the object (from 6\arcsec\,
to 8\arcsec\, for  WFC and from $5\farcs6$ and  $6\farcs5$ for HRC) in
order to estimate the ``true'' sky level if a smaller  annulus is chosen, 
closer to the  center of the object,
the sky level will include more  light from the wings of the stars and
therefore the sky-subtracted counts will be lower.  From the tabulated
EE in  this paper  or from  PSFs created with  TinyTim and  assuming a
stable PSF, it  is possible to calculate the correction  to apply to a
measurement with any choice of sky  annulus to correct it for the true
sky level.

Finally,  we must  stress  that accurate  aperture  corrections are  a
function of time and location on the chip.  A blind application of the
data tabulated  in this paper  should be avoided, especially  at small
radii, and when using small apertures for the photometry, the aperture
correction should be derived for  each frame on which measurements are
made. Avoiding this necessary  step can introduce systematic errors in
the  photometry up  to  several percents.   The  tables provided  here
should be used to estimate approximate aperture corrections when it is
otherwise impossible  to determine such corrections  directly from the
image.

\subsection{Aperture Correction In The Near-IR} \label{acorred}
In order  to assess  when the broadening  of the  PSF due to  the long
wavelength  halo should  be taken  into  account and  to estimate  the
aperture correction for red stars we combined the information from the
EE profiles of all the stars used for the characterization of the long
wavelength  light  scattering.   We  first  calculated  the  effective
wavelength in  all filters for each  star. 
The effective  wavelength can  change significantly  within the
same filter; for instance, in  the WFC/F850LP there is a difference of
$\sim$760  \AA\ between the  effective wavelength of  the hot white
dwarfs  and  the  cold  T  dwarf.  Then  we  calculated  the  aperture
correction from a few selected  aperture radii to the nominal infinite
aperture using the EE profile of each star.  We finally correlated the
aperture   correction   and   the   effective  wavelength   for   each
filter/aperture combination  as shown in  Fig.~\ref{efflam_apt_wfc} and
\ref{efflam_apt_hrc} for WFC and  HRC respectively, for a few selected
apertures  (3,5,  and  10  pixel  radii  equivalent  to  $0\farcs125$,
$0\farcs25$,  and  $0\farcs5$,  for   WFC  and  3,5,10  and  20  pixel
radius,equivalent  to  $0\farcs075$,  $0\farcs125$,  $0\farcs25$,  and
$0\farcs50$ for  the HRC).  The  values for each star  are represented
with a  different symbol.  Not  all the stars  have a complete  set of
observations in all filters.  For each star we display only the points
relative  to  broad-band and  narrow-band  filters  for  which it  was
possible to determine  an EE profile up to  $5\farcs5$. For any choice
of aperture radii  there is a fairly smooth  variation of the aperture
correction as a function of wavelength.  For both the WFC and the HRC,
the aperture correction is fairly uniform in the visible but increases
in the blue  and in the near-IR.  The  increase of aperture correction
in the  blue is due to  charge diffusion (see \S  \ref{ee}), while the
increase  in the  red  is due  to  the long  wavelength  halo (see  \S
\ref{halo}).  Naturally,  the smaller  the aperture, the  stronger the
variation in  aperture correction  with the effective  wavelength.  In
Tables \ref{WFC_acl} and \ref{HRC_acl} we list the aperture correction
to the fiducial  aperture for a regular grid  of effective wavelengths
and several aperture radii obtained after interpolation of the plotted
data.

 In order to use these tables, the effective wavelength for a specific
 observation need  to be  calculated.  If an  empirical spectrum  or a
 model of the SED of the object is available, the effective wavelength
 can be  easily calculated for any  filter, for example  with the task
 CALCPHOT  in  SYNPHOT.  Table  \ref{app_efflam}  lists the  effective
 wavelength of  a sample of stars  of different spectral  types in the
 near-IR filters for WFC and HRC.

If  the SED  of the  object is  totally unknown  but 
observations in at  least two filters exist with the same  ACS camera, it is
still possible to get an  estimate of the effective wavelength.  First
the count rates of the  target should be measured  in both filters
using the same aperture radii and in such a way that  the instrumental color for
the selected apertures can be calculated:
\begin{displaymath}
 color_r \, =\, -2.5
\,  log((count\, rate )_{flt1,r} /  [count \, rate]_{flt2,r})
\end{displaymath}
where flt1 and flt2  are the two filters and {\em r}  is the radius of
the aperture where  both fluxes have been measured.   Then a $color_r$
{\em vs} effective wavelength relation can be created with SYNPHOT for
any  combination of  filters  and  aperture radii  using  an atlas  of
synthetic spectra  covering a large range of  spectral types.  Several
spectral atlases  are available in STSDAS  but any atlas  can be used.
Both  effective wavelength  and  $color_r$ can  be  computed with  the
routine CALCPHOT in SYNPHOT.  In fact, a new keyword ``aper'' has been
implemented to call for the encircled energy tables in the OBSMODE for
ACS.  This keyword allows one to  estimate the flux of a star within a
selected aperture and not just the total flux in an infinite aperture,
which is the default for SYNPHOT.
As  an example  we show  the $color_r$  {\em vs.}  effective wavelength
relation for  three near-IR  filters for the  WFC using  the $color_r$
F814W-F850LP and an aperture of r=$0\farcs5$ in Fig.~\ref{efflam_colIR}.  
In  order to produce  this plot, we  ran CALCPHOT
using  the Bruzual,  Perrson, Gunn  and Stryker  (BPGS)  stellar atlas
available  in STSDAS  to calculate  the effective  wavelength  and the
synthetic  color$_{r}$ with r=$0\farcs5$.   We also  plot in  the same
figure   the   position  of   the   five   stars  used for  the EE study. This 
approach  can be applied to  any
camera, color and aperture.  From plots of this type it is possible to
estimate the  effective wavelength of  an object and get  the aperture
correction  from  any  choice   of  aperture  radius  to  the  nominal
``infinite'' aperture from Tables \ref{WFC_acl} and \ref{HRC_acl}.

In  order to  better quantify  the impact  that the  near-IR scattered
light  has  on  ACS  photometry  we  present  two  cases  in  Appendix
\ref{halo_example}, the first  one for a high redshift  galaxy and the
second for point source photometry of an extremely red star.

\section{GAIN, LINEARITY AND SATURATION} 

Both WFC and HRC can  be operated at multiple gain settings, nominally
at  1, 2,  4 and  8 e$^-$/ADU.   However only  two settings  are fully
supported for each camera, gain 1 and  2 for WFC, and 2 and 4 for HRC.
On-orbit calibration  is not available for  unsupported gain settings,
and therefore  they will not be  covered in this paper  (see AIHv5 for
more information  on all available  gain settings).  The  default gain
values of  $\sim$ 1 e$^-$/ADU for  WFC and $\sim$ 2  e$^-$/ADU for HRC
have been used  to establish the adjustment to  the quantum efficiency
curves.

The absolute  gain value for the  WFC and HRC were  derived during the
ground  calibration  campaign  at  GSFC  in  the  spring/summer  2001.
Details on  the procedure and analysis  is available in  Martel et al.
(2001a, 2001b) respectively  for WFC and HRC. The  measured gain had a
typical  accuracy  of  0.6\%.   Bohlin  et al.   (2002)  improved  the
determination of the  relative gain between the pair  of amplifiers on
the same  CCD chip and the gain  ratio between the two  WFC chips. The
authors used pre-flight flat fields and maintained the same mean value
over all amplifiers  at a given gain setting  by imposing a continuity
constraint across quadrant boundaries on each chip and at the adjacent
edges of the gap between  the two chips.  This provided an improvement
in amplifier-to-amplifier gains to better than 0.1\%.

Since all the photometric calibration  data are taken with the default
gains, it  is important to  know the accurate ratio  between different
gain settings to transform the current calibration to non-default gain
observations.   Gilliland (2004) performed  an accurate  adjustment of
the major gain values relative  to the default gain levels through the
analysis  of repeated  observations in  the  same filter  of the  same
stellar  field,  at  different  gain settings.   Gilliland's  findings
improved  the relative  calibration  of gains  to  better than  0.1\%,
removing  errors that average  about 1\%  for both  HRC and  WFC.  The
corrections to the gain value for WFC  gain 2 and 4 and HRC gain 1 and
4 were implemented in CALACS on  January 6, 2004.  The new values (see
Table \ref{gainfix})  are being used  for the calibration of  any data
retrieved from the HST archive after that date.  Table\, \ref{gainfix}
shows the  corrective coefficient  to add to  the zeropoints  for data
processed before January 6, 2004.

Absolute errors of $\sim$ 0.6\% are still possible in the default gain
values.   However,  these  absolute  errors  have  no  impact  in  the
photometric calibration of ACS because the adjustments to the detector
quantum efficiency,  and therefore the zeropoints, were  based on data
acquired with the  default gain settings.  The use  of the appropriate
relative gain ratios between quadrants and between the default and the
supported  settings  will  assure  that  the  photometric  calibration
accuracy is maintained in all modes.  Since the output of CALACS is an
image already converted to electrons (FLT files) or electrons/sec (DRZ
files), there is no  need to differentiate the photometric calibration
for different  gains. For the current  paper we assume  that the image
has already been converted to electrons and therefore 1 ADU $\equiv$ 1
e$^-$.

The default gain settings for WFC  and HRC do not sample the full well
depth of  the CCDs.   Both WFC and  HRC employ  16-bit analog-to-digit
converters (ADC), which  can produce a maximum of  2$^{16}-1$ = 65,535
ADU.  The largest  number of electrons representable by  these ADCs is
therefore  given by $gain  \, \times\,  2^{16}-1$.  Any  charge packet
containing  more  electrons   than  $gain  \,\times\,  2^{16}-1$  will
therefore reach  digital saturation.  At the default  gain setting the
digital saturation  occurs for charge  packets of $\sim$  65,535 e$^-$
and $\sim$ 131,072 e$^-$ for the WFC and HRC, respectively.

The physical  full well of  the CCDs has  been measured on  the ground
with  flat  field  illumination  (Sirianni \&  Clampin,  1999a,b)  and
on-orbit  with observations  of  star fields  (Gilliland, 2004).   The
measured  full well  is $\sim$  84,000 e$^-$  for the  WFC  and $\sim$
165,000 e$^-$  for the HRC, with  variations of the order  of 10\% and
18\% in  the FOV for the  WFC and HRC  respectively (Gilliland, 2004).
Once the physical saturation is  reached, pixels lose their ability to
collect additional  charges.  The  additional charge will  then spread
into adjacent pixels along the  same column (blooming).  Since in both
cameras  the  digital  saturation   is  reached  before  the  physical
saturation, the use of higher gain settings can be beneficial for very
bright objects.

Gilliland (2004)  also analyzed the  linearity of the  response beyond
the physical saturation level for both ACS CCDs.  They found that with
the  use of  a gain  setting  that samples  the full  well depth,  the
response  of the  CCD remains  linear well  beyond saturation.   It is
therefore  possible   to  perform  $<$   1\%-accuracy  {\em  aperture}
photometry of  isolated saturated stars  by selecting an  aperture big
enough to contain  all pixels that received charge  from the saturated
pixels.  On the  contrary, if gain = 1 is set,  any information on the
charge  content  of pixels  that  reached  the  digital saturation  is
lost. In  some cases,  however, is still  possible obtain  an estimate
(with  an  accuracy  of  2-10\%  )  of the  magnitude  of  the  object
(Ma\'iz-Apell\'aniz, 2002).

\section{THE ABSOLUTE FLUX SCALE OF ACS}\label{DQE}
Considerable effort has been  devoted to the determination of accurate
response functions for the photometric passbands of ACS.  These curves
are the product of the  characteristic functions of all the individual
components in  the optical  path of the  combination of  telescope and
scientific instrument.   In particular, in  the case of the  ACS, this
includes the  reflection efficiency of the  optical telescope assembly
(OTA) and of the internal  ACS mirrors, the transmission efficiency of
the filters, the transmittance of  the dewar window(s) and the quantum
efficiency of the detectors.  This means that although the HRC and WFC
channels  share the same  filters, the  different optimization  of the
optics and detectors (see \S  \ref{intro}) can yield a quite different
response function for the same photometric passband.  These curves are
implemented in SYNPHOT, which is  used by the ACS pipeline to populate
the image header, to estimate exposure times for ACS observations, and
to provide  synthetic photometry calibration.   The sensitivity curves
are also  used to translate  physical quantities from  stellar models,
such  as isochrones,  into the  observational plane.   The  ACS system
response curves  represent the basis of the  ACS synthetic photometric
system.

Estimates  of  the  transmission,  reflectance, and  response  of  all
components of  ACS were made  on the ground  (see Ford et  al.  1998).
Modification to the pre-launch response curve, made by comparison with
the  observed count  rates of  spectrophotometric standard  stars, are
often  needed to  improve the  accuracy of  the  predicted sensitivity
curves.
 
The first check of the on-orbit sensitivity of the ACS was made during
SMOV (Sirianni et al. 2002b).  The throughput of the WFC was found to be
higher than expected  from ground measurements: from a  few percent in
the red up to $\sim$ 20 \% in the blue.  Similarly the HRC sensitivity
was higher  than predicted in  the visual and  in the red but  it also
showed an  unexpected dip in the  UV at $\sim$ 3200  \AA.  These first
observations were  used to derive rough corrections  to the pre-flight
sensitivity curves for a preliminary SYNPHOT update in August 2002.

Additional  observations  have  permitted   a  fine  tuning  of  these
corrections  with   an  average   improvement  of  $\sim$2\%   in  the
uncertainty.  The data  consist of an extended set  of observations of
the spectrophotometric standard  stars GRW +70 5824 and  GD71, both DA
dwarves, taken between  May, 2002, and January, 2003.   Of these, only
GD71  is a  primary HST  spectrophotometric standard.   Its  data were
collected in May,  2002, during the SMOV program,  in all filters with
both  the WFC  and HRC  cameras.  GRW+70  5824 was  observed  in July,
August and  September of  2002 through most  of the filters  with both
channels.  Additional observations were collected in the course of the
HST  cycle 11:  GRW+70 5824  was observed  though all  WFC  filters in
January,  2003,  and  all  HRC  filters in  February,  2003.   Finally,
observations  of GRW+70  5824 in  the  three bluest  HRC filters  were
gathered in December, 2002 and January, 2003. In all cases, WFC images
were  collected by reading  out a  512$\times$512 subarray  around the
target,  whereas   for  the   HRC  the  default   was  a   full  frame
configuration.  In order  to remove cosmic ray hits,  at each epoch, a
pair of images was obtained for each combination of camera and filter.
The default gain  settings were used for WFC and HRC,  namely 1 and 2.
A synopsis  of the observations  used, with the measured  count rates,
and a more detailed explanation of  the data reduction are given in De
Marchi et al. (2004).

All  the  data have  been  been  processed  with the  standard  CALACS
pipeline.  The analysis was carried out on the geometrically corrected
files.  Aperture photometry has  been performed on the reduced dataset
as described in \$ \ref{ee}.  

In order to  calculate precisely the total flux  of the standard stars
and check simultaneously the  stability of the photometric performance
of the  camera we chose to  perform aperture photometry  in all frames
from all  epochs in  a radius smaller  than the  ``infinite'' aperture
($5\farcs5$). We selected an aperture of 1\arcsec\ radius (20 pixels
for WFC,  40 for  HRC).  This  aperture is large  enough to  be fairly
independent of changes in the PSF  core with focus and position on the
chip, and  to include most of the  flux, yet not so  large that errors
are  dominated by  the background.  The  observed count  rates in  the
1\arcsec\ radius aperture were converted to an ``infinite'' aperture by
applying the aperture correction calculated from the mean EE profiles.

This analysis has also permitted a first check on the stability of the
photometry  with time.   Since  launch the  ACS  has proved  to be  an
extremely stable and repeatable  instrument.  In the selected aperture
radius the stability is better than  0.4\% for the WFC and better than
0.5\% for the HRC. The larger scatter  in the HRC is limited in the UV
where the image quality is  very sensitive to focus changes especially
in the PSF core.  No significant throughput variations are observed in
the UV (Boffi et al. 2004).

The observed count  rates were compared with the  ones predicted using
the currently  available sensitivity curves of the  HST+ACS system, so
that the latter could be  updated.  Given the photometric stability of
the instrument,  and in order to reduce  the statistical uncertainties
associated with  the measurement  process, all the  data for  the same
combination of  detector and filter  were averaged together.   We also
averaged the results of WFC-1 and WFC-2 because the variations between
the two are  consistent with the temporal variation  of the count rate
in each chip.  We therefore  use the standard deviation of the various
epochs/chips as a measure of the observational uncertainty.

Although it  is not possible to independently  measure corrections for
any of  the sensitivity  curves, we decided  to hold fixed  the mirror
reflectivity and  window transmission curves and  apply any correction
to  the  detector  responsivity  quantum efficiency  (RQE)  or  filter
throughput.   We assumed  the wavelength  dependence of  each filter's
throughput curves to  be as measured on the ground  and allowed for an
overall  scaling factor for  each filter.   Since the  same wavelength
range is  often covered by more  than one filter, it  was necessary to
simultaneously  compare observed  and predicted  count rates  over the
whole wavelength domain and  derive incremental sensitivity updates to
be applied  simultaneously to all  modes.  However, since  the filters
are shared  between the two  cameras, these scaling factors  have been
taken  as  the  average of  the  two  corrections  (see De  Marchi  et
al. 2004).  Finally, although we  decided to apply all the large scale
corrections to the  RQE curves, even if not  just detector-related, we
required the solution  to be smooth, without bumps  and wiggles.  This
procedure  has  been  repeated  iteratively  until  convergence.   The
derived RQE curves (Fig.~\ref{newdqeplots}) can now reproduce the count
rates in broad  band filters with an accuracy  better than 0.5\%.
The  new RQE curves  and the  corrected filter
curves were implemented in the ACS pipeline on December 10, 2003.  The
new files can be downloaded  at the STScI/ACS reference files Web page
and were included in the last release of STSDAS.

\section{ZEROPOINTS} \label{zp}
The photometric system based on the ACS filters is, by definition, the
natural photometric system  of ACS.  The knowledge of  the response of
the  instrument allows  us to  define a  synthetic  system.  Synthetic
photometry refers  to magnitudes and  fluxes computed from  a spectral
energy flux  distribution and the  response function of  a photometric
system.  The approach  taken here deals with the  determination of the
absolute flux scale of the ACS, as opposed to a conversion from ACS to
standard   filters.   The   transformation  between   ACS   and  other
photometric systems is discussed in \S \ref{transf}.

If $P(\lambda)$  is the  dimensionless passband transmission  curve in
the  interval [$\lambda_1,\lambda_2$],  the apparent  magnitude  of an
object with energy flux distribution $f_{\lambda}\,(\lambda)$ is
\begin{equation}\label{eqnn}
 m_P \:  = \:-2.5\, log  \,\frac{\int_{\lambda_1}^{\lambda_2} \lambda
f_{\lambda}  P(\lambda)  d\lambda}{\int_{\lambda_1}^{\lambda_2}\lambda
f_{\lambda}^{*}P(\lambda) d\lambda} + m_{P}^{*}
\end{equation}
where  $m_{P}^{*}$ is  a  known apparent  magnitude  from a  reference
spectrum   $f_{\lambda}^{*}$.    $f_{\lambda}^{*}$   and   $m_{P}^{*}$
completely   define  the   zero-point  of   a   synthetic  photometric
system\footnote{The energy flux distribution $f_{\lambda}$ in equation
  \ref{eqnn} is multiplied by $\lambda$/hc in order to convert it into
  photon  flux  distribution  as  appropriate  for  a  photon-counting
  detector.}.

Several  different photometric  systems can  be defined  for  the same
instrument.  For  example, the photometric systems based  on the WFPC2
filters are  the WFPC2  flight system and  the synthetic  system.  The
WFPC2 flight system  (Holtzman et al. 1995a) is  defined so that stars
of zero color in the Landolt  UBVRI system have color zero between any
pair  of  WFPC2  filters,  and  have  the  same  magnitude  in  V  and
WFPC2/F555W.  In the synthetic photometric system, the zeropoints were
determined so  that the magnitude  of Vega, when observed  through the
appropriate WFCP2 filters would be identical in the closest equivalent
filters  in  the  Johnson-Cousins   system.   This  system  was  later
implemented in the STSDAS SYNPHOT  package as the VEGAMAG system where
the zeropoints are defined by the magnitude of Vega being exactly zero
in  all filters.   The VEGAMAG  system, although  also related  to the
absolute  flux scale,  is designed  to resemble  closely  the standard
Johnson-Cousins system.

For  the ACS,  we  present  three different  versions  of a  synthetic
photometric system. The first one is the VEGAMAG system, the analog of
the   WFPC2  synthetic   system   implemented  in   SYNPHOT  and   the
standard-based system.  The other two, STMAG and ABMAG, are flux-based
systems.  Their zeropoints are defined in terms of a reference flux in
physical units  rather than  the flux of  Vega.  These  two flux-based
systems  define  an  {\it  equivalent  flux  density}  for  a  source,
corresponding to the  flux density of a source  of predefined spectral
shape that would produce the observed count rate.  The equivalent flux
is then converted  into a magnitude with the use  of a constant chosen
such that the  magnitude in the V band corresponds  roughly to that in
the Johnson  system.  The advantage  of the ABMAG and  STMAG magnitude
systems is that  the magnitude is directly related  to physical units,
and they are  therefore much simpler and cleaner.   The basic equation
in  their  definition  is  the   {\it  mean  flux  density}  per  unit
wavelength,  $f_{\lambda}$ (STMAG), or  per unit  frequency, $f_{\nu}$
(ABMAG). The  choice between  flux-based and standard-based  system is
mostly a matter of personal preference.

The zeropoint of an instrument,  by definition, is the magnitude of an
object that produces one count  per second. We use this definition and
SYNPHOT to calculate the zeropoints in this paper.  Any zeropoint must
refer to a count rate measured  in a specific way.  In this paper they
refer  to  a  nominal  ``infinite''  aperture  of  $5\farcs5$  radius.
Independently  of  the different  definitions,  the  magnitude in  the
bandpass P in any of the ACS systems is given by:

\begin{equation}
\label{mag}
ACSmag(P) =  -2.5\;log\,(total\, count\, rate \,[e^-/s]) + Zpt(P) = OBMAG(P) + Zpt(p)
\end{equation}
and  the choice  of the  zeropoint Zpt(P)  determines  the photometric
system of ACSmag.

\subsection{VEGAMAG  Magnitude System}  

The VEGAMAG  system makes use of  Vega ($\alpha$ Lyr)  as the standard
star.   Calibrated  empirical   spectra  are  available  covering  the
wavelength range from 3300\,\AA\ to 1.05 $\mu$m (Hayes 1985).  Bohlin
et al. (1990) have extended this spectral coverage down to 1150\,\AA\
using  IUE spectra.   Since its  initial release,  SYNPHOT  adopted as
reference  the synthetic spectrum  of Vega  computed by  Kurucz (1993)
models with  a normalization at V  = +0.034 mag, assuming  3670 Jy for
V=0.  Composite  spectra of Vega  have been constructed  by assembling
empirical and synthetic spectra (e.g.  Colina et al.  1996).  The most
recent absolute  calibration of Vega  from the far-UV  to the IR  is a
composite spectrum with IUE data  from 1140 to 1700 \AA\ and HST/STIS
from 1700 \AA\  to 4200 \AA\ (Bohlin \&  Gilliland 2004).  Longward of
4200 \AA\ the Kurucz (2003)  model of Vega is used. The normalization
has    been     done    to     the    observed    flux     level    of
$3.46\,\times\,10^{-9}\,erg\,cm^{-2}\,s^{-1}\,$\AA$^{-1}$     at     5556     \AA,
corresponding to  V=0.026.  Since this new spectrum  has been adopted
as the Vega reference spectrum in SYNPHOT, we decided
to use it to calculate the  zeropoints for the ACS VEGAMAG system.

In  the VEGAMAG system,  by definition,  Vega has  magnitude 0  in all
filters.  Since the zeropoints in  all VEGAMAG systems are tied to the
observed  Vega fluxes,  their  synthetic absolute  magnitude may  have
systematic errors (probably up  to 2\%).  Smaller errors, however, are
expected in the colors.  The  values of the VEGAMAG\_Zpt(P) are listed
in  Tables \ref{wfczpt}  and \ref{hrczpt}  for  both the  WFC and  HRC
filters.

\subsection{STMAG  Magnitude System} 

The  STMAG magnitude  system (Koorneef  et al. 1986) is  based  on the
 definition of mean flux density per unit wavelength as:
\begin{equation}
f_{\lambda}(P) \, \equiv \,\frac{\int f_{\lambda}\;P(\lambda)\lambda 
d\lambda}{\int P(\lambda)\lambda d\lambda}
\end{equation}
and  the selection  of  a flat  reference  spectrum in  $f_{\lambda}$.
$f_{\lambda}$ is expressed in $erg\,s^{-1}\,cm^{-2}\,$\AA$^{-1}$.  The
magnitudes in the STMAG system are given by
\begin{equation}
STMAG(P) =-2.5  \; log\,( f_{\lambda}(P)) + K
\end{equation}
 The zeropoint K is defined by  setting the magnitude of a source that
 has a flux density of 1 $erg\,s^{-1}\,cm^{-2}\,$\AA$^{-1}$ to -21.10.
 Another way to express this zeropoint is to say that an object with a
 flux                            density                            of
 $3.631\times10^{-9}\,erg\,cm^{-2}\,s^{-1}\,$\AA$^{-1}$    will   have
 magnitude Stag = 0 in every filter.

Historically, the spectral flux density per unit wavelength that would
generate  1 count/sec  (within the  nominal infinite  aperture defined
above) is stored  in HST image headers as  the keyword entry PHOTFLAM.
Calling PHOTFLAM(P) the value of  PHOTFLAM for the passband P, then to
convert directly count rates to average flux density:
\begin{displaymath} 
f_{\lambda}(P)=  PHOTFLAM(P) \,\times\,  (count \, rate).  
\end{displaymath}
where the count rate is in $e^{-} \,s^{-1}$.
The value of PHOTFLAM(P) can be derived directly by the
passband function $P(\lambda)$:
\begin{displaymath}
PHOTFLAM(P)\,=\, \frac{hc/A}{\int P(\lambda)\,\lambda\,d\lambda}
\end{displaymath}
where  A is  the telescope  collecting area  in square  centimeters. We
listed  the  value  of  PHOTFLAM(P)  for all  ACS  filters  in  Tables
\ref{wfilter} and  \ref{hfilters}.  These values are  also reported in
the photometry keyword  section in the header of  the SCI extension of
the calibrated ACS FITS file.

Thus, following  the standard
 definition of any zeropoint, the STMAG zeropoint is:
\begin{displaymath}
  STMAG\_Zpt(P)= -2.5 \; log\,(PHOTFLAM(P)) - 21.10
 \end{displaymath}
and the STMAG is therefore defined  as: 
\begin{eqnarray}
STMAG(P)& =&-2.5  \; log\,( f_{\lambda}(P)) - 21.10 \nonumber \\ 
        & =& -2.5\;log\,(totalcounts \,\times s^{-1}) + STMAG\_Zpt(P)
\end{eqnarray}
The  $STMAG\_Zpt(P)$ are listed in Tables \ref{wfczpt} and \ref{hrczpt}
for both  WFC  and HRC filters.

\subsection{ABMAG  Magnitude System} 

The ABMAG  magnitude system  is the analog  of STMAG for  a constant
flux density  per unit frequency  $f_{\nu}$ (Oke 1964).  The reference
spectrum      is     flat      in      $f_{\nu}$     expressed      in
$erg\,s^{-1}\,cm^{-2}\,Hz^{-1}$.  The  magnitudes in the  ABMAG system
are given by
\begin{equation}
ABMAG(P) =-2.5  \; log\,( f_{\nu}(P)) - 48.60
\end{equation}
The constant is  chosen so that ABMAG matches  the Johnson V magnitude
for  an object  with a  flat spectrum.   Another way  to  express this
zeropoint  is  to   say  that  an  object  with   a  flux  density  of
$3.631\times10^{-20}\,erg\,cm^{-2}\,s^{-1}\,Hz^{-1}$     will     have
magnitude  ABMAG =  0 in  every filter.   There is  a  simple relation
between  the  STMAG  and  the   ABMAG  due  to  their  own  definition
($f_{\lambda}  =  f_{\nu}  \,c/\lambda^2$)  and the  pivot  wavelength
definition:
\begin{displaymath}
\lambda_{p}(P)\,\equiv\,\sqrt{\frac{c\;f_{\nu}(P)}{f_{\lambda}(P)}} \,=\, \sqrt{\frac{\int{P(\lambda)\lambda
d\lambda}}{\int{P(\lambda)d\lambda/\lambda}}}
\end{displaymath}.
Thus, STMAG can be converted into ABMAG as: 
\begin{equation}
\label{ab1}
ABMAG(P)=STMAG(P)-5\;log\;\lambda_{p}(P)+18.6921
\end{equation}
where $\lambda_{p}\,$ is in \AA. This can also be rewritten as
\begin{equation}
\label{ab1}
ABMAG(P)=STMAG(P)-5\;log\;(\lambda_{p}(P)/5475.4)
\end{equation}
therefore the two  flux-based systems will give the  same magnitude at
$\lambda$ = 5475.4  \AA.  This transformation provides an  easy way to
calculate   the  zeropoints   in  the   ABMAG  system   starting  from
PHOTFLAM(P):
\begin{displaymath}\label{eq9}
ABMAG\_Zpt(P)= -2.5 \; log\,(PHOTFLAM(P)) - 21.10 -5\;log\;\lambda_{p}(P)+18.6921
\end{displaymath}
Eq. \ref{ab1} can therefore be rewritten as: 
\begin{equation}
ABMAG(P) =  -2.5\;log\,(totalcounts \,\times s^{-1}) + ABMAG\_Zpt(P)
\end{equation}
ACS  $ABMAG\_Zpt(P)$ are listed in Tables \ref{wfczpt} and \ref{hrczpt}
for both WFC  and HRC filters.

\section{PHOTOMETRIC TRANSFORMATIONS}\label{transf}

Since  the ACS filters  do not  have exact  counterparts in  any other
``standard'' filter  set, we  strongly recommend that  ACS photometric
results be referred to a system based on its own filters.  Sensitivity
curves  can be  used  to translate  physical  quantities from  stellar
models,   such  as  surface   luminosity  and   effective  temperature
T$_{eff}$, into  the ACS observational plane. This  conversion is done
by  means of  a bolometric  correction and  T$_{eff}$-color relations.
For example Girardi et al. (2002) have published theoretical bolometric
corrections and  color transformations for  any broad-band photometric
system and have  converted several sets of Padova  isochrones into the
ACS photometric  system (Girardi 2003,  private communication).  Bedin
et al.   (2004) transformed the  entire set of evolutionary  models by
Pietrinferni  et al.   (2004) into  the WFC  observational photometric
system.     The     total    throughput    curves     of    the    ACS
cameras\footnote{available                                           at
  http://acs.pha.jhu.edu/instrument/photometry/}   can  be
implemented  in the  isochrone synthesis  code of  Bruzual  \& Charlot
(2003) to  compute  the  spectral  evolution  of  stellar  populations
directly in the ACS observational plane.

Transformation  between  even  nominally similar  photometric  systems
presents  difficult and  poorly-understood problems (Young  1993).
Manfroid \& Sterken (1992) identify two kinds of error associated with
the transformations:  {\em Conformity errors} that arise from the fact
that passbands in  two systems are different, and  there is no rigorous
way to evaluate the corrections needed to properly transform data from
one system  to another,  and {\em reduction  errors}, caused  by wrong
values in the estimated  transformation coefficients due to inevitable
measurement errors.

In  the  traditional  empirical  approach  the  difference  between  a
magnitude in the natural system and the ``standard'' system is plotted
against a  convenient color  index, and the  fit is performed  using a
straight line  or a low-order polynomial.  This  approach has numerous
problems.  It  requires many accurate observations  of standard stars,
which  might  not  be feasible  (for  example  in  the case  of  HST).
Moreover the  validity of the obtained transformations  will always be
limited to an homogeneous group of stars.  Different fitting-functions
are required for reddened stars, for metal-poor (or metal-rich) stars,
and  for   such  effects  as  age,  rotation,   magnetic  field,  etc.
Furthermore,  in  order to  increase  the  throughput  at the  central
wavelength and  make the entire  system more efficient,  modern filter
transmission curves are steep-sided and ripple-topped (see for example
the comparison between  ACS and UBVRI filters or  even between ACS and
WFPC2  in  Figs.~\ref{filters_w2h}).  This adds  enormously  to  the
complexity of the  color-transformation problems (Sterken \& Manfroid,
1992),  because  the transformations  become  even  more sensitive  to
spectral features.

A  photometric transformation theory  has been  developed in  the last
thirty  years  trying  to   find  a mathematical    relation  between
integrals involving  various functions, such as  the spectral response
of  the instrument,  and  the spectral  distribution  of stellar  flux
(Young,   1992a,   1992b,    Sterken   \&   Manfroid,   1992,   Young,
1994).  Although these methods  have been  very useful  in identifying
several causes  of conformity  errors, they have  shown that  an exact
solution is not possible for the existing photometric systems.  It is,
however,  important  to  note  some of  their  conclusions.   Although
high-order   terms  are  usually   absent  from   low-order  empirical
transformations,  they  are  responsible  for  significant  systematic
errors  (Young, 1992a).   These high-order  terms  are not  linear
functions of  a color index, but  involve both higher  power and cross
products of color, curvature  and derivative-like indices.  Those terms
are  different  for objects  with  different  spectra.  To  accurately
estimate  the high  order terms  a better  sampling of  the wavelength
space  is required,  with overlapping  bands.  Two  ideal  systems are
exactly transformable if the  passbands of one are linear combinations
of the passbands  of the other, i.e. bandpasses  need to overlap with
each filter peak  close to the steepest slope  of the neighbors (Young
1992a).   All  existing photometric  systems,  including  the ACS,  are
undersampled and consequently non-transformable.

Finally,  the   errors  that   arise  from  the   use  of   any  color
transformation for  stars having color  indices that fall  outside the
range of stars used to  derive the transformations have been discussed
by Manfroid et al.  (1992). These authors show the inherent difficulty
of peculiar objects such as Wolf-Rayet stars, luminous blue variables,
cataclysmic variables, supernovae and quasars.

Although  some of  the passbands  of ACS  will eventually  establish a
standard  in  their own  right,  the  scientific  community, at  least
initially,  will  sometimes  require  the comparison  of  observations
obtained through  ACS passbands with those obtained  in other systems.
In the  following pages we  provide the coefficients to  translate ACS
photometry  to other well  known photometric  systems. They  should be
used with  extreme caution.  Internal consistency between  the WFC and
HRC photometry  is granted by the provided  transformation between the
two systems.   We also provide conversion  from ACS to  WFPC2 and from
ACS to the Landolt UBVRI photometric system.  ACS has also a subset of
the SDSS filters; the conversion to this system will be presented in a
separate paper.

In general, transformations  between two different photometric systems
rely upon the spectral energy  distribution of the targeted objects as
well as the absolute response  function of the instruments.  There are
two routes to establishing  the transformations.  One approach is to
make direct use of observations of the same object (typically stars in
well-studied  globular clusters)  in the  two systems.   In principle,
this straightforward determination is  believed to provide an unbiased
approximation to the  real shape of the transformation  curve (for the
selected stars) and can be  directly used without any knowledge of the
instrumental throughputs.   However, because  of the limited  HST time
that can be allocated to calibration programs, the current database of
ACS observations is still insufficient to produce transformations that
are valid  over a  large color range.  In the future,  this deficiency
will be remedied as more and more ACS data become public.

The other method is to deduce transformation curves based on synthetic
photometry.   This  method  takes  advantage of  the  well-established
response  curves  of   the  ACS  and  the  BPGS   catalog  of  stellar
spectra. The transformations obtained in  this way can span much wider
color  ranges,  but their  accuracy  is,  of  course, limited  by  the
fidelity of the  ACS throughput curves and the  quality of the spectra
in  the  atlas.  In  this  paper, we  present  the  results from  both
approaches and show that in most cases the transformations using these
two approaches  are consistent with each  other.  Therefore, synthetic
transformations based on larger color range can be safely employed and
should be considered the norm, unless otherwise indicated.

Although the  current WFC  and HRC throughput  curves can  predict the
observed count rates of  the spectrophotometric standard stars with an
accuracy  of  $\sim$ 1\%, strictly  speaking  they  are  valid for  the
specific  stellar  properties of  the  standard  star  used. When  the
transformation  coefficients  presented in  this  paper are  employed,
account must be  taken of the different spectral  shapes (e.g.  due to
surface  gravity,  metallicity, extinction,  redshift,  etc.)  of  the
targets under investigation.  Passband  differences may not be evident
from the photometry of the  object used to derive the transformations.
If these  transformations are  used for bluer  or redder  objects, for
objects  with   different  gravities  or  for  any   object  not  well
represented by  the targets used  to derive the  transformations, then
systematic magnitude  differences are possible (see  also Bessel 1990,
Holtzman et al. 1995a).

\subsection{General Notes }
In  the  following  sections  we   define  {\em  SOURCE  (S)}  as  the
photometric system of the observed data that we want to transform, and
{\em   TARGET    (T)}   as   the   photometric    system   after   the
transformation. Following this definition SMAG is the magnitude in the
SOURCE system,  and TMAG and TCOL  are the magnitude and  color in the
TARGET system.  Since several photometric  systems can be  defined for
each instrument, we need to  clarify the default magnitude system used
in the  following transformations.  In order to  minimize the possible
confusion  and reduce  the redundancy  among different  zeropoints, we
only  provide OBMAG-to-OBMAG  (see  Eq.\,\ref{obmag}) transformations,
except  for  the transformation  to  the  Landolt photometric  system.
The  OBMAG-to-OBMAG  transformations  can be used to  convert
magnitudes to  any photometric  system by simply  subtracting (adding)
the  corresponding zeropoints before  (after) the  transformation.  We
present an example in \S\, \ref{example}.

The transformations have the following format:
\begin{equation}
\label{transeq}
TMAG\, =\, SMAG\, +\, c0\, +\, c1\,\times\,TCOL\,+\,c2\,\times\,TCOL^2
\end{equation}
where SMAG is the OBMAG in the source system and TMAG and TCOL are the
OBMAG magnitude and color (as  difference of two OBMAGs) in the target
system,   and  c0,   c1,  and   c2  are   the  coefficients   for  the
transformation.  Please  note that in the  transformation equation the
color term is in the TARGET  system and not in the SOURCE system.  The
same  approach  was   used  by  Holtzman  et  al.    (1995a)  for  the
transformations  from WFCP2  to UBVRI  because this  formalism carries
several advantages as discussed  by Stetson (1992).  Consequently TCOL
must  be derived  iteratively  using ACS  observations  in two  colors
unless the target color is known.

The  number  of objects  used  for  the observational  transformations
varies  greatly depending  on the  area  of the  FOV, the  photometric
quality  of the  data and  the brightness  of the  stars.  Photometric
errors  have been used  as initial  criteria to  select stars  for the
transformations.   Depending on  the numbers  of stars  available, the
error threshold  was set  between 0.02 and  0.06 mag.  We  divided the
color  range in  equally  sized  bins and  calculated  the median  and
standard deviation for each bin.  Outliers have been rejected based on
the  local median  and standard  deviation, using  a $\sigma$-clipping
criterion.  In transformation plots  where observed stars are used, we
show only  representative points which  are just local medians  in any
given color bin.

To  estimate the  uncertainties of  the transformation  parameters, we
followed  the  approach employed  by  Holtzman  et  al. (1995a).   For
transformations derived from the observational data, we weighted stars
equally  and  iterated  the  $\chi^2$ minimization  by  adjusting  the
magnitude  errors until  $\chi^2$ reached  unity.  In  order  to avoid
over-estimation of the uncertainty of synthetic stars, we measured the
scatter  from the  fit and  used  this as  a weight  for the  $\chi^2$
minimization.   This procedure not  only prevents  the fit  from being
partially weighted by a few  bright stars, but also provides realistic
errors which are not modeled by synthetic photometry.

In  deriving the  transformation between  two photometric  systems the
functional form  of the transformation should  be addressed.  Although
in principle, a quadratic least-square fit is preferable, we found that
for observational  transformations, given the limited  color range and
limited statistics, a linear fit was more appropriate.  In the case of
observational transformations between the  CCD cameras of the ACS that
use the same filters we  also decided to perform a linear least-squares
fitting with  iterative rejection.  For  all synthetic transformations
we performed a quadratic least-squares fitting.

A single  set of coefficients  does not always  fit the data  over the
large color  range of  the stellar atlas.   Sudden differences  in the
response  curves  can  produce   discontinuities  in  the  color  term
dependencies and force us to limit  the color range for which a set of
coefficients can be used. When this happens, we divide the color range
into two regions and provide separate coefficients for each segment.

Finally, the precision of the synthetic transformations depends on the
the  accuracy  of  the  total  response function  of  the  photometric
passbands  of  ACS.  As  we   discussed  in  \S  \ref{DQE},  a  fairly
significant revision of the pre-launch RQE has been necessary in order
to reproduce  the observed on-orbit sensitivity.  Although the derived
RQE curves can now reproduce the count rates in all broad band filters
with an accuracy  better than 0.5\%, it is possible  that the shape of
the  total response  of  the bluest  and  reddest passbands  is not  a
perfect  representation  of  the  real  response  curve.   This  could
introduce systematic  errors in  the color transformations.   In these
cases (F435W  and F850LP  for the  WFC, and F220W  and F850LP  for the
HRC), we suggest to use the observed transformations, if available.


\subsection{ACS Internal Transformations}

The WFC  and the HRC share all  the visual and near-IR  filters of the
ACS.  The transformation between the two systems is therefore expected
to be quite simple.  However the different optimization of the cameras
(mirrors  and  CCDs) produces  different  total  throughput curves  in
particular   in  the  blue   and  near-IR   (Fig.~\ref{filters_w2h}).
Therefore  the  transformation  between  filters in  these  wavelength
regions could have a significant dependence on the color term.

For these ``internal'' transformations  between the two CCD cameras we
used  observations  of  two   globular  clusters,  NGC  104  (Programs
9018,9656,9666) and NGC 2419 (Program 9666).  These globular clusters
have very  different metallicities ([Fe/H]~$\sim -0.7$ for  NGC 104 and
[Fe/H] ~$\sim -2.2$  for  NGC 2419),  essentially  spanning the  whole
metallicity distribution of the Galactic globular cluster system.

The raw data have  been processed  through the CALACS  pipeline at  the STScI.
For moderately crowded fields ($\sim$6$'$ west of the core) of NGC~104,
we  simply  performed 5-pixel  radius  aperture  photometry using  the
IRAF/APPHOT  package.   Because   the  NGC~104  data  were  originally
acquired  for  the assessment  of  the  geometric  distortion and  low
frequency  residuals  in  the  flat  field,  each  filter  observation
consists of  multiple pointings offsets by large  dithers.  We decided
to select  only stars present in  at least three pointings  and to use
the  sigma-clipped   averages  of   the  magnitude  in   the  multiple
observations as an estimate of the brightness of the star.

The photometry  of NGC~2419 required  more complicated field-dependent
handling.  Due to the larger distance of the cluster a single pointing
of WFC covers  both the compact core and  the moderately crowded outer
regions.  In  the external regions  of NGC~2419 we can  safely perform
5-pixel radius  aperture photometry.  However,  PSF-fitting photometry
is required for  the crowded region of NGC~2419 common  to the HRC and
WFC.  While  the PSF of the HRC  can be assumed to  be nearly constant
across the field, the PSF varies significantly across the WFC FOV (see
\S \ref{ee}), thus preventing us from applying in the crowded area the
PSF  extracted  from  isolated  stars  in the  external  regions.   We
confirmed this fact by  performing our initial PSF photometry assuming
a spatially constant PSF and the result showed systematic errors up to
$\sim$4\%.    Although  in  principle  a  PSF   can  be  iteratively
constructed even for  an extremely crowded field, the  task to achieve
calibration-quality  photometry becomes  prohibitively  impractical if
not  impossible.  An  alternative is  to use  TinyTim (Krist  2003c) to
model  the  PSF on  the  crowded area.  However,  it  is difficult  to
determine the  time-varying behavior of  the PSF, mainly due  to focus
offset changes.  Besides, most  PSF photometry packages cannot readily
make use of PSF images from  Tiny Tim.  We therefore decided to create
position-dependent  PSF templates  using the  sparser  observations of
NGC~104.  The  PSFs were easily  sampled from isolated stars  over the
entire field of view and then applied to the NGC 2419 field.  In order
to avoid  effects due to PSF  time variations, when  possible, we used
the  NGC~104 images which  were closest  to the  date of  the NGC~2419
observations.  After  checking the residuals  in PSF-subtracted images
as  well  as  various  goodness-of-fit  criteria,  such  as  $\chi^2$,
best-matching  PSF templates  were determined  for each  filter.  This
technique allowed us to remove any systematic effect on the photometry
of the crowded regions.

Examining the NGC~104 and NGC~2419 data separately, one does  not see strong
evidence for a metallicity dependence in the transformation relations.
Due  to the  different 
metallicities and distance moduli of the two clusters ((m-M)=13.27 for
NGC~104 (Zoccali et al. 2001)  and (m-M)=19.88 for NGC~2419 (Harris et
al. 1997)) the  range of colors that our  selection criteria allow are
not the  same in both clusters.   In particular, for  NGC~2419 we used
stars on the horizontal branch, the asymptotic giant branch and on the
red and  sub-giant branches.  The NGC  104 data allow  an extension to
the red  since we can  use main sequence  stars at least  4 magnitudes
fainter than the turn-off point.

Fig.~\ref{cmds_acs}  shows the observed  color-magnitude diagrams for
the  two  clusters  for  the  WFC  and HRC.   In  order  to  look  for
metallicity effects,  we must compare  stars of similar colors  in the
two clusters.  In  most transformations, the data of  the two clusters
overlap  only  in a  quite  narrow  color  range spanning  $\leq$  0.6
magnitudes.  We therefore used  the model atmospheres of Kurucz (1993)
to  investigate  the metallicity  dependency.   We compared  synthetic
transformations   using  three   different   metallicities,  [Fe/H]~$=0$,
[Fe/H]~$=-0.5$ and [Fe/H]~$=-2.0$  and find that in the  color range covered
by  the observational  data,  the systematic  differences between  the
three  metallicities  are at  most  0.3\%.   We  therefore decided  to
combine the data  for both clusters and also  include the observations
of  the  two  spectrophotometric  standards  to  determine  the  final
transformations.

Even  when  combining all  the  observational  data,  the color  range
covered by all the stars (WDs  $+$ NGC2419 $+$ NGC 104) were generally
limited          to          $0.0\,<\,F435W-F555W\,<\,1.2$          or
$-0.4\,<\,F555W-F814W\,<\,1.2$   in  the   HRC  VEGAMAG   system;  the
transformation from  observed stars cannot be assumed  to be corrected
for stars outside this color range.  Synthetic transformations cover a
larger color range and should be used instead.

We list  the transformation coefficients  to use in  Eq.~\ref{transeq}
for different combinations of most of the full-sized filters in Tables
\ref{wfc2hrc} and  \ref{hrc2wfc} for  the transformations from  WFC to
HRC and from HRC to WFC, respectively. The transformations between the
primary filters from WFC to HRC are shown  in Figs.~\ref{tranw2h1}. 
 
These figures  show that the  synthetic system matches  the observations
fairly  well, usually within  1\% for  all stars  within the  range of
colors in  which our  observations were made.   There are  however a few
cases, F435W and F625W for  example, where the discrepancy can rise up
to a couple of percent for the reddest stars.
In  principle, one  could argue  that since  the long  wavelength light
scattering is  more severe  in the  HRC than in  the WFC  an incorrect
aperture correction could be the reason for this discrepancy. However,
this is not  the case, because these filters are  not affected by this
problem, which  starts at  $\lambda > $  7000 \AA.  Also,  the near-IR
filters, which suffer from extended  PSF wings, and therefore could be
more  affected  by  aperture   correction  errors,  do  not  show  any
systematic differences.

Although  the BPGS  atlas does  not  include stars  with high  surface
gravities, but only main-sequence stars and giants, the agreement with
the two spectrophotometric white dwarfs does not show any systematics.
This analysis  suggests that we have  a good understanding  of the ACS
synthetic system and that  the derived transformations are accurate to
1-2\% or better in all filters.  The synthetic curves can thus be used
in  place of  the observational  ones to  produce  transformations for
objects with spectra different from the observed stars.

\subsection{Transformations from ACS to UBVRI}\label{a2g}
ACS filters  differ significantly from  the UBVRI bandpasses.  The ACS
filters are usually narrower,  start at shorter wavelengths and, apart
from F814W, extend less  to the red (see Fig.~\ref{filters_w2h}).
  As  a  consequence the  transformations
between  the two  systems  may be  very  sensitive to  details in  the
underlying   stellar  spectrum   and  show   strong   dependencies  on
metallicity and surface gravity.

We compared ACS  observations with ground-based photometry  of NGC 2419.
In particular, we  retrieved BVRI  data    from      Stetson      (2002)
and an  independent BVRI  dataset was provided  by Saha  et al. (2005).
The two  NGC~2419 photometric control  datasets are the  same recently
used  to  examine the  accuracy  of  the  WFPC2 zeropoints  (Heyer  et
al. 2002,  2004).  The two  observational CMDs from the  two different
sources  are shown  in Fig.~\ref{cmds_ground}.   In order  to  have a
large number of common stars in  the HRC and WFC FOV, ACS observations
were centered in a region fairly  close to the center of NGC~2419.  As
a consequence, the match with ground observations~- centered in a more
sparse  region~-  resulted in  only 30  to 60  stars depending  on the
filter and camera.

We compared the  two NGC~2419 control datasets and  find a significant
color  term between them:  red stars  tend to  be brighter  in Stetson
(2002) than in Saha et al. (2005) with differences up to 3-4\% for stars with
V-I $>$ 1.0  in the I band.  Such differences,  also discussed in Saha
et al.  (2005), could  be due to  different filter  throughput curves,
red-leak control and systematic effects in the transformation from the
instrumental system to the standard system for the low surface gravity
stars in NGC~2419.

For the  synthetic UBVRI  photometry we first  used the  Landolt UBVRI
filter set  (i.e., Johnson  UBV + Cousins  RI) defined in  SYNPHOT. In
particular the throughput  data for the Johnson UBV  bands are the U3,
B2 and V synthetic bandpass data  of Buser \& Kurucz (1978), while the
Cousins R and  I throughputs are taken from  Bessel (1983). However we
got  poor agreement  with  the observed  transformation  using the  B2
filter. We then  tried using other B synthetic  bandpasses and found a
much better agreement  when using the B filter from  the Harris set in
use at KPNO and WIYN\footnote{The curves used for the synthetic UBVRI
are available at \url{http://acs.pha.jhu.edu/instrument/photometry/}}.

For both  ACS cameras, we compared the  observed transformations using
the  two NGC~2419  control datasets  with the  corresponding synthetic
transformations.   We obtained  poor  agreement with  the  I bands  of
Stetson's  dataset  with  systematic offsets  of the  order of  a few
percent between  the two  curves.  The observed  R-band transformation
also shows some departures for  the reddest stars.  Finally the V band
shows  a good  match between  observed and  synthetic transformations.
Using Saha's dataset we have significantly better results in V, R and I
available bands.   The B  band show a  similar agreement  using either
Stetson's or Saha's dataset. The observed and synthetic transformation
provide similar results for stars  with B-V $>$ 0.5, whereas for bluer
stars the discrepancy between the two transformation can be as high as
5\%. Given the possible uncertainties in the shape of the blue side of
the total response curve with the F435W filter, our bluest filter with
WFC, we suggest  to use the observed transformations  when this filter
is involved.

We decided  to use only  the Saha photometric  dataset of NGC  2419 to
calculate  the  coefficients  for the  observational  transformations.
However,  in  determining  the  new  zeropoints for  WFPC2,  Heyer  et
al.(2002),  find that  Stetson's photometry  produces  values slightly
closer to  the historical zeropoints of WFPC2  than Saha's photometry.
Unfortunately, no color term was taken into account.

We list the observational and synthetic transformation coefficients to
use in Eq.~\ref{transeq} for different combinations of most filters in
Tables  \ref{wfc2ground}  and  \ref{hrc2ground}  for  WFC-to-BVRI  and
HRC-to-UBVRI,   respectively.  The  transformations   between  primary
filters are shown in Figs.~\ref{tranw2gr} and \ref{tranh2gr}.
An example on how to calculate the transformation is provided in \ref{example}

As  expected,  Figs.~\ref{tranw2gr}  and \ref{tranh2gr}  show  large
scatter in the synthetic transformation. The BPGS atlas includes stars
with a  wide range  of surface gravity  and possibly  metallicity.  We
used  the synthetic  system to  investigate dependencies  of  the main
transformations on  gravity and metallicity.   For the gravity  we use
the Bruzual Spectrum Synthesis Atlas (BZ77) implemented in SYNPHOT. It
provides  solar  metallicity  spectra  for main-sequence  stars  (from
spectral class O5 to M6), giant  stars (from O8 to M6) and supergiants
(from  O9 to M1).   For the  metallicity dependence  we use  the model
atmosphere  of Kurucz  (1993). We  selected models  to cover  the main
sequence from  O3 to M2  for the solar  metallicity, and for  [Fe/H]~=
+0.5,   $-1.0$   and   $-2.0$.    The   results   are   shown   in   
Figs.~\ref{tranw2gr_prob}  and  \ref{tranh2gr_prob}  for  the WFC  and  HRC,
respectively.   In both  figures  the  column on  the  left shows  the
dependence on the surface gravity; filled circles represent data for a
synthetic main-sequence.  The sequences  of giant and supergiant stars
are represented  with gray stars and boxes,  respectively.  The column
on the right shows the dependence on the metallicity for main sequence
stars:  black dots show  the solar  metallicity, filled  triangles are
[Fe/H]~=~0.5, gray-filled stars are [Fe/H]~$=-1$ and squares are [Fe/H]~$=-2$.

These figures show that the dependency of the transformations on these
quantities can  be quite  large.  Gravity provides  a large  spread in
short and long wavelength filters  for stars around zero color and for
low  temperature stars.   Metallicity  causes more  spread for  redder
stars.  Significant departures from  the relation derived from stellar
spectra are  expected for  the transformation of  non-stellar objects,
especially for galaxies at high redshift.

The  derived transformations  from the  ACS  systems to  UBVRI can  be
applied to main-sequence and giant stars without introducing errors of
more than a few percent.  However,  it should not be expected that the
1-2\% accuracy of ACS photometry  will be preserved on the transformed
data.  Transformation  to U, B and  I filters are  highly dependent on
the   object  spectral   details  and   Tables   \ref{wfc2ground}  and
\ref{hrc2ground}  should  be used  with  extreme  caution for  objects
different  from  those  used  to  derive  the  relations.   Therefore,
whenever possible, we strongly recommend working with the ACS system.

\subsection{Transformations from ACS to WFPC2}\label{a2wfpc2}

Most ACS broad-band filters have, at least nominally, a close match in
the vast WFPC2 filter  complement.  However, the different response of
the  two instruments  and specific  filter designs  create significant
differences    between   the    ACS   and    WFPC2,   as    shown   in
Fig.~\ref{filters_w2h}.

Since the installation of WFPC2 on HST a decade ago, a large amount of
WFPC2 data have become public  and therefore available through the HST
archives for  calibration purposes.   Now many characteristics  of the
instrument are  well understood, although  the refinement is  still in
progress (Baggett  et al.  2001; Casertano \&  Wiggs 2001;  Whitmore \&
Heyer 2002; Dolphin 2002a).

In order to derive transformations between ACS and WFPC2, we retrieved
WFPC2  observations  of  NGC~104  (programs  7465,6114,6660,6114)  and
NGC~2419 (programs 7268,7630,9601) from the STScI archive.
Accurate  stellar  photometry  on  WFPC2  images  requires  a  careful
treatment  of the  undersampling effect,  of  position-dependent pixel
size  variations,   and  of  the  charge   transfer  efficiency  (CTE)
degradation.   We  found  HSTPHOT,  an  automatic  stellar  photometry
software  specifically designed  for  WFPC2 data  (Dolphin 2000b),  to
account  for   all  these   issues  with  acceptable   accuracy.   The
dependability  of   the  package   was  tested  by   first  performing
PSF-fitting photometry with  IRAF DAOPHOT on a subsample  of the data,
applying aperture corrections measured from isolated stars in the same
images,  and CTE  correction from  Dolphin (2002a)  and  comparing the
results  with  HSTPHOT results.   Since  the  results  from these  two
methods  did not  show any  systematic difference,  we decided  to use
HSTPHOT to  automatically process all the WFPC2  observations.

 The  detailed description  and  usage  of HSTPHOT  can  be found  at
  \url{http://www.noao.edu/staff/dolphin/ hstphot/}.
We  used the  ``multiphot"  photometry feature  of  HSTPHOT instead  of
performing  photometry  on  combined  images.  This  method,  although
reducing the  detectability of faint stars,  increased the photometric
accuracy  for  the   bright  stars  that  we  used   to  estimate  the
transformation  coefficients.   Finally,  the  HSTPHOT  output  flight
system  magnitudes were  converted into  OBmags by  subtracting flight
system zeropoints  (Dolphin 2000b).  Fig.~\ref{cmds_wfpc2} shows the
observed color magnitude diagram for the two clusters.

There  is a subtlety  involved in  the determination  of OBMAG  in the
WFPC2  photometric  system,  because  WFPC2  zeropoints  are  slightly
different depending on the chip. We choose the WF3 chip as a basis for
our OBMAG  definition. For the  synthetic transformation, we  used the
sensitivity curves for all WFPC2/WF3 components currently available in
SYNPHOT (Bagget et al. 1997).

We list the observational and synthetic transformation coefficients to
use in Eq.~\ref{transeq} for different combinations of most filters in
Tables  \ref{wfc2wfpc2}  and   \ref{hrc2wfpc2}  for  WFC-to-WFPC2  and
HRC-to-WFPC2  conversions,  respectively.   The appropriate  WFPC2/WF3
zeropoints  for  gain~=  15  must  be  added  when  transforming  the
WFPC2\_OBMAG into  any WFPC2 photometric  system.  The transformations
between  primary  filters are  shown  in  Figs.~\ref{tranw2wfpc2}  and
\ref{tranh2wfpc2}.

\section{The HRC UV RED LEAK} \label{leak}
We  investigated the  red leak  of the  UV filters  of the  HRC.  When
designing  a UV  filter,  the  high suppression  of  the red  off-band
transmission  has  to be  traded  with  the  in-band throughput.   The
elevated quantum efficiency of the  HRC makes it possible to design UV
filters  with limited red  leak. 
 We defined  all contributions from  $\lambda > $ 4000  \,\AA,  as  red  leak.
 In  order  to  quantify  the  red-leak contribution  as a function  of the  stellar 
color,  we used  the BPGS library  of stellar  spectra and  the HRC  
throughput  curves.  Figure~\ref{rleak2} shows  the amount of light  contributed 
by red  leak as a function of color for the three filters.  Clearly there is no red leak
in F330W, but it becomes important in F250W and particularly in F220W.
Accurate UV  photometry in  the F250W and  F220W filters  will require
red-leak corrections for objects with a  spectrum as red as that of an
F5V star or colder and will be essentially impossible for objects with
a spectrum  later than  M.  In  addition to the  majority of  the flux
being off-band, the red photons  will contribute to create an extended
halo (see \S \ref{halo}) which will superimpose on the sharper UV PSF.

\section{REDDENING IN THE ACS SYSTEM} \label{red}
The  ACS filters  can  differ significantly  from the  Johnson-Cousins
UBVRI bandpasses  (see \S\ref{a2g}), and  these differences need  to be
taken  into  account  when  calculating  reddening  corrections.   ACS
magnitudes should be corrected with extinction coefficients calculated
in the native photometric system.   For the same reason, any reddening
correction  should  be  applied   before  transforming  to  any  other
photometric system.   We calculated a grid  of extinction coefficients
for all  ACS filters  as a  function of E(B-V)  up to  E(B-V)=5.5. The
extinction  was  computed using  the  synthetic  throughput curves  in
SYNPHOT and adopting the Galactic extinction law presented by Cardelli
et  al. (1989).   For medium-  and broad-band  filters  the extinction
depends on  the object's color.  We therefore  selected three template
stars (O5V, G2V,  M0V) from the BPGS atlas  and three galaxy templates
(E, Sc,  Im) from Ben\'itez et  al.  (2004) and  computed the extinction
coefficient for the grid of E(B-V) values.  Fig.~\ref{redd1} shows the
extinction  in a  few  primary ACS  filters  as a  function of  E(B-V)
computed for  three stars of different types.   Tables \ref{est_w} and
\ref{est_h}  list the  extinction A(P)  normalized to  the photometric
measurement  of   E(B-V)=A(V)/3.1  for  all  ACS   filters  using  the
interstellar  extinction  of Cardelli  et  al.   (1989).   Due to  the
difference in the filter  transmission curves between the ground-based
system and the  ACS systems, there could be  systematic differences in
the extinction  coefficients for  the two systems.   Fig.~\ref{redd2}
shows that the extinction in  F330W, F435W and F555W is systematically
higher than  in U, B and  V because the effective  wavelength in these
ACS band is shorter than  in the corresponding Johnson filters. On the
contrary the extinction in F814W is systematically lower than in the I
band because the  effective wavelength in the F814W  is longer than in
I.

\section{CONCLUSIONS}

We have  presented the current status and  the photometric calibration
of the  two CCD channels  of ACS.
The overall performance is as expected from pre-launch testing
of the  instrument.  A  positive surprise was  the discovery  that the
ground throughput prediction underestimated the overall sensitivity of
the camera by a few percent in  the red, up to $\sim$ 20\% in the blue
for WFC and in the visual and the near-IR for HRC.  HRC also showed an
unpredicted dip in the UV response at $\sim$ 3200 \AA.

Initial signs of degradation due  to the HST radiative environment are
already  visible  in terms  of  dark  rate  increase, charge  transfer
efficiency  degradation  and  growth   of  the  permanent  hot  pixels
population.   The detailed  effect  of  CTE losses  as  a function  of
brightness and background level  is being determined with a continuous
effort at  STScI. 
The correction  formula to be  applied to photometric  measurements is
constantly kept up  to date on the ACS web server  at STScI.

  The  long wavelength  light scattering  in  ACS CCDs  has a  twofold
  effect: the width of the  PSF increases significantly in the near-IR
  filters and  the aperture correction  in the near-IR depends  on the
  SED  of the object.   Recipes to  calculate the  aperture correction
  have been presented.

Photometric calibration data of spectrophotometric standard stars have
been  obtained  for  all  ACS  filters to  provide  transformation  to
physical fluxes.   The HRC and  the WFC synthetic  photometric systems
have been constructed and  reproduce count rates of spectrophotometric
standards  to within 0.5\%  in all  broadband filters.  Zeropoints are
provided in three different magnitude systems.

Transformations  between HRC  and WFC  photometric systems  are given.
Although ground  observation and  synthetic photometry have  been used
for  the  determination  of  the  transformation to  UBVRI  and  WFPC2
photometric  systems,  we  strongly  recommend  that  ACS  photometric
results be referred  to a system based on  its own filters.  Synthetic
photometry can be  used to construct isochrones and  to safely convert
models into the  ACS observational plane.  In some  instances a direct
comparison with previous results in a different photometric system may
still be needed.  We warn the reader that transformations to UBVRI and
WFPC2 depend on details of the stellar spectra and should be used with
caution.  Generally,  transformed data will  be accurate within  a few
percent, but  differences could  be significantly larger  for peculiar
spectra. We also provide tables for the color transformation for a few
galaxy templates as a function of the redshift.

The effective  wavelength of the ACS filters  can differ significantly
from  their  counterparts  in  the   WFPC2  and  UBVRI  systems.  As  a
consequence,  accurate reddening  corrections must  be made  in  the ACS
system,   before   transforming the   magnitude  to   other   photometric
systems.  Extinction  curves  for   the  ACS  filters  for  a  standard
interstellar reddening law have been presented for a sample of stellar
and  galaxy templates.  Extinction  for other  reddening  laws can  be
derived using the synthetic ACS system.

The  process of understanding  and calibrating  ACS is  ongoing.  More
calibration  programs are currently  planned and  more data  are being
analyzed.  ACS users are encouraged  to stay posted for future reports
from  the STScI  and to  contact  the ACS  Instrument Scientists  with
questions (help@stsci.edu).

\acknowledgments

Many people have contributed to  this work.  ACS was engineered, built
and tested  by a  dedicated team at  Ball Aerospace, Boulder,  CO.  We
thank the  entire ACS Investigation Definition Team  and all personnel
at GSFC who have  supported ground calibration and pre-launch testing.
Special thanks  to the entire crew  of STS-109 who  did a  superb job
during an extraordinarily demanding servicing mission.

Thanks  to ACS  personnel at  STScI, the  ACS  photometric calibration
group  and the  ACS  science team.   They  have prepared  many of  the
calibration proposals  and were  always available to  discuss results.
Among these people,  we specially appreciate the comments  and work we
received  from  Ralph  Bohlin,   Francesca  Boffi,  Nick  Cross,  David
Golimowski, Inge Heyer, John Krist, Adam Riess and Roeland van der Marel.

We are indebted  to Abhijit Saha for providing  unpublished magnitudes and
data relating to  the UBVRI system.  We also  thank Tom Brown, Stefano
Casertano, Leo  Girardi, Brad Whitmore  and Manuela Zoccali  for their
useful suggestions and insightful discussion.

The entire paper has been modelled after  equivalent WFPC2 calibration
papers by John  Holtzman and collaborators. We are  indebted with them
for setting a bench mark in comprehensive calibration paper
making  our job much easier.

MS is particularly  grateful to the ACS IDT team  and all personnel of
the JHU  Physics and Astronomy  department for five years  of pleasant
and  productive  work.   Finally,  we  are  grateful  to  K.~Anderson,
S.~Busching,  A.~Framarini,  S.~Barkhouser,  and  T.~Allen  for  their
invaluable contributions to the ACS project at JHU.

We have much appreciated  the many useful suggestions and constructive
criticism of Gordon Walker, the  referee of this paper, whose valuable
help has considerably improved the presentation of our work.

ACS was developed under NASA  contract NAS 5-32865, and this research
has been supported by NASA grant NAG5-7697 and by equipment grant from
Sun  Microsystems,  Inc.  The  Space  Telescope  Science Institute  is
operated by AURA Inc., under NASA contract NAS 5-26555.

\bigskip

\appendix

\section {Acronyms}
\begin{tabular}{ll}
 ACS  & Advanced Camera for Surveys \\
 ADHv4 & ACS Data Handbook version 4.0\\
ADU  &  Analog-to-Digital Unit\\
AIHv5 & ACS Instrument Handbook version 5.0\\
BPGS & Bruzual, Perrson, Gunn and Stryker stellar atlas\\
CALACS & CALibration pipeline for ACS data\\
CCD  & Charge Coupled Device\\
CTE  & Charge Transfer Efficiency\\
EE   & Encircled Energy\\
FOV  & Field Of View\\
FWHM & Full Width at Half Maximum \\
GSFC   &Goddard Space Flight Center\\
HRC &  High Resolution Channel\\
HST  &Hubble Space Telescope\\
IR  & Infra Red\\
MPP & Multi Pinned Phase\\
OTA  &  Optical Telescope Assembly\\
PSF & Point Spread Function\\
RQE & Responsive Quantum Efficiency\\
SED & Spectral Energy Distribution\\
SMOV &  Servicing Mission Orbital Verification\\
STIS & Space Telescope Imaging Spectrograph\\
STScI & Space Telescope Science Institute\\
SBC  &Solar Blind Channel\\
TinyTim & Program which produces simulated HST point spread functions\\
UV  &  Ultra Violet\\
WFC & Wide Field Channel\\
WFPC2 &Wide Field and Planetary Camera 2\\
\end{tabular}
\clearpage

\section{PHOTOMETRIC CALIBRATION COOKBOOK}
Here we briefly summarize the steps required to perform photometry
on an ACS image. See also ADHB03.

\subsection{Point Source Photometry}\label{cook_star}

\begin{enumerate}
\item Retrieve the  image(s) from the HST archive  or proceed with the
  manual  recalibration of  the  data following  the  recipe given  in
  ADHv4.  Be  sure to use  the best reference files  usually available
  a few weeks after the data acquisition.

\item After running CALACS on  the association file the main output is
  the  flat fielded  corrected {\em  FLT} image  and/or  the geometric
  corrected {\em DRZ} image.
\begin{itemize}
\item If CR-SPLIT observations are available, the {\em FLT} images are
  combined to produce a {\em CRJ}  frame free of cosmic rays.  We draw
  the reader's attention to the fact that the default settings for the
  CR rejection routine  are quite conservative.  If the  images in the
  CR-SPLIT pair do not show  any significant offset ($< $ 0.05 pixel),
  the {\em  scalenoise} parameter may  be reduced a more  efficient CR
  rejection  (see   \S  \ref{crsplit}).   However,  in   the  case  of
  undersampled images, one  should verify that the peak  of the PSF is
  not  affected (i.e.  truncated)  by a  more ``aggressive"  CR-removal
  technique.
\item  If instead  the observation  consisted of  dithered  images but
  without  CR-SPLIT the resulting  FLT images  will still  have cosmic
  rays  which  are removed  during  the  combination  of the  dithered
  observations. {\em  MultiDrizzle} performs a  better CR-rejection on
  non   CR-splitted    observations   than   PyDrizzle.\footnote{Since
    September 2004  {\it MultiDrizzle} is part of  CALACS.  See ADHB03
    for more information on running {\em MultiDrizzle} on ACS images}.
\end{itemize}
\item  Perform  photometry  of objects  in  the  field  using any  choice  of
  photometry technique.
\begin{itemize}
\item If  work is carried  out on the  {\em FLT} image instead  of the
  distortion  corrected {\em DRZ}  file, it  is necessary  to multiply
  each SCI  (and ERR) frame of  the {\em FLT} file  by the appropriate
  pixel area  map file (see  \S \ref{geometric}). This  operation will
  ensure that the  total integrated flux of the  star is conserved. We
  underline here  that even  after the application  of the  pixel area
  map,  the PSF  of the  stars  will be  distorted and  the amount  of
  distortion  is position  dependent.  The  encircled  energy profiles
  published in this paper were calculated from non-distorted PSFs.  As
  a consequence the  aperture correction that can be  derived from the
  EE  profile cannot be  directly applied  to non-drizzled  images, at
  least for  small radius apertures. The zeropoints  published in this
  paper  refer to  an ``infinite"  aperture and  can be  used  once the
  aperture correction to a ``total" flux has been applied.

\item The  units of the {\em  FLT} and {\em DRZ}  images are electrons
  and  electrons/sec respectively.  Many  photometry packages  for PSF
  fitting use an optimal weighting scheme which depends on the readout
  noise, gain and the true counts  in the pixels. As a consequence, in
  order  to compute  the weights  correctly, the  image should  not be
  divided by the  exposure time. Therefore the {\em  DRZ} image should
  be  multiplied by the  exposure time  before running  such packages.
  The total number of electrons, instead of the electron rate, is also
  needed in  order to properly  calculate the CTE correction.  For the
  same reasons  the background sky  should not be subtracted  from the
  image.

\item  When performing  aperture  or PSF  fitting photometry,  possible
  systematic errors due to the  spatial variation of the PSF should be
  addressed.  This is  particularly important  when the  photometry is
  carried out using a small radius  aperture ($< $ 5 pixels). For very
  small  aperture (r$  <  $  2 pixels)  the  systematic variations  in
  aperture correction  can be as high  as 10-15\%.  An  example of the
  analysis of the optimal aperture is provided in Appendix B.
\end{itemize}

This step should give you a  set of measurements in units of electrons
or electron/sec  in your selected  aperture for all your  objects. The
instrumental   magnitude  can   be  calculated   as  $   -2.5  \times
log(electrons \times s^{-1})$.

\item Consider  correcting the photometry for  CTE degradation. Unless
  your data  were acquired in the  first semester of 2002,
  the  effects  of  CTE degradation  are
  probably present in your data. Check  on the STScI ACS web page for
  updates on the CTE degradation correction.  There are several things
  to keep in mind when calculating the correction for CTE degradation
  in your image:
\begin{itemize}
\item the  flux level and the  background level to use  in the formula
  are  in electrons and  not electrons/second.  When working  with DRZ
  images  it is  necessary  to  multiply the  measured  value by  the
  exposure time.
\item when applying the CTE correction to combined images, each of the
  combined images must have the  same or very  similar exposure
  times. The  output of CALACS,  when CR-SPLIT or DRZ  observations are
  available, is  the sum of the  combined images. In this  case, if all
  the combined images have the  same exposure time the measured values
  in  the equation,  such as  SKY and  FLUX should  be divided  by the
  number of images used  to create the add-combine image.  Determining
  the  CTE  correction  for  a  combined image  created  with  unequal
  exposure  times  (for  example,  one  long exposure  and  one  short
  exposure) cannot easily be done.
\item In  particular for WFC observations  the FOV is  big enough that
  the background sky might present gradients. In general, the use of a
  constant  sky  value for  the  determination  of  CTE correction  in
  different  regions   of  the  FOV  is  not   recommended.   This  is
  particularly true for all filters  where the PSF wings are broadened
  by  the long  wavelength halo.  Depending on  the brightness  of the
  star, the  signal in the PSF  wings can enhance the  "local" sky and
  reduce the amount of lost signal within the aperture where the flux
  is measured. In such instances a global sky value would overestimate
  the CTE correction.
\end{itemize}

\item  Apply  the aperture  correction    to all  photometric
measurements to transform the instrumental magnitudes into OBMAG:
\begin{enumerate}
\item for filters bluer than  F775W: determine the offset between your
  photometry   and   aperture   photometry   with   a   $0\farcs5$   radius
  aperture.  This is  usually done  by measuring a few  bright isolated
  objects  in the  field.  Apply  this offset  to all  the photometric
  measurements. Use Table \ref{ee05} to get the value of AC05 for each
  filter. Add AC05  to the instrumental magnitude or  insert the count
  rate at  r=$0\farcs5$ and the  value of AC05  in Eq. \ref{obmag2}  to get
  OBMAG.
\item for all  near-IR filters: if your object is  a blue star proceed
  as in (a).  For all other objects you need to estimate the impact of
  the {\em red  halo}: first estimate the effective  wavelength of the
  observation   using  the  recipe   in  \S   \ref{acorred},  SYNPHOT
  or  Table \ref{app_efflam}.  Then  calculate the aperture
  correction from  a $0\farcs5$ aperture radius to  infinity using the
  relation in Figs.~\ref{efflam_apt_wfc} and \ref{efflam_apt_hrc} and
  Tables \ref{WFC_acl} and \ref{HRC_acl} and compare the results
  with  the values in  Table \ref{ee05}.   If the  calculated aperture
  correction  is similar,  within a  few tenths  of a  percent  to the
  corresponding  value in  Table  \ref{ee05}, then  the  red halo  has
  little or no  impact in your observations and  you should proceed as
  in case (a).   If however the difference is  larger, then you should
  use    the    relation    in   Figs.~\ref{efflam_apt_wfc}    and
  \ref{efflam_apt_hrc}     and     Tables     \ref{WFC_acl}     and
  \ref{HRC_acl} to directly  calculate the aperture correction from
  your aperture radius to infinity.
\end{enumerate}
\item  Apply the  appropriate zeropoint  from Tables  \ref{wfczpt} and
  \ref{hrczpt} to transform OBMAG to any choice of magnitude system.
\item If  your observations were processed with  CALACS before January
  6, 2004  and the gain is  different from the default  setting (1 for
  WFC, 2 for HRC), use Table \ref{gainfix} to correct the photometry.
\item The magnitude can be corrected for interstellar extinction using
  Tables~\ref{est_w}  and \ref{est_h}.  Given  E(B-V), the  extinction
  A(P) in  the passband P is  calculated by multiplying  E(B-V) by the
  value  in the  table which  correspond to  the correct  passband and
  spectral type.
\item If  you want  to transform the  data to a  different photometric
  system use Eq.\ref{transeq} with the coefficients in \S \ref{transf}
  which are appropriate for the source and target systems.  You should
  apply the  correction for interstellar  extinction (see \S\ref{red})
    before applying the transformations.

\end{enumerate}

\subsection{Resolved Object Photometry}

\begin{enumerate}
\item Retrieve the  image(s) from the HST archive  or proceed with the
  manual  recalibration of  the  data following  the  recipe given  in
  ADHv4.  Be sure to use  the best reference files usually available a
  few weeks after the data acquisition.

\item After running CALACS on  the association file the main output is
  the flat fielded corrected  {\em FLT} image and/or the geometrically
  corrected {\em DRZ} image (see Point 2 in \S\ref{cook_star}).

\item Perform photometry  of objects in the field  using any choice of
  photometry technique (see Point  3 in \S\ref{cook_star}).  This step
  should  give you  a set  of measurements  in units  of  electrons or
  electron/sec  in your selected  aperture for  all your  objects. The
  instrumental  magnitude   can  be   calculated  as  $   -2.5  \times
  log(electrons \times s^{-1})$

\item Consider  correcting the photometry for  CTE degradation. Unless
  your data  were acquired in the  first semester of 2002,  due to the
  large  amount  of transfers,  the  effects  of  CTE degradation  are
  probably present in  your data. Check on the STScI  ACS web page for
  updates  on   the  CTE  degradation  correction  (see   Point  4  in
  \S\ref{cook_star}).

\item
Apply   the  aperture  corrections   to  transform   the  instrumental
magnitudes into  OBMAG: these will  of course be different  in general
for extended objects  than for stars, so the  corrections given in the
tables should  not be applied  blindly.  However, for  barely resolved
galaxies, the aperture corrections may  be similar to the stellar ones
if  the apertures  are at  least several  pixels in  radius.   If high
accuracy  is   required,  consider  simulating   galaxy  profiles  and
convolving  them  with  the  PSFs  of the  appropriate  bandpasses  to
determine the amount  of light lost through the  adopted aperture.  If
one is only interested in galaxy  colors for a given aperture, and the
PSFs  of  the  two  bandpasses  are nearly  identical,  this  step  is
unnecessary;  however,  for  colors  involving  the  near-IR  filters,
particularly F850LP, there will in  general be more light lost through
the redder  filter.  In this case,  one can determine the  size of the
differential  aperture  correction for  the  two  bands by  convolving
galaxy models with the respective PSFs or else deconvolving the images
and measuring the color change for the given aperture.

\item  Apply the  appropriate zeropoint  from Tables  \ref{wfczpt} and
  \ref{hrczpt} to transform OBMAG to any choice of magnitude system.
\item If  your observations were processed with  CALACS before January
  6, 2004,  and the gain is  different from the default  setting (1 for
  WFC, 2 for HRC), use Table \ref{gainfix} to correct the photometry.
\item The magnitude can be corrected for interstellar extinction using
  Tables~\ref{est_w}  and \ref{est_h}.  Given  E(B-V), the  extinction
  A(P) in the  passband P is calculated by  multiplying E(B-V) by the
  value  in the  table which  correspond to  the correct  passband and
  spectral type.

\item If  you want  to transform the  data to a  different photometric
  system see  \S\ref{galcoltx}.  You  should apply the  correction for
  interstellar  extinction  (see  \S\ref{red} )  before  applying  the
    transformations.

\end{enumerate}

\subsection{Surface Photometry}
\begin{enumerate}
\item Retrieve the  image(s) from the HST archive  or proceed with the
  manual  recalibration of  the  data following  the  recipe given  in
  ADHv4. Be sure  to use the best reference  files usually available a
  few weeks after the data acquisition.

\item After running CALACS on  the association file the main output is
  the flat fielded, corrected {\em FLT} image and/or the geometrically
  corrected {\em DRZ} image (see Point 2 in \S\ref{cook_star}).

\item Measure the surface brightness in units of e$^\_$/sec/pixel.  If
  one  is interested in  absolute surface  photometry within  a single
  bandpass,  we recommend using  the distortion-corrected  (DRZ) image
  and converting  from flux per pixel  to flux per  arcsec$^2$ via the
  output pixel  scale used in  the drizzling process.  The  FLT images
  may  also be  used,  and since  the  pixel area  variation has  been
  divided out by applying the flat field, it is not necessary to apply
  the pixel area  map.  However, in this case,  the appropriate factor
  to use  in the conversion from  pixel area to  arcsec$^2$ depends on
  the mean pixel scale over  the detector region used in normalization
  of the flat fields.  Of course, if one is only interested in surface
  colors,  this factor cancels  (since the  flat fields  for different
  bandpasses were normalized  in the same way), and  the colors can be
  determined simply from the flux per pixel in the two bands.

\item The  effect of  the PSF  red halo should  also be  considered in
  doing surface photometry of extended objects in near-IR filters (see
  discussion in \S\ref{halo}).

\item  Convert   the  count  rate  into   OBMAG/arcsec$^2$  using  the
  appropriate pixel scale.

\item  Apply the  appropriate zeropoint  from Tables  \ref{wfczpt} and
  \ref{hrczpt}   to  transform  OBMAG/square   arcsec  to   a  surface
  brightness in your choice of magnitude system.

\item If  your observations were processed with  CALACS before January
  6, 2004  and the  gain was different  from the default  setting, use
  Table \ref{gainfix} to correct the photometry.

\item If  you want  to transform the  data to a  different photometric
  system use Eq.\ref{transeq} with the coefficients in \S \ref{transf}
  which are  appropriate for the  source and target systems.   For the
  color   conversion  of   galaxies   use  the   instructions  in   \S
  \ref{galcoltx}.   You should apply  the correction  for interstellar
  extinction before applying the transformations.

\end{enumerate}

\section{Aperture Correction for Red Objects: EXAMPLES}
\label{halo_example}

{\bf CASE 1: Photometry of a  High-Redshift Galaxy }\\ 

The red  ``halo'' can  affect both  the magnitude and  the color  of a
galaxy.  The most commonly  used software for performing photometry of
galaxies  is  SExtractor  (Bertin  \& Arnouts,  1996)  which  provides
several  measurements of the  magnitude.  The  most commonly  used for
faint galaxy  photometry is MAG\_AUTO, an  aperture magnitude measured
within an elliptical  aperture adapted to the shape  of the object and
with a width scaled {\em  n} times the isophotal radius.  Other common
choices  are  MAG\_ISO,  an  isophotal  magnitude  that  measures  the
integrated light above a  certain threshold, and MAG\_APER, a circular
aperture. MAG\_AUTO is the best choice for estimating total magnitudes
but even for filters not  affected by long wavelength light scattering
it excludes a significant amount of the light (Ben\'itez et al. 2004).
In  order to  accurately measure  the color  of galaxies  it  is quite
common to adopt  a single aperture defined by  a detection image
\footnote{A detection image is either an image from one chose band, or
  the  weighted  sum  of  images  obtained  in  different  bands}  and
selecting  MAG\_ISO or MAG\_APER.   If all  filters have  similar PSFs
then the magnitude measurements in all filters will be affected by the
same systematic errors that cancel out when subtracting the magnitudes
to calculate the colors.  If  the PSFs are different, then an aperture
correction ought to be applied to the magnitudes in each filter.

For this simulation  we use a model of a SB2  galaxy at redshift z=5.8
normalized  to a magnitude  ABMAG =  26.0 in  the F850LP  filter.  The
color of this galaxy is F775W-F850LP  = 1.65 (1.71) in ABMAG WFC (HRC)
system.  Using the data of the characterization of the long wavelength
light  scattering  and   SYNPHOT,  we  estimate  that the  effective
wavelength  $\lambda_{eff}$  and  the  fraction  of  the  total  light
enclosed in a 0.3'' aperture radius at the calculated $\lambda_{eff}$.

\footnotesize
\begin{tabular}{l p{0.75in}| l l l}

               &                 & F775W    &  F850LP\\
\hline
WFC            &   $\lambda_{eff}$(\AA)& 8026.2   &  9125.4      \\
               &      EE(r=$0\farcs3$) & 0.879     &  0.816\\
HRC            &   $\lambda_{eff}$(\AA)& 7997.0   &  9210.8      \\
               &      EE(r=$0\farcs3$) & 0.809     & 0.674 \\
\end{tabular}   
\normalsize

If we remove  the AB zeropoints and scale the total  counts for the EE
for an aperture of $0\farcs3$ radius we obtain the following OBMAG:

\footnotesize
\begin{tabular}{l l l}
               &    F775W    &  F850LP\\
\hline
WFC            &    2.14   &  1.36      \\
HRC            &    2.99   &  2.04      \\
\end{tabular}  
\normalsize

At this point there are two options: treat the measured counts
as  total counts,  or  decide  to apply  an  aperture correction.  The
aperture  correction can  be calculated  from different  sources, from
stars in the FOV, from the  tabulated EE for the standard stars, using
SYNPHOT    or    calculating     the    effective    wavelength    and
Tables \,\ref{WFC_acl} and\,\ref{HRC_acl}.   In Table \ref{halo_ex1} we
report the different cases and show the different results.

\bigskip

{\bf CASE 2: Photometry of a L-dwarf Star}\\ 

For  this  example  we  simulate  observations of  a  L3.5  star  with
magnitude VEGAMAG = 22.00 in the  WFC F775W filter.  The color of this
star is F775W-F850LP  = 1.84 (1.99) in VEGAMAG  WFC (HRC) system.  For
this example we  perform photometry in a $0\farcs2$  (4 pixels) radius
aperture for WFC, and a $0\farcs1$ (4 pixels) radius aperture for HRC.
Using the observed  spectra of the L3.5 star  2M0036+18 and SYNPHOT, we
estimate the effective  wavelength $\lambda_{eff}$ of the observations
and the fraction of the  total light enclosed in the selected aperture
radius. \\
 \footnotesize
\begin{tabular}{l p{0.75in}| l l l}

               &                 & F775W    &  F850LP\\
\hline
WFC            &   $\lambda_{eff}$(\AA)& 8003.7   &  9383.5      \\
               &      EE(r=$0\farcs2$) & 0.829     &  0.718\\
HRC            &   $\lambda_{eff}$(\AA)& 7980.4   &  9445.2      \\
               &      EE(r=$0\farcs1$) & 0.629     & 0.414 \\
\end{tabular}   
\\
\normalsize

If we remove the VEGAMAG zeropoints and scale the total counts for the
EE an aperture of 4 pixels radius we obtain the following instrumental
magnitude: \footnotesize
\\
\begin{tabular}{l l l}
               &    F775W    &  F850LP\\
\hline
WFC            &    $-$3.05   &  $-$3.81      \\
HRC            &    $-$2.00   &  $-$2.70     \\
\end{tabular}  
\\
\normalsize

At this  point it is  important to address  the issue of  the aperture
correction.  The aperture correction  can be calculated from different
sources, from stars in the FOV, from the tabulated EE for the standard
stars, using SYNPHOT or calculating the effective wavelength and using
Tables\,\ref{WFC_acl} and\,\ref{HRC_acl}.   In Table \ref{halo_ex2} we
report the different cases and show the different results.

\section{Supporting material for color transformations}
All figures and tables in this appendix are in support of
sections \ref{a2g} and \ref{a2wfpc2}. Readers should read
the appropriate section before using the coefficient
listed in these tables in Eq. \ref{transeq}.

\subsection{Example}\label{example}
Here we present,  as an example, a transformation  between the HRC and
the WFC photometric systems.  Suppose we observed the globular cluster
NGC~2419  with  ACS/HRC  in  F555W  and  F814W  and  have  stored  the
photometric tables  in the ABMAG  HRC photometric system.  We  want to
compare this result with previous WFC observations of the same cluster
in the VEGAMAG system.  This is  the procedure that we need to follow,
step by step:

\noindent 1) Transform the HRC ABMAG photometry to OBMAG\\
This is simply done by subtracting the ABmag zeropoints (Table \ref{hrczpt} )
from our dataset:\\
\\
\begin{tabular}{c}
$F555W_{HRC,OBMAG} = F555W_{HRC,ABMAG} -  25.248  $ \\
$F814W_{HRC,OBMAG} = F814W_{HRC,ABMAG} -  25.287  $  
\end{tabular}\\

\noindent 2) Compute the first estimate of TCOL \\ 
This is the less critical  step and  can be done by simply assuming\\
\\
 $F555W_{WFC,OBMAG}$ = $F555W_{HRC,OBMAG}$
and\\
 $F814W_{WFC,OBMAG}$ = $F814W_{HRC,OBMAG}$ \\
$TCOL\,\equiv\, F555W_{HRC,OBMAG}\, -\, F814W_{HRC,OBMAG}\,\simeq\, F555W_{WFC,OBMAG}\, -\, F814W_{WFC,OBMAG} $

\noindent  3)  Transform  $F555W_{HRC,OBMAG}$  to  $F555W_{WFC,OBMAG}$
\\ \\ From Table \ref{hrc2wfc}, we select the coefficients in the {\em
  synthetic} column for SMAG=F555W TMAG=F555W and TCOL=F555W-F814W and
use  them   in  eq.   \ref{transeq}:\\  $F555W_{WFC,OBMAG}  \,   =  \,
F555W_{HRC,OBMAG}   -0.470    -   0.004   \phantom{x}TCOL    +   0.000
\phantom{x}TCOL^2 $\\

\noindent  4)  Transform  $F814W_{HRC,OBMAG}$  to  $F814W_{WFC,OBMAG}$
\\  
We proceed  similarly to  Step 3  for the  F555W but  in  the {\em
  synthetic}  column for  SMAG=F814W, TMAG=F814W  and TCOL=F555W-F814W
there  are  two  sets  of  coefficients  depending  on  the  value  of
TCOL:\\   for  $TCOL   <   -0.2$  \\   $F814W_{WFC,OBMAG}   \,  =   \,
F814W_{HRC,OBMAG}   -0.655    -   0.012   \phantom{x}TCOL    -   0.031
\phantom{x}TCOL^2 $\\  for $TCOL > -0.2$\\ $F814W_{WFC,OBMAG}  \, = \,
F814W_{HRC,OBMAG}   -0.651    +   0.003   \phantom{x}TCOL    +   0.003
\phantom{x}TCOL^2 $\\

\noindent 5) Now we have  better estimate of $F555W_{WFC,OBMAG}$ and $
F814W_{WFC,OBMAG}$ from steps (2)  and (3).  Obviously, TCOL should be
updated.\\ $TCOL = F555WW_{WFC,OBMAG}\, -\, F814W_{WFC,OBMAG}$\\

\noindent 6) Repeat Steps 3 trough 5  until convergence is reached.
Usually not more than 10 iterations are needed.
  
\noindent 7) Convert  $F555W_{WFC,OBMAG}\,$ and $\, F814W_{WFC,OBMAG}$
into   the    VEGAMAG   by   adding   the    zeropoints   from   Table
\ref{wfczpt}:\\  $F555W_{WFC,VEGAMAG} =  F555W_{WFC,ABMAG} +  25.711 $
\\ $F814W_{WFC,VEGAMAG} = F814W_{WFC,ABMAG} + 25.487 $ \\

In Fig.~\ref{tranexp} we show  the comparison between  the HRC-to-WFC
transformed  magnitude  and  the  WFC  photometry  over  plotting  the
transformed magnitudes in the observed WFC color magnitude diagram and
plotting the differences in magnitude in the two filters as a function
of the color.

\subsection{Transformation from ACS to UBVRI}
{\bf  NOTE FOR EDITORS: insert here:}\\
Fig.~\ref{cmds_ground}\\
Tables  \ref{wfc2ground}  and  \ref{hrc2ground}\\
Figs.~\ref{tranw2gr} and \ref{tranh2gr}.
Figs.~\ref{tranw2gr_prob}  and  \ref{tranh2gr_prob}

\subsection{Transformation from ACS to WFPC2}
{\bf  NOTE FOR EDITORS: insert here:}\\
 Fig.~\ref{cmds_wfpc2}\\
Tables  \ref{wfc2wfpc2}  and   \ref{hrc2wfpc2} \\
 Figs.~\ref{tranw2wfpc2}  and \ref{tranh2wfpc2}.

\section{Transformation of Galaxy Templates}\label{galcoltx}

The color  transformations described  in the previous  sections cannot
offer accurate  results for all  spectral shapes, especially  when the
effects  of  redshift  and  intergalactic absorption  are  taken  into
account.  Therefore it is desirable to describe the colors of observed
galaxies at different redshifts in the different systems as accurately
as possible.   To do  so, we  use a variation  of the  CWWSB templates
library,  introduced  in  Ben\'itez  (2000) to  calculate  photometric
redshifts and  formed by the main  types (E/SO, Sbc, Scd  and Im) from
Coleman,  Wu and  Weedman 1980  and two  starbursts (SB2  and  SB3) of
Kinney  et al.   1996.  Ben\'itez  et al.   (2004; hereafter  B2004)
presented a variant  of this library, which has  been calibrated using
spectroscopic  redshift  catalogs with  good  quality photometry.   By
definition,  the new  templates agree  much better  with  the observed
colors  of real  galaxies,  and for  instance  reduce the  photometric
redshift scatter by a factor of 25\% in the HDFN, although they have
not been optimized for that goal.

We generate estimated colors for the B2004 library in intervals of 0.05
in redshift, using the ACS WFC F606W filter as reference, for the WFC,
HRC,  WFPC2 and  Johnson-Cousins  systems.  We  apply corrections  for
intergalactic  hydrogen extinction  using the  prescriptions  of Madau
(1995). All colors, except the Johnson-Cousins ones, correspond to the
AB system. The colors were  calculated using the functions included in
the  module   bpz\_tools  from   the  BPZ  package   (Ben\'itez  2000,
\url{http://acs.pha.jhu.edu/$\sim$txitxo/bayesian.html})    which   offers
similar accuracy to  SYNPHOT.  The Vega reference spectrum  is that of
Bohlin \& Gilliland (2004).

We present here only the tables relative to the E/S0, Sbc and Im templates
for the WFC and Johnsons-Cousins system. Tables for other templates are
available on-line at \url{http://acs.pha.jhu.edu/instrument/ photometry/}

Tables  \ref{tx_w_el}-\ref{tx_lan_sbc}  must  be  used  with  caution,
however, because obviously these  6 ``atomic" spectral templates cannot
represent all  the possible variations  of real galaxy  spectra except
for some limited populations.  Moreover, the colors at given redshifts
obtained here are evaluated  without the consideration of the spectral
evolution,  whose   absence  may  introduce   significant  systematics
particularly  at  high  redshifts.   Nevertheless, despite  all  these
limitations, one  can still use these  tables as a  guide to transform
galaxy  colors  between  different  photometric systems  for  a  given
spectral type and redshift.  To this  aim, we need to find the closest
possible match in redshift and spectral type in the electronic tables,
using as many colors as  possible.  Then, we look up the corresponding
color in  the table corresponding to  the spectral type  and filter of
interest, using interpolation between  spectral types and redshifts if
needed.

For  example,  let  us  suppose  we  have  ACS/WFC  photometry  of  an
elliptical galaxy at  redshift z = 1 with ACS/WFC ABmag  F606W - F814W =
1.82.  If we want to know the F606W - F814W color in the
HRC ABmag system,  from Table \ref{tx_h_el} we can  read the following
multi-instrument  colors for  z = 1:\\ 
HRC$_{F606W}$  -  WFC$_{F606W}$ = 0.04 and \\ 
HRC$_{F814W}$ -  WFC$_{F606W} = -1.87$ \\
 Therefore we can
directly calculate  the ACS/HRC ABmag  color for this galaxy:  F606W -
F814W =  1.91.  Moreover, if we  need to compare  the ACS observations
with  ground-based observations, for  example in  the R-I  color, from
Table \ref{tx_lan_el} we  derive for z = 1:\\ 
R -  WFC$_{F606W} = -0.59$ and \\ 
I - WFC$_{F606W} = -1.59$ \\ 
and therefore R-I = 1.00.

\section{Optimal Aperture Size}
\label{appA}

There are several factors to take into account when selecting the size
of  the  aperture when  measuring  the  flux  of stars.   For  stellar
photometry,  the  largest signal-to-noise  ratios  (i.e., the  smallest
photometric  errors)  are  generally  obtained with  relatively  small
apertures (Howell  1989). Large  apertures can have  large photometric
errors when the total signal  from the source become comparable to the
total background signal. They  also can suffer from measurement errors
from the background flux.  On the other hand, very small apertures can
suffer  from  small-number  statistics  because of  the  low  measured
signal.

There  are a  number  of  trade-offs to  consider  when selecting  the
optimal  aperture  size.  The  signal  level  of  the source  and  the
background noise  are the main components.  Several  other factors can
impact  the choice of  the aperture  radius, such  as the  PSF spatial
variations, the EE profile, the read noise and the crowding.  In order
to provide  a more quantitative analysis of  the various dependencies,
we can  develop a case  study.  Consider an  isolated star that,  in a
given exposure,  produces $\sim$1000 electrons in an  image taken with
WFC/F606W.   The  default  gain  is  1  e$^-$/ADU  and  the  background
(sky+dark  current) is  $\sim$  10 ADU.   The  read noise  is $\sim$  5
e$^-$. We  can predict the S/N  ratio in different  aperture radii, by
calculating the fraction of source flux enclosed in the aperture area,
and the total noise associated  with the measurement.  The total noise
is  obtained by adding  in quadrature  the Poisson  noise of  the star
signal, the  read noise, the  Poisson noise of the  background signal,
plus  two more  noise  contributions.   The first  one  is related  to
breathing and  focus variations on  orbital timescales that  can cause
small variations in the aperture corrections.  As an initial estimate,
we used  the data  published for WFPC2  (Suchkov \&  Casertano, 1997).
The  second noise  contribution,   $\sim$  10\% of  the aperture
correction, is related to uncertainties in the aperture correction and
its variation  due to the PSF  spatial variation across  the chip.  In
both cases, we  express their impact as percentage  of the source flux
in the  aperture.  They are included  in the noise  calculation in the
form of additional Poisson noise.

According to this simulation, the  highest S/N ratio is obtained with a
4  pixel aperture radius.   Small variations  from the  optimal radius by
$\pm$ 1 pixel make little  difference.  We repeated the same test for
  the F435W  and  F850LP filter  and  we find  similar results.   The
slightly larger PSF  in the F850LP tends to produce  the highest S/N at
slightly bigger  aperture.  The comparison for these  filters is shown
in Fig.~\ref{OAW}.   A similar test for HRC shows  that an aperture of
4-5 pixels is also the optimal  choice for the HRC where the effect of
the  broadening   of  the  PSF   in  the  near-IR  is   more  evident
(Fig.~\ref{OAH}).   Brighter stars have larger  optimal aperture sizes
than do fainter stars.  For a  point source object with signal $>$ 20000
ADU, the optimal choice is a radius of $\sim$1$''$.  If one must use only
one aperture  size, it  is clearly advantageous  to chose  an aperture
size that  produces the smallest  photometric errors for  the sources
of most scientific interest.

\clearpage

\begin{figure}
\epsscale{.75}
\plotone{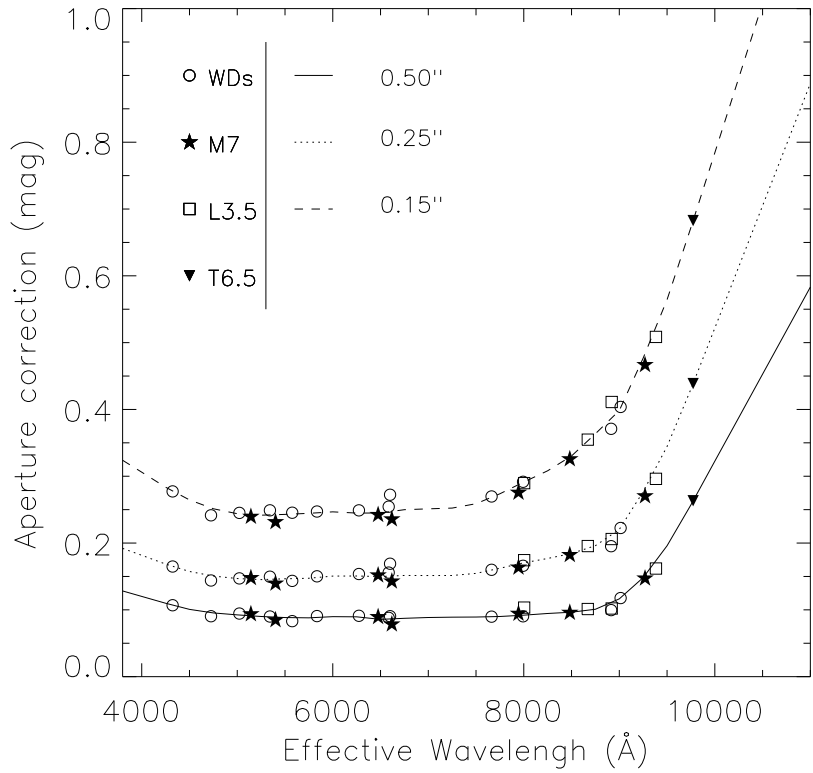}
\caption{Aperture  corrections {\em vs}  effective wavelength  for the
  WFC.  Three  different aperture radii are shown  here.  The aperture
  correction  is from the  selected aperture  to the  nominal infinite
  aperture.  Different stars are  represented with  different symbols.
}
\label{efflam_apt_wfc}
\end{figure}

\begin{figure}
\epsscale{.75}
\plotone{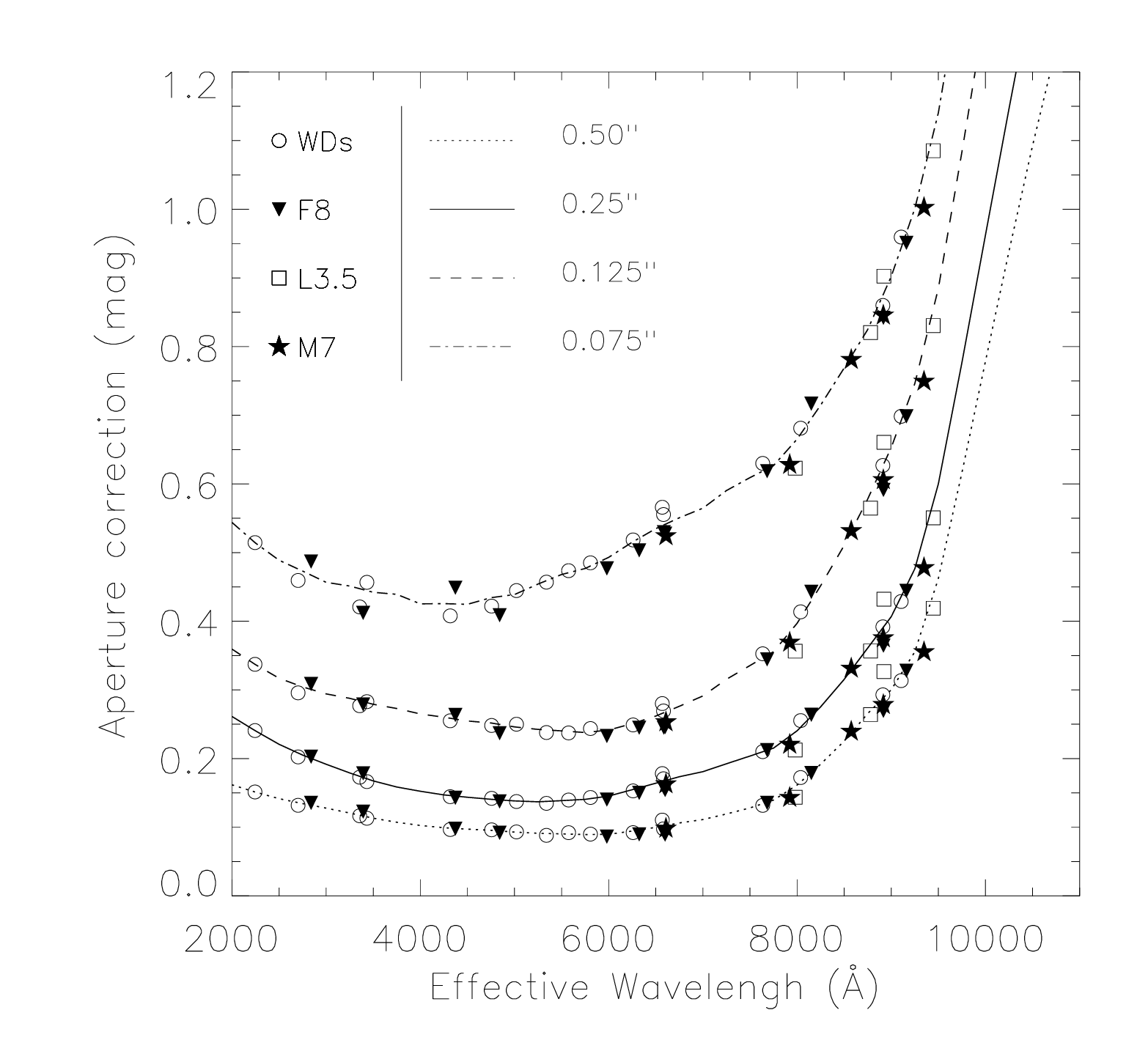}
\caption{Same as Fig. \ref{efflam_apt_wfc} but for the HRC.}
\label{efflam_apt_hrc}
\end{figure}

\clearpage 

\begin{figure}
\epsscale{.7}
\plotone{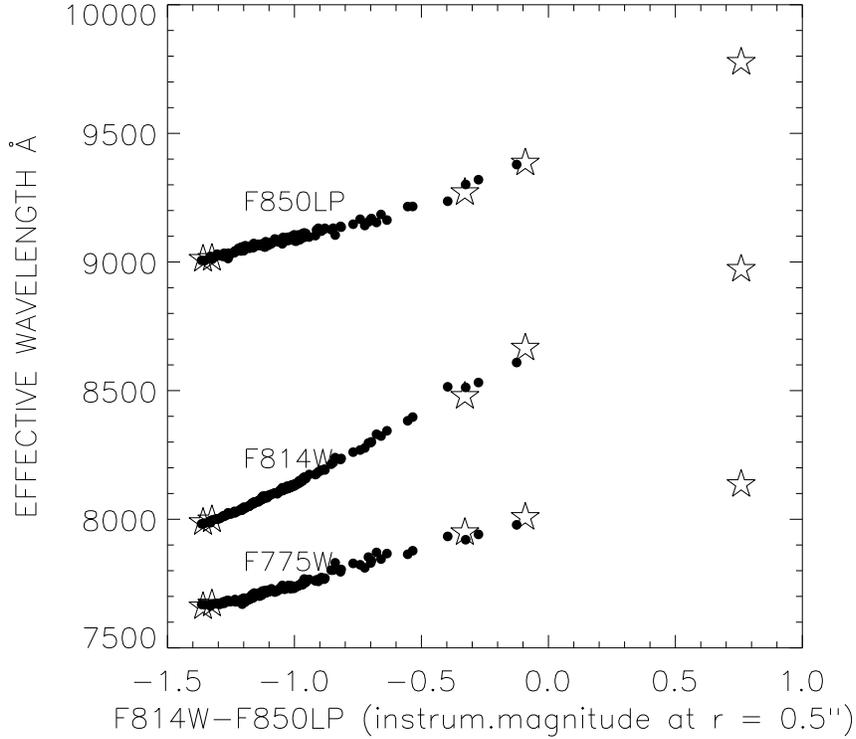}
\caption{Effective  wavelength  {\em  vs}  color  relation  for  three
  near-IR  filters  of WFC.   The  color  F814W-F850LP  is defined  as
  $-2.5\,
  log([counts/sec]_{F814W,0\farcs5}\,/\,[counts/sec]_{F850LP,0\farcs5})$,
  where $[counts/sec]_{filter,0\farcs5}$  is the count  rate within an
  aperture  radius  of $0\farcs5$.  The  dots  represent the  Bruzual,
  Persson, Gunn \& Stryker stellar atlas. The stars are the five stars
  used for the EE study.}
\label{efflam_colIR}
\end{figure}

\begin{figure}
\plotone{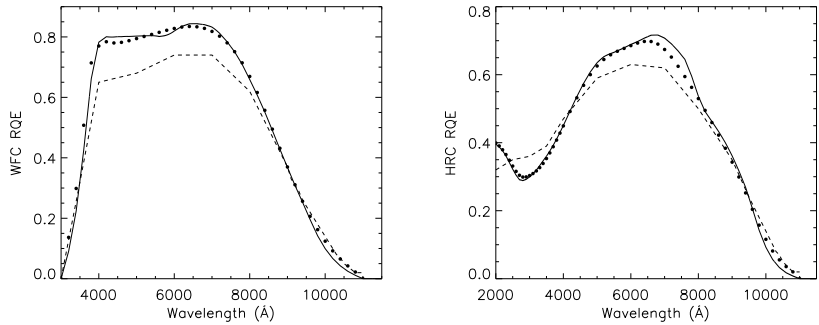}
\caption{Responsive quantum efficiency curves for  the WFC (left) and HRC (right)
  CCDs.  The dashed lines  show the pre-flight measurements, while the
  dotted  lines show the  first on-orbit  correction (Sirianni  et al.
  2002). The final RQE curves resulting from the iterative process are
  shown as a solid line (De Marchi et al. 2004). \label{newdqeplots}}
\end{figure}

\clearpage
 
\begin{figure}
\plotone{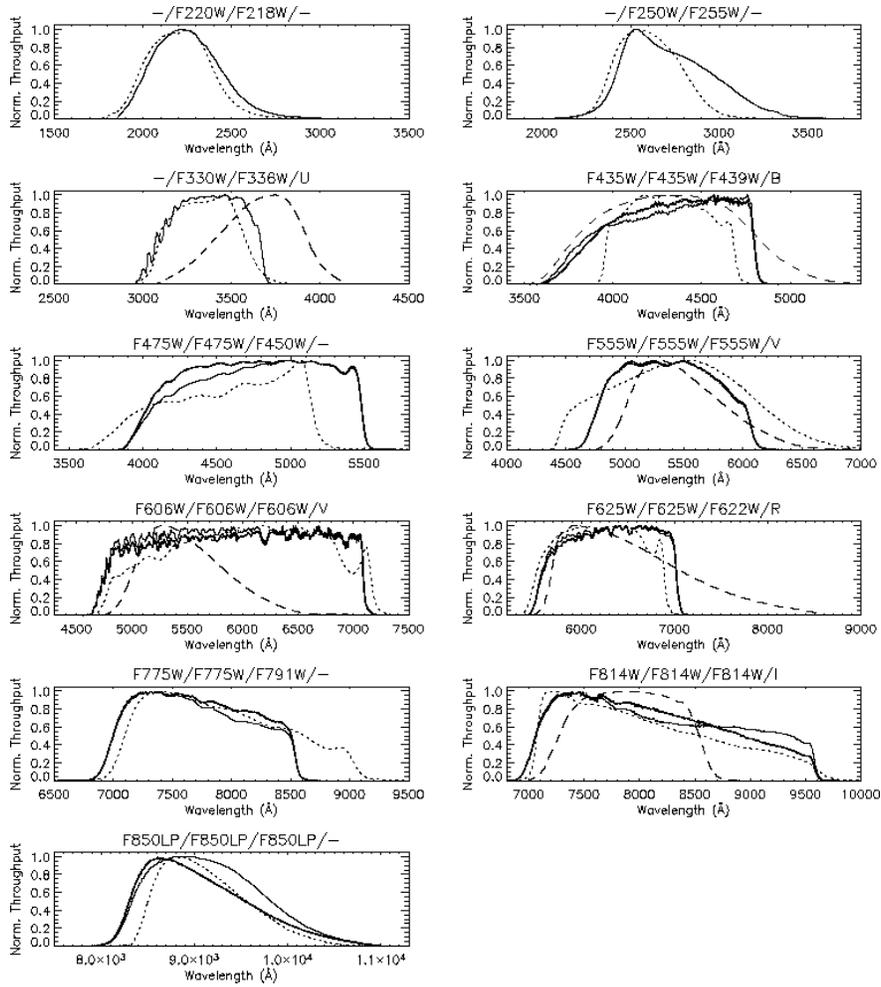}
\caption{Comparison  between  the  total  response curve  for  several
  filters  of WFC  (solid thick  line), HRC  (solid thin  line), WFPC2
  (dotted  line) and the  LANDOLT filters  (dashed line).   All curves
  have been normalized  to their peak value. The  title of each panel
  shows the name of the filter WFC/HRC/WFPC2/LANDOLT. }
\label{filters_w2h}
\end{figure}

\begin{figure}
\centering
\plottwo{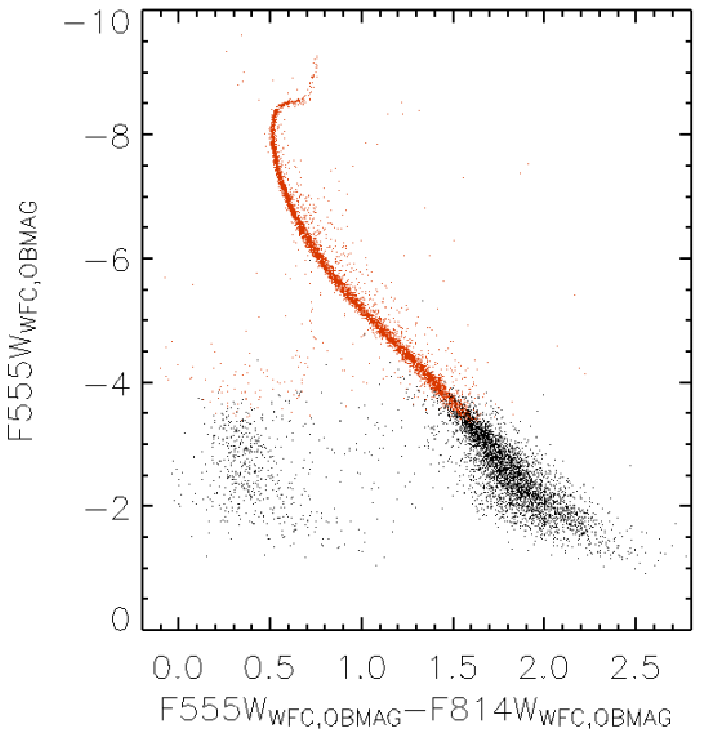}{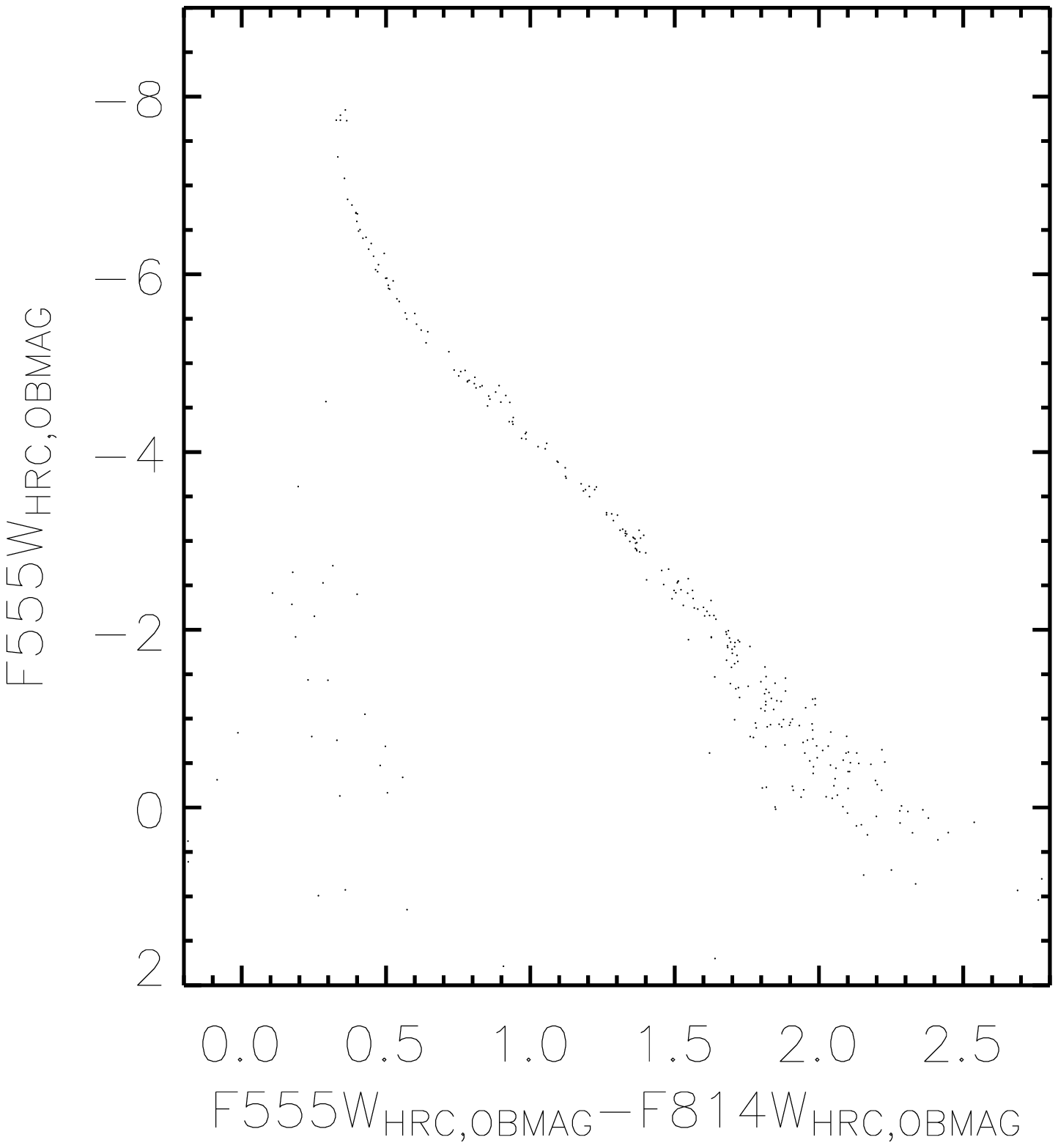}\\
\plottwo{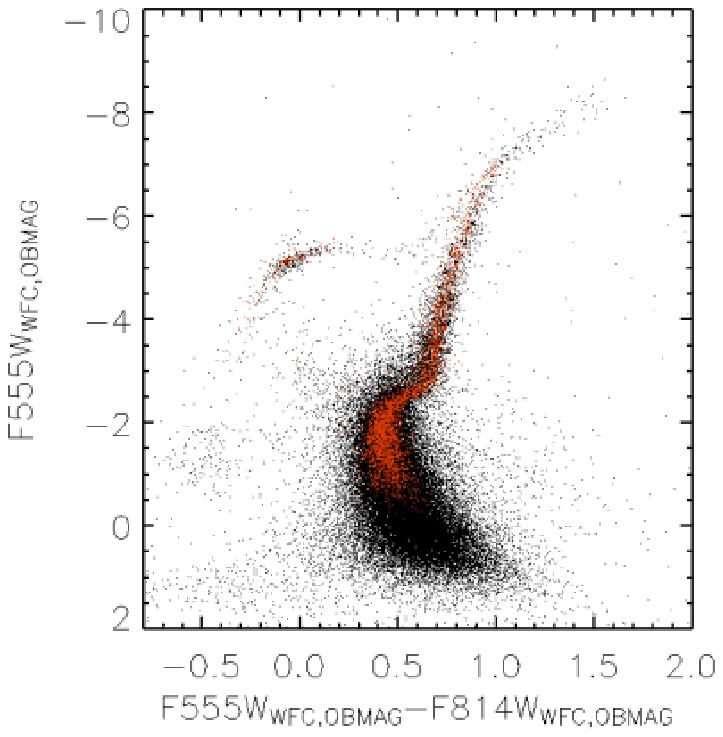}{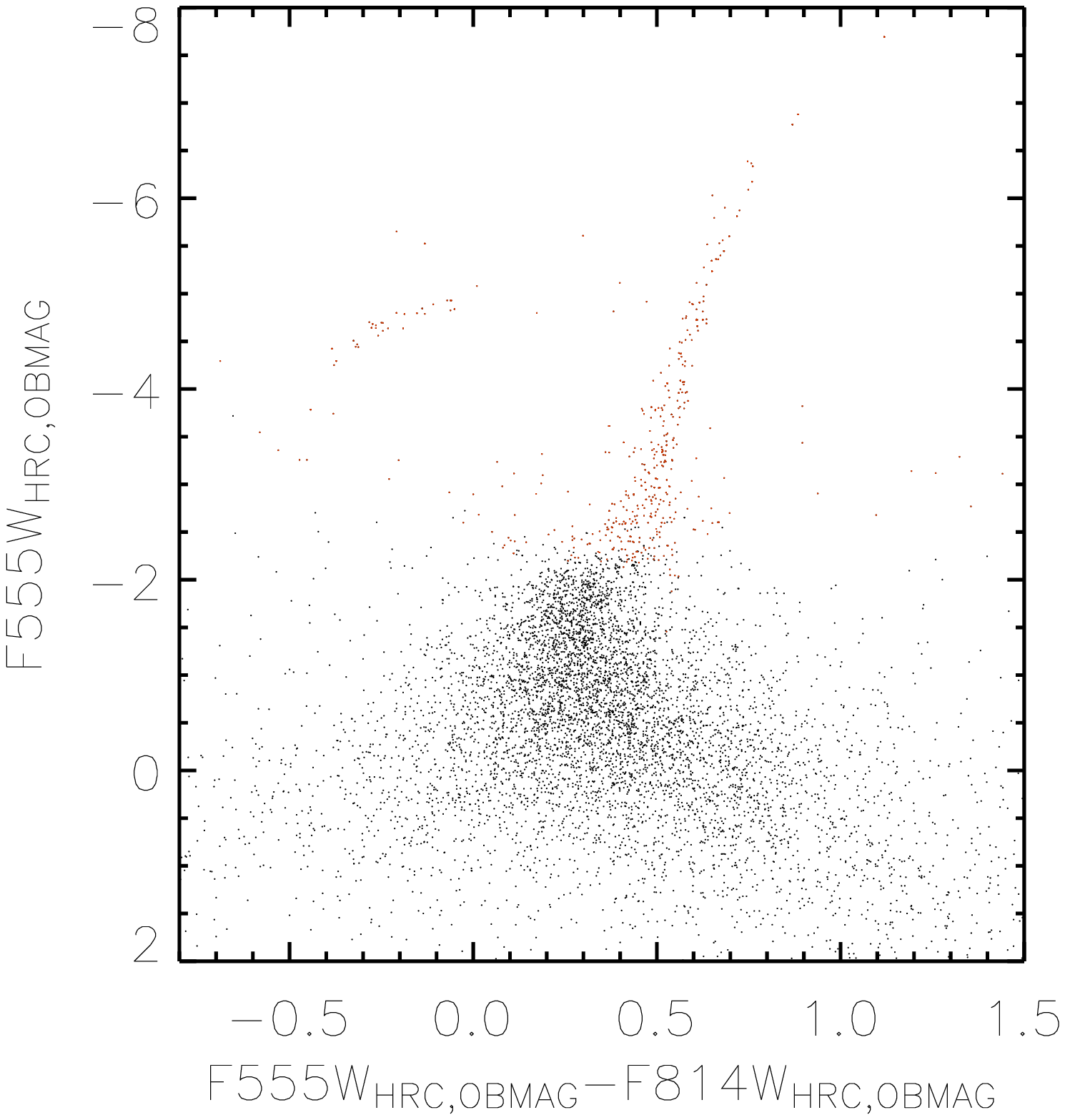}
\caption{Observed  color  magnitude  diagrams  for NGC~104  (top)  and
NGC~2419 (bottom) with WFC (left) and HRC (right).}
\label{cmds_acs}
\end{figure}

\clearpage

\begin{figure}
\begin{center}
\includegraphics[angle=0,height=7.5in,clip=true]{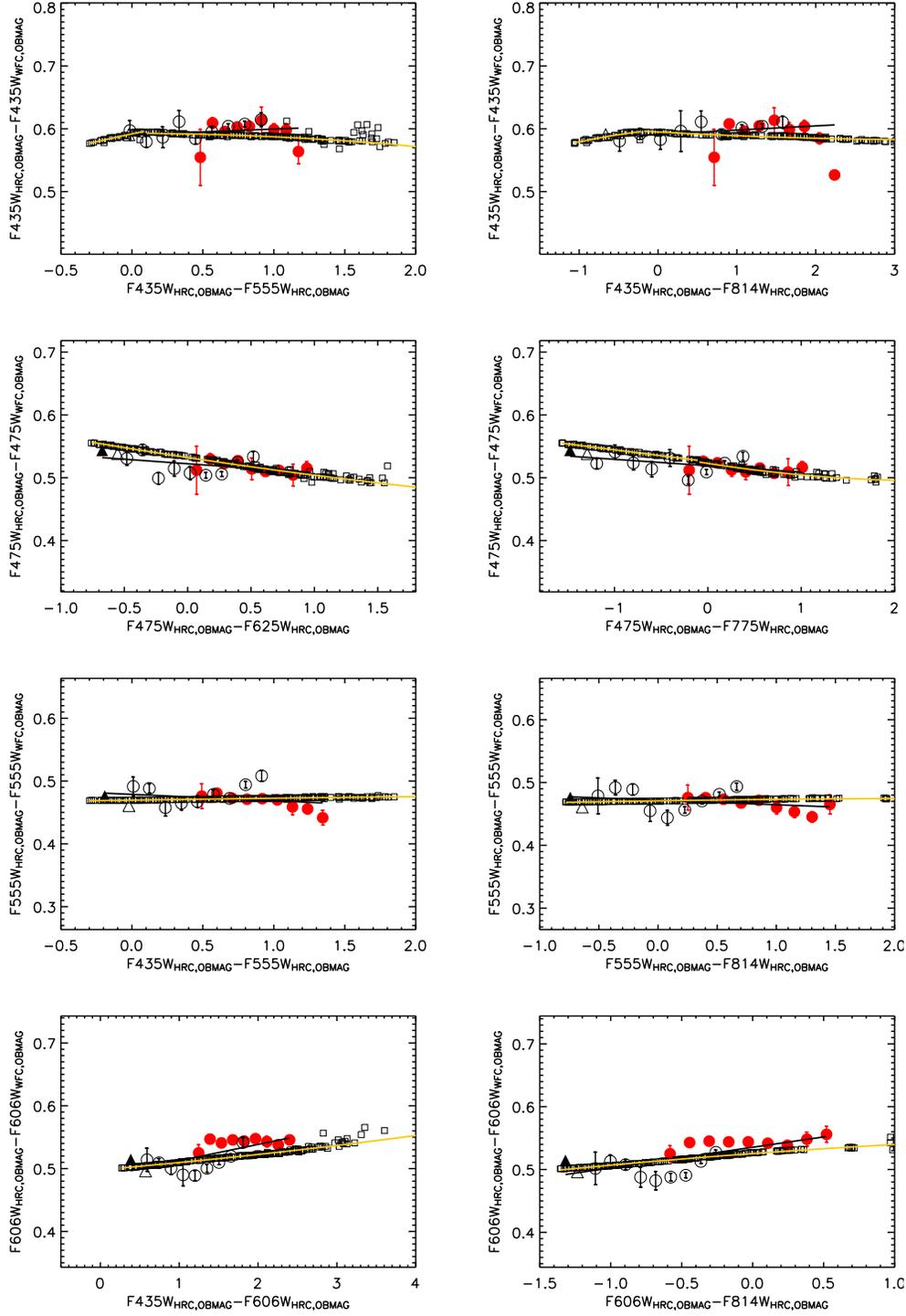}
\end{center}
\caption{Observed and synthetic WFC-to-HRC transformations for primary
  photometric filters. Squares show the synthetic measurements for the
  BPGS atlas.  The circles  are observational data (filled for NGC104,
  open  for   NGC2419).   The  two  standard  stars   are  shown  with
  triangles. The black line shows  the linear fit to the observational
  points, the  yellow curve (light  gray in B\&W) shows  the synthetic
  transformations.\label{tranw2h1}}
\end{figure}

\setcounter{figure}{14}
\clearpage
\begin{figure}
\figurenum{\ref{tranw2h1}}
\begin{center}
\includegraphics[angle=0,height=7.5in,clip=true]{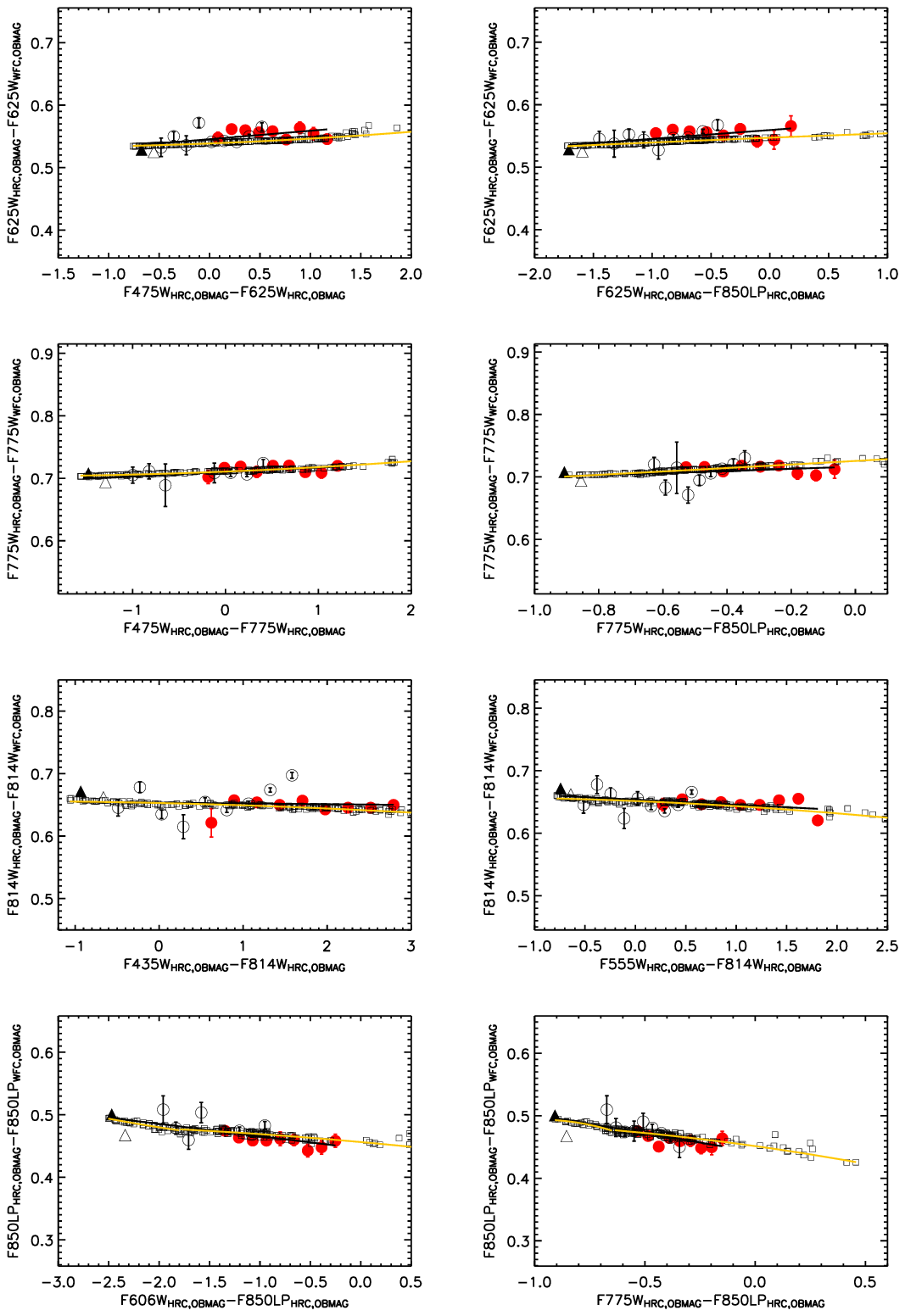}
\end{center}
\caption{Continued.}
\end{figure}

\setcounter{figure}{15}
\clearpage

\begin{figure}
\begin{center}
\includegraphics[width=4in,clip=true]{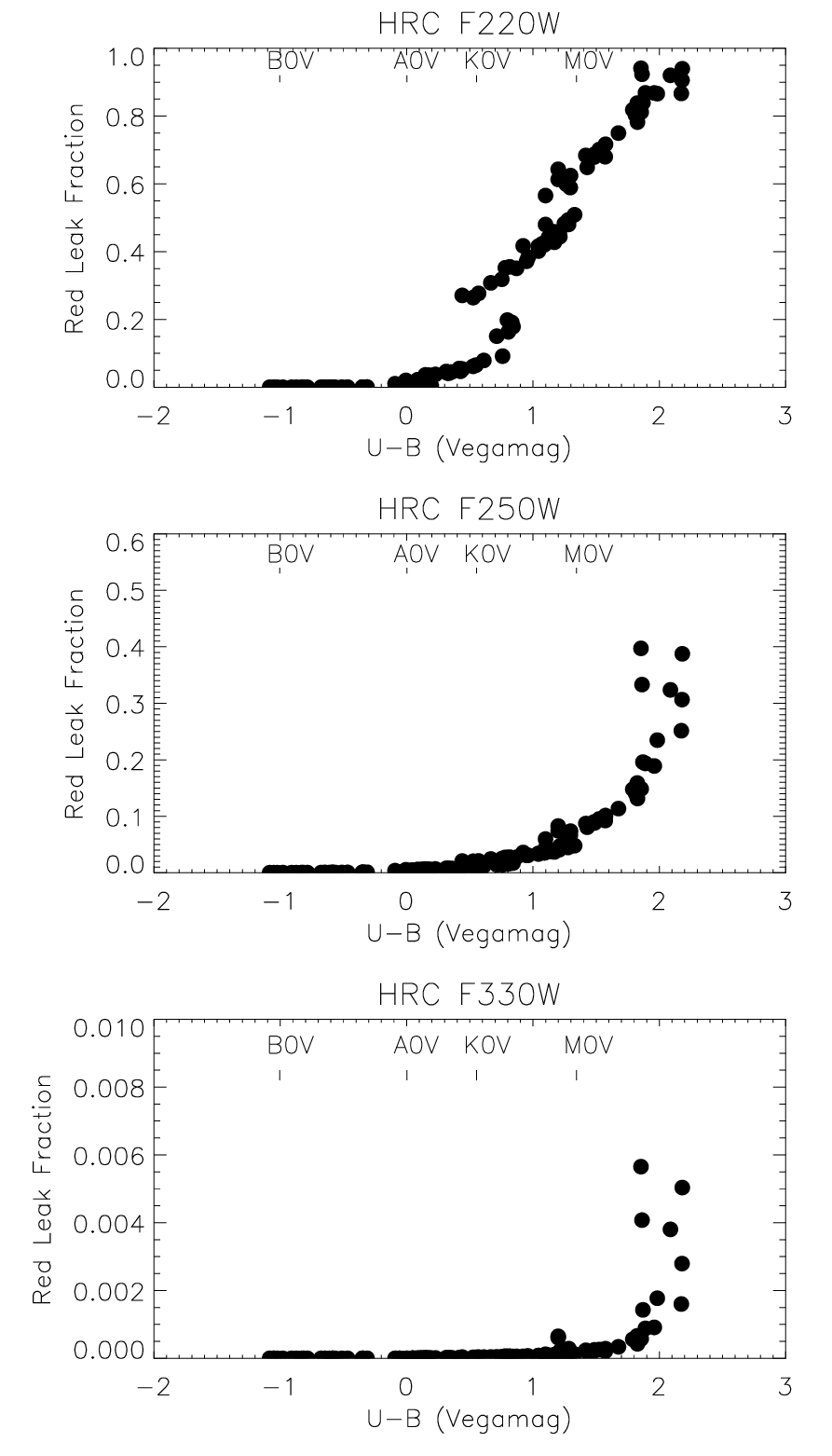}
\end{center}
\caption{Amount  of light contributed  by red  leak ($\lambda  > $4000
\AA) as a function of color.}
\label{rleak2}
\end{figure}

\clearpage
\begin{figure}
\begin{center}
\includegraphics[angle=0,width=5in,clip=true]{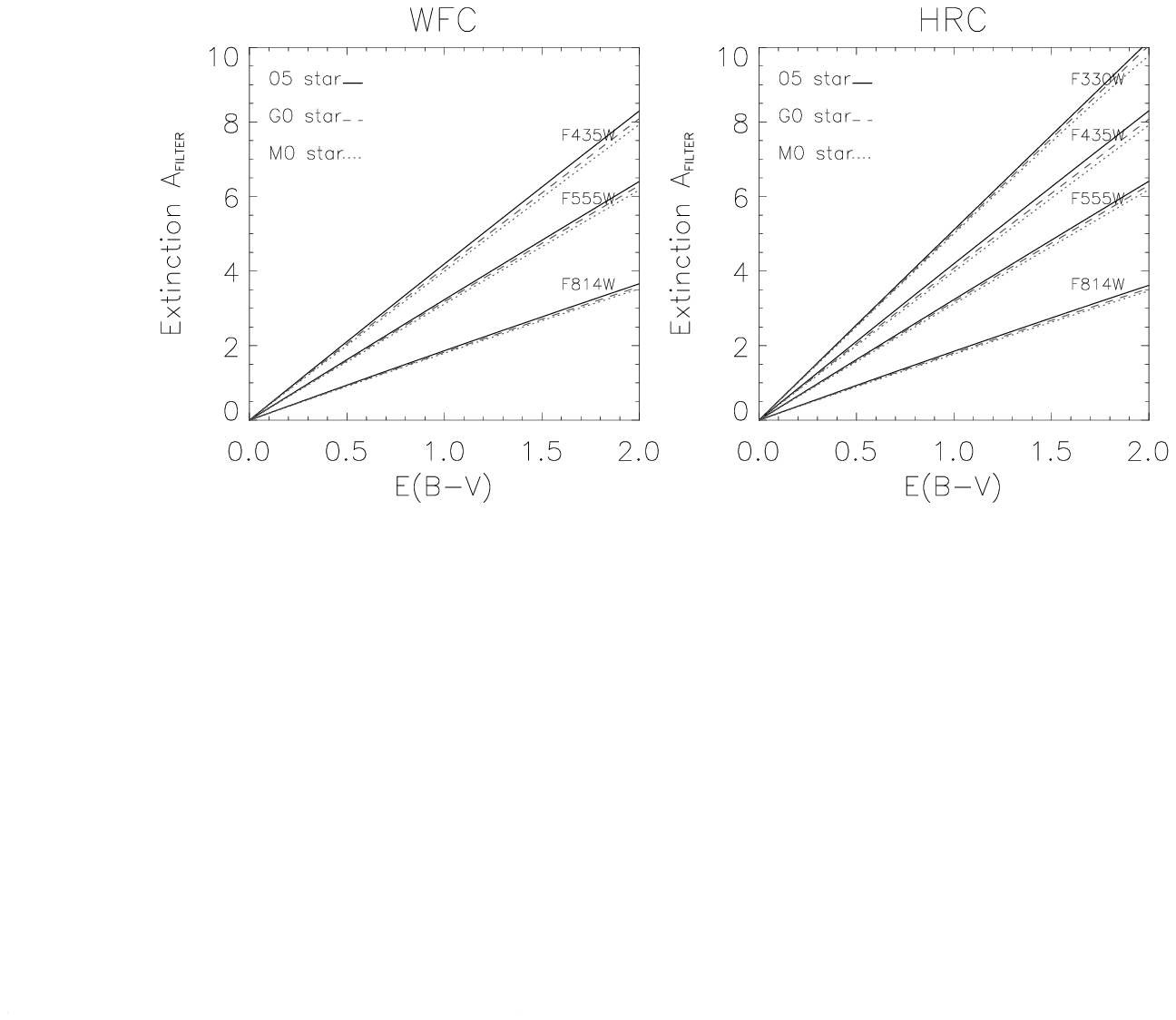}
\end{center}
\caption{Extinction in ACS WFC and HRC filters as a function of E(B-V)
for three stars of different  spectral types. We used the interstellar
extinction curve from Cardelli et al. (1989) with R$_v$=3.1 }
\label{redd1}
\end{figure}

\begin{figure}
\begin{center}
\includegraphics[angle=0,width=5in,clip=true]{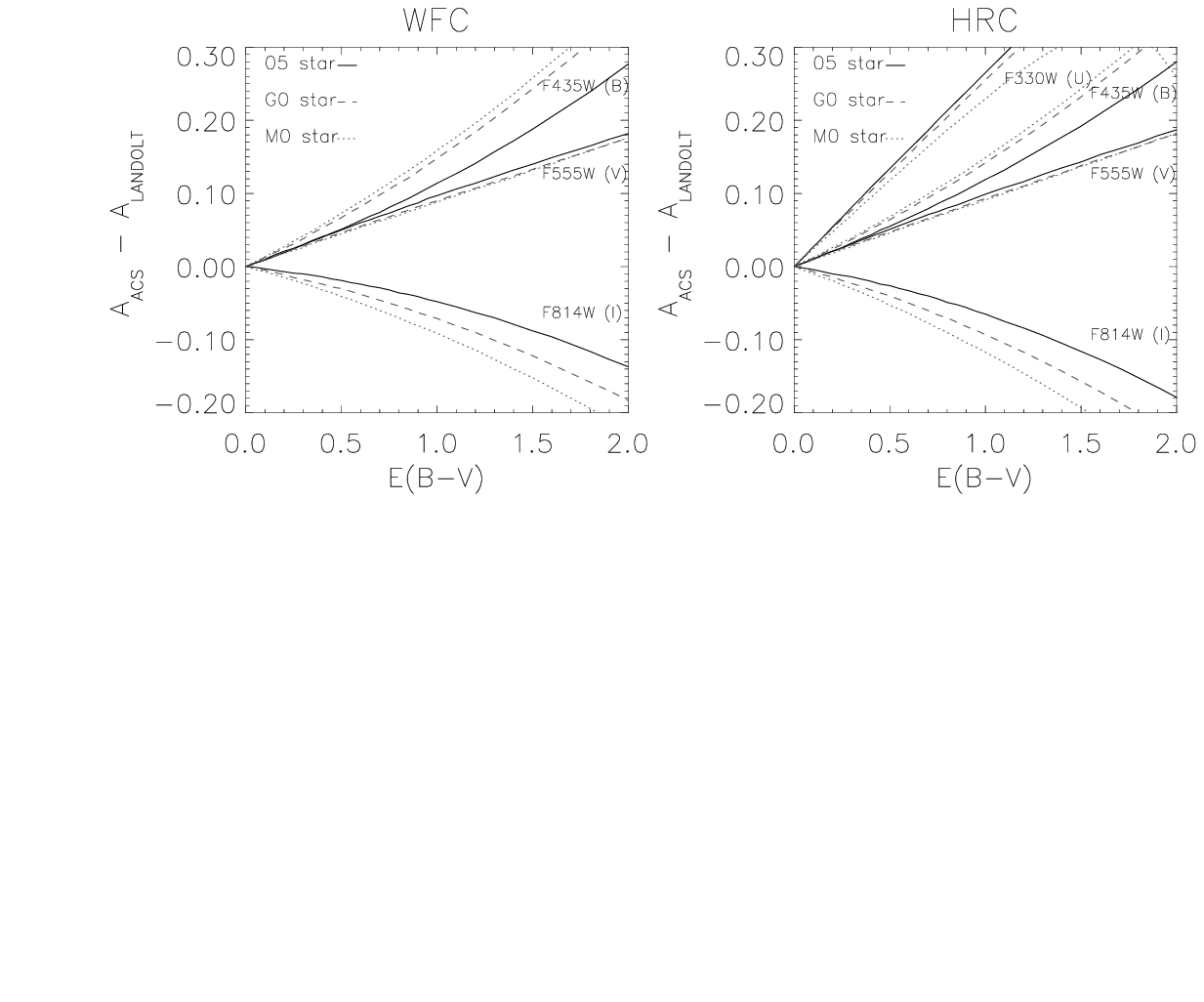}
\end{center}
\caption{Difference  in  extinction between  the  ACS  system and  the
ground system as a function of E(B-V).}
\label{redd2}
\end{figure}

\clearpage

\begin{figure}
\centering
\plottwo{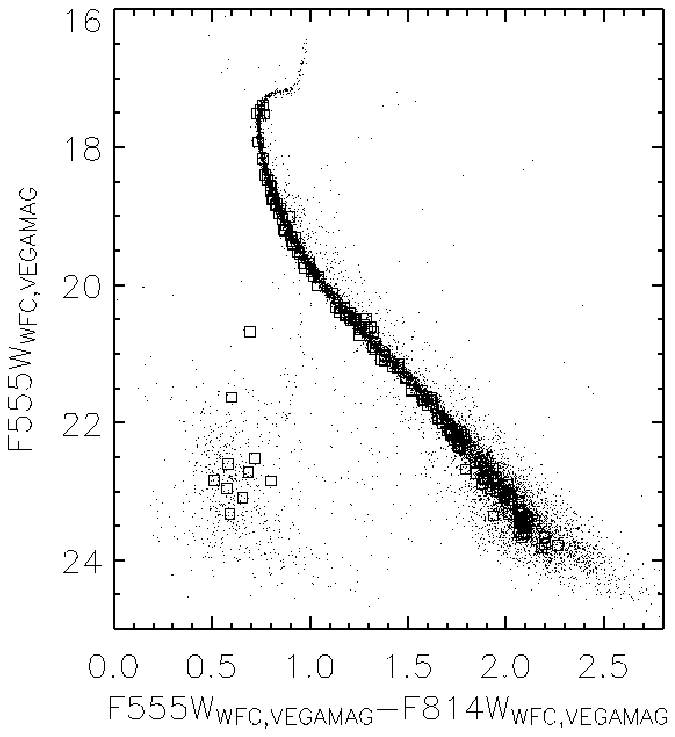}{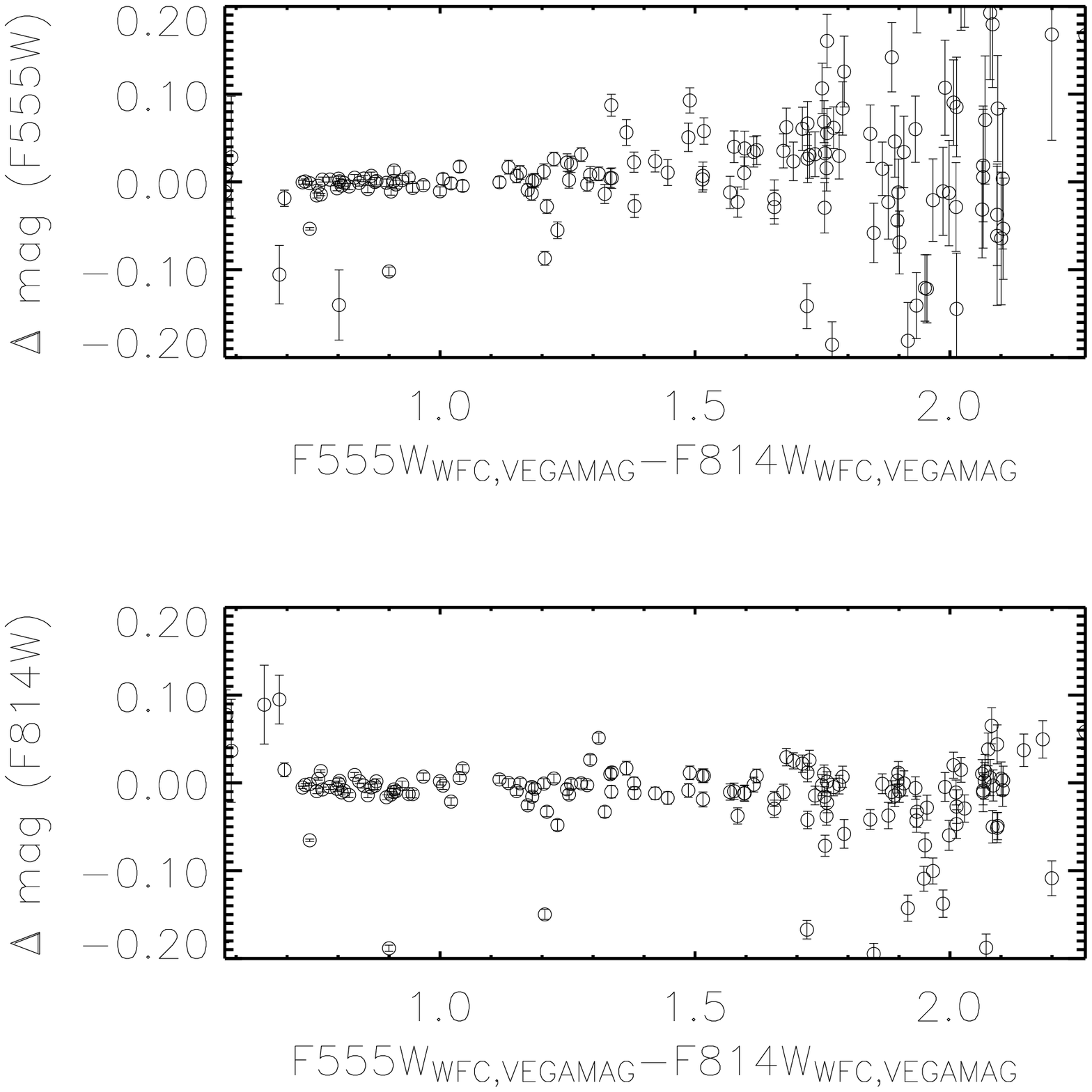}
\caption{Left:  HRC photometry  of NGC~104  (open  square) transformed
  into  the  WFC observational  plane  of  WFC.  Right: Difference  in
  magnitude (Transformed HRC  - WFC) in the two  filters as a function
  of color.  }
\label{tranexp}
\end{figure}

\begin{figure}
\centering
\plottwo{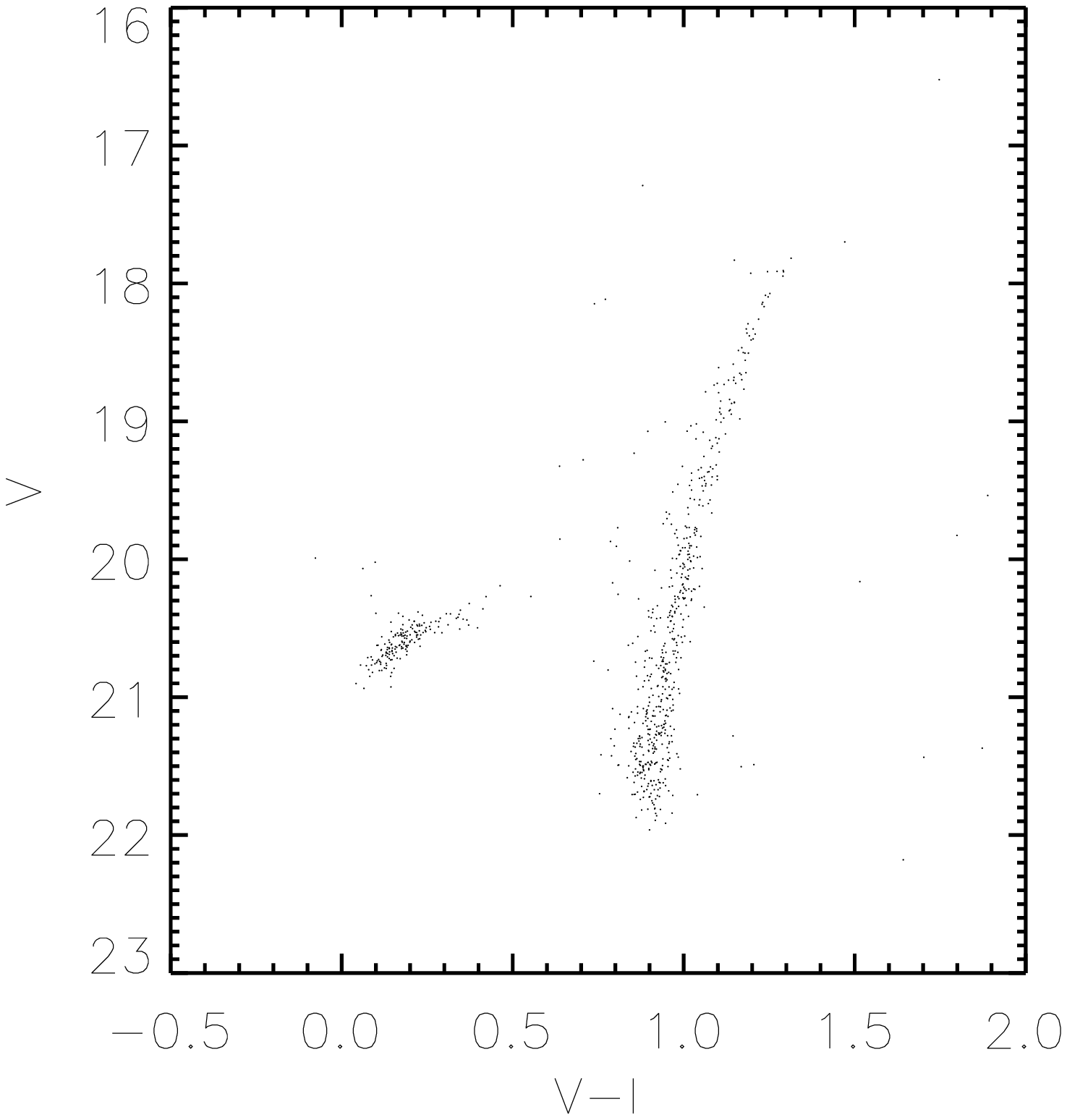}{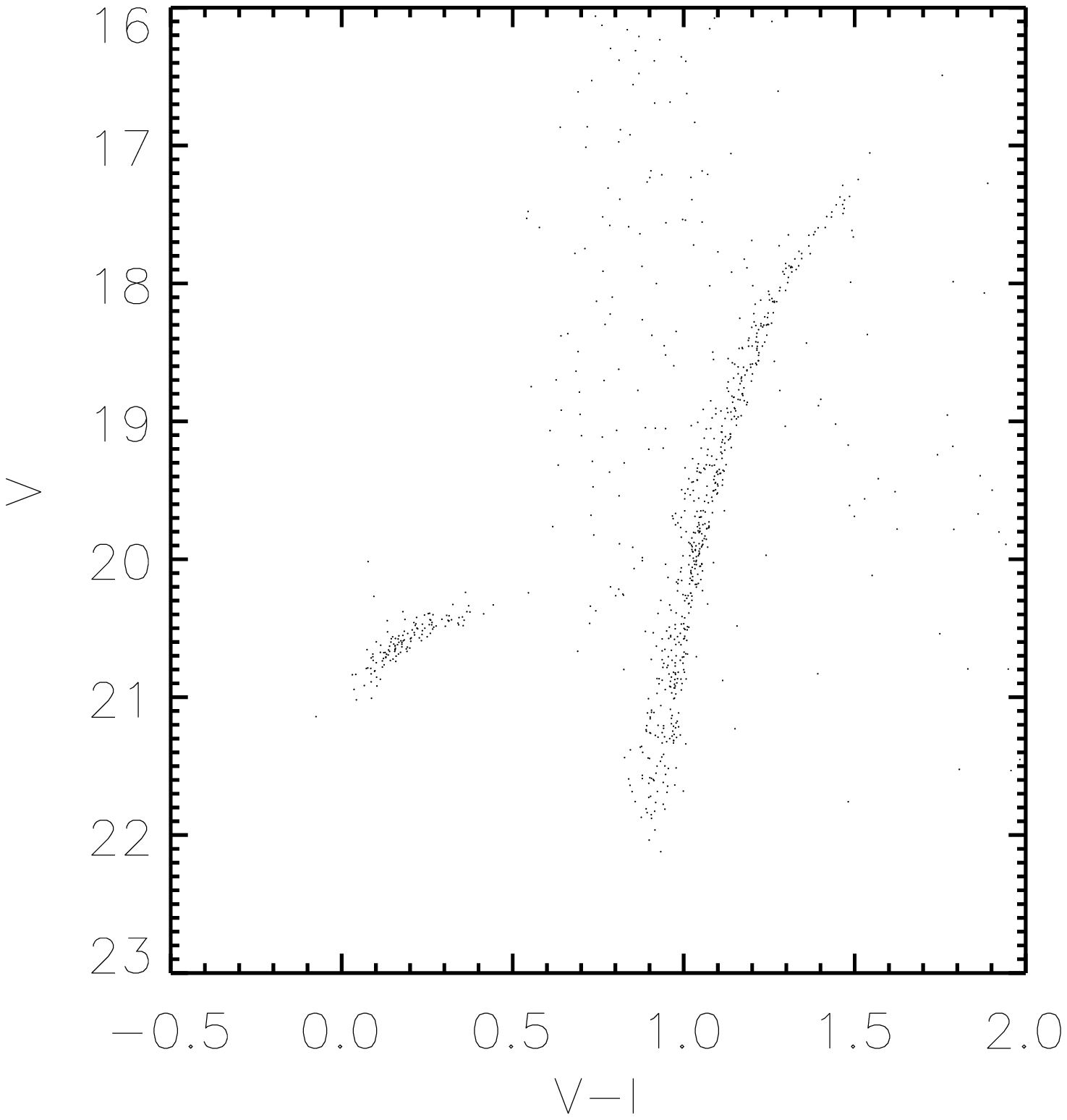}
\caption{Observed  color  magnitude diagrams  for  NGC~2419 from  Saha
  (2005) (left) and Stetson (2002) (right) }
\label{cmds_ground}
\end{figure}

\clearpage

\begin{figure}
\begin{center}
\includegraphics[angle=0,height=7.5in,clip=true]{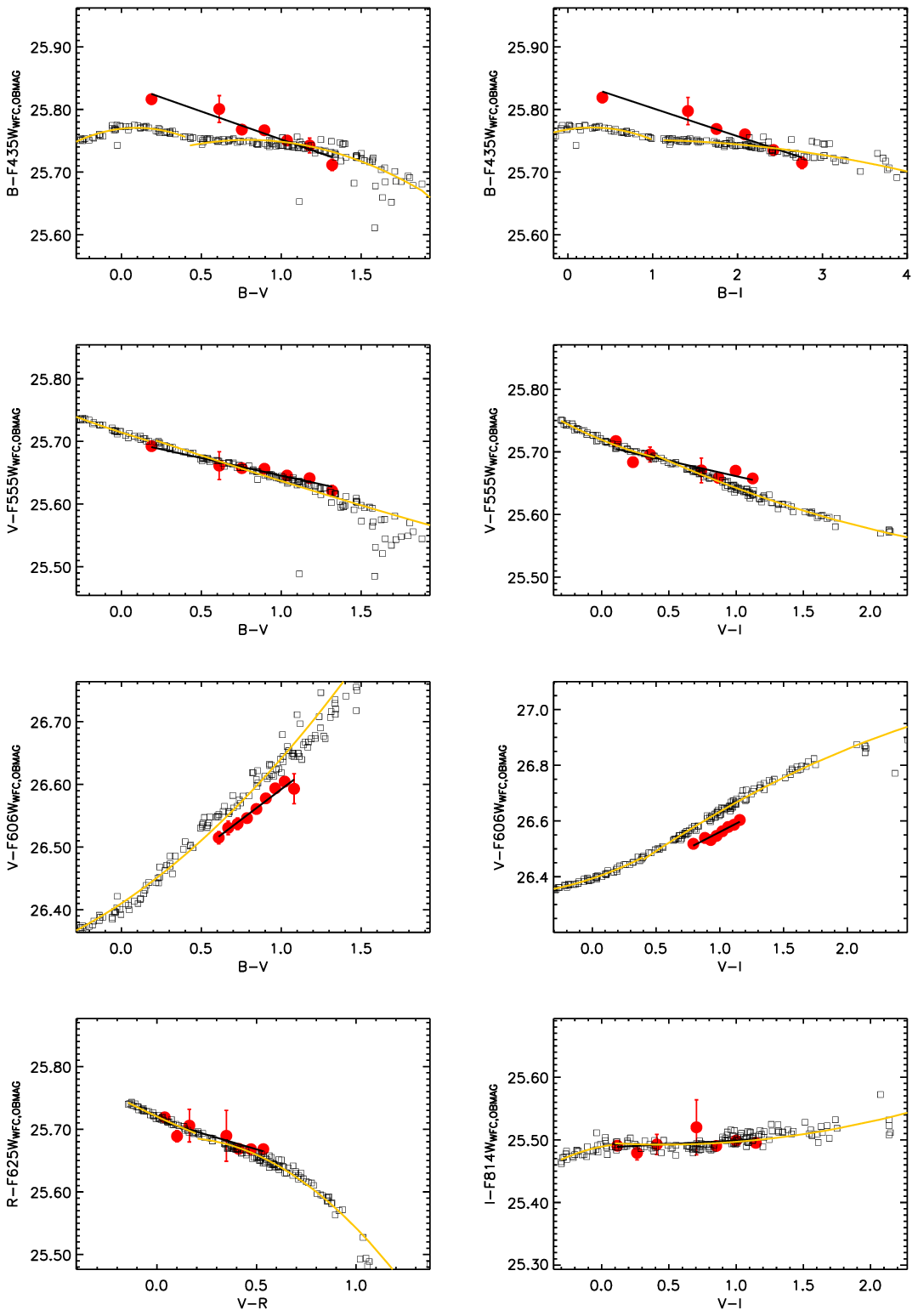}
\end{center}
\caption{Observed   and  synthetic  WFC-to-BVRI   transformations  for
  primary photometric filters. Squares show the synthetic measurements
  for the BPGS atlas.  The  circles are observational data (filled for
  NGC~104, open for NGC~2419).  The black line shows the linear fit to
  the  observational  points,  the  light curve  shows  the  synthetic
  transformations.}
\label{tranw2gr}
\end{figure}
\clearpage

\begin{figure}
\begin{center}
\includegraphics[angle=0,height=7.5in,clip=true]{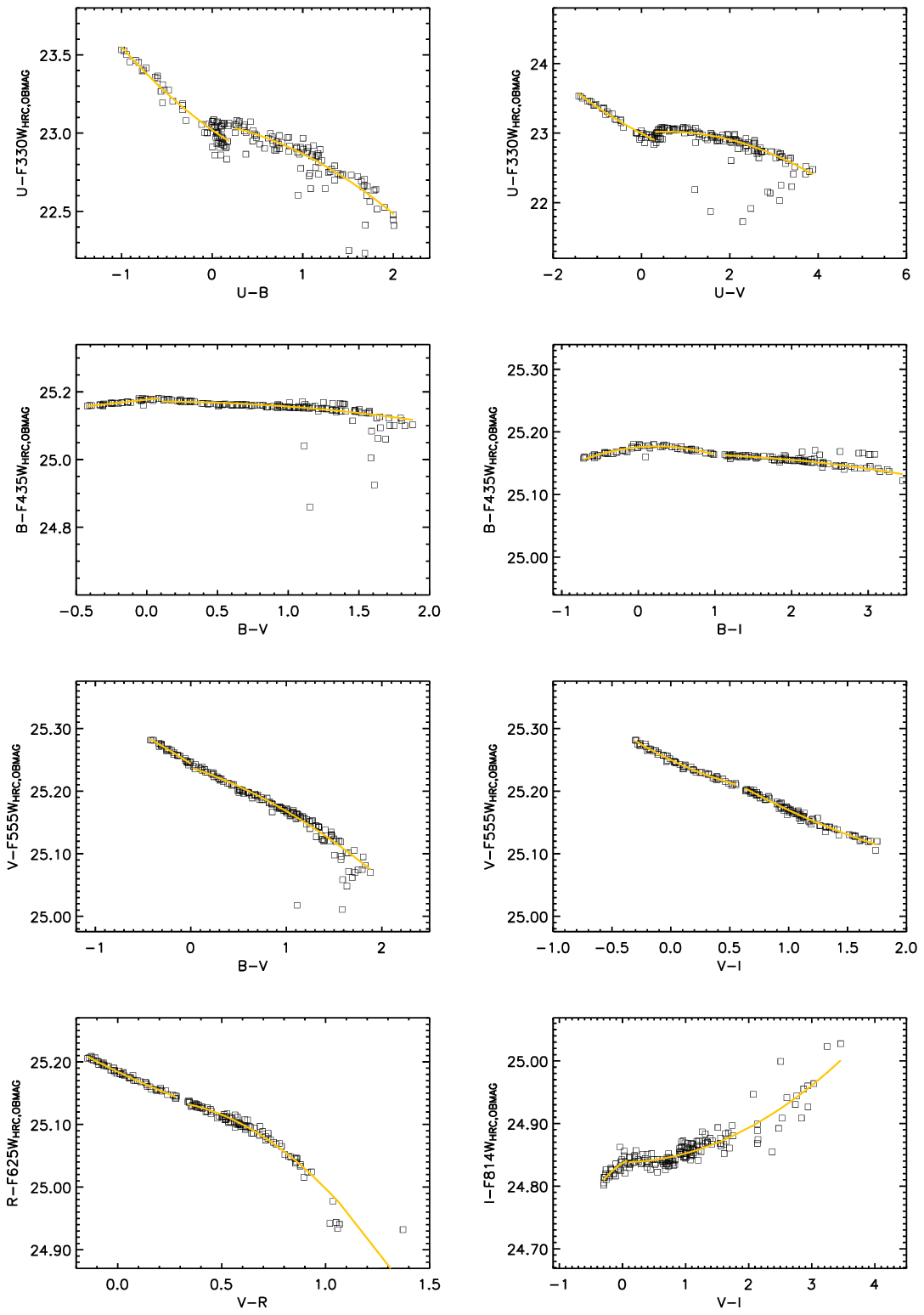}
\end{center}
\caption{Observed  and   synthetic  HRC-to-UBVRI  transformations  for
 primary photometric filters.  Squares show the synthetic measurements
 for the BPGS  atlas.  The circles are observational  data (filled for
 NGC~104, open  for NGC~2419).  The black  line shows the  linear fit to
 the observational points, the light curve  shows
 the synthetic transformations.}
\label{tranh2gr}
\end{figure}

\begin{figure}
\begin{center}
\includegraphics[angle=0,height=7.5in,clip=true]{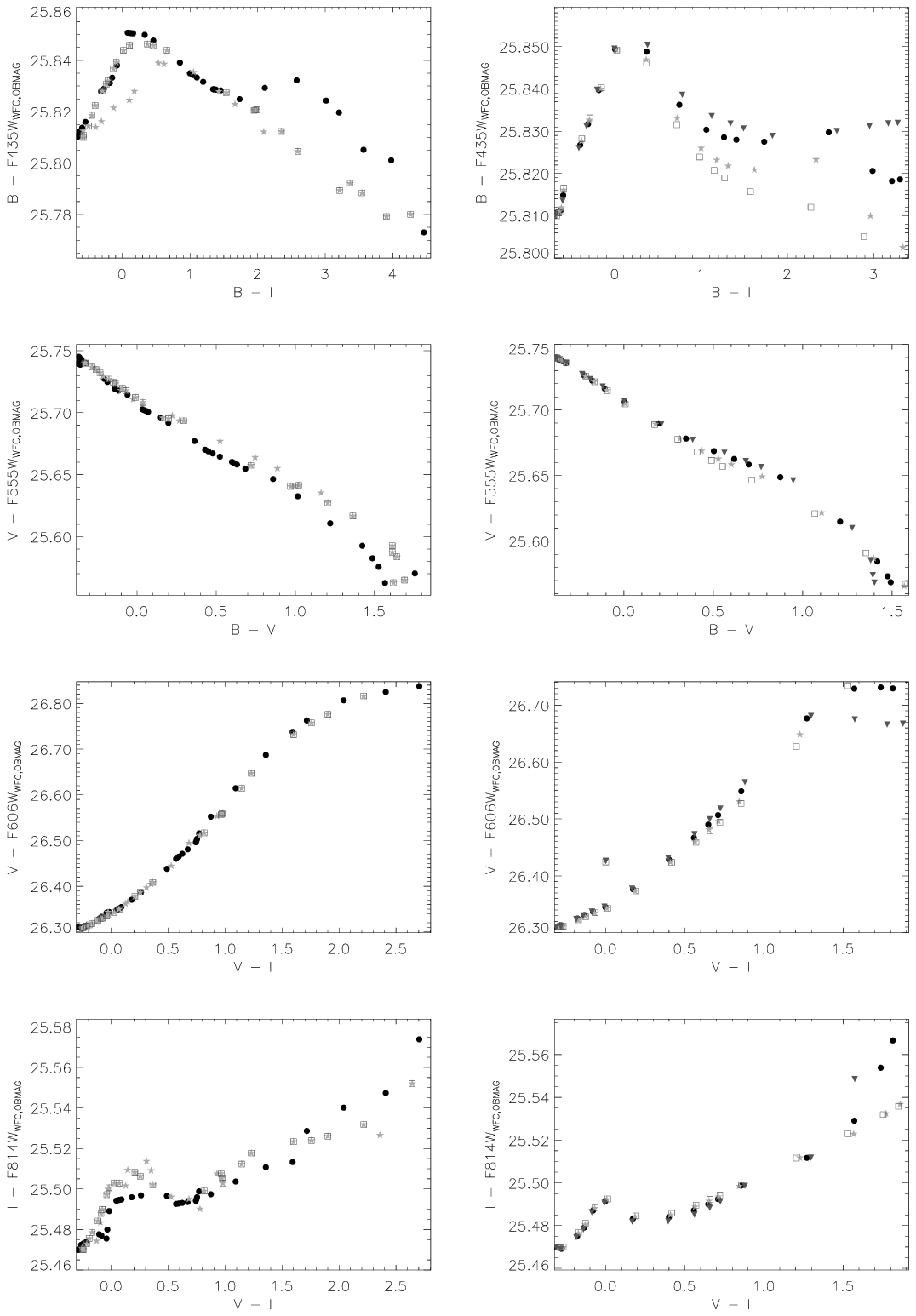}
\end{center}
\caption{Effect of  surface gravity (left) and  metallicity (right) on
  synthetic transformation  WFC-to-BVRI for a  few primary photometric
  filters. Left: black circles are for main-sequence stars, gray stars
  and  boxes are for  giants and  supergiants, respectively  (all from
  BZ77).   Right:   Black  circles  are   for  MS  stars   with  solar
  metallicity, filled triangles  are [Fe/H]~=~0.5, gray-filled stars are
  [Fe/H]~$=-1$ and squares are [Fe/H]~$=-2$ (all from Kurucz).  }
\label{tranw2gr_prob}
\end{figure}

\clearpage

\begin{figure}
\begin{center}
\includegraphics[angle=0,height=7.5in,clip=true]{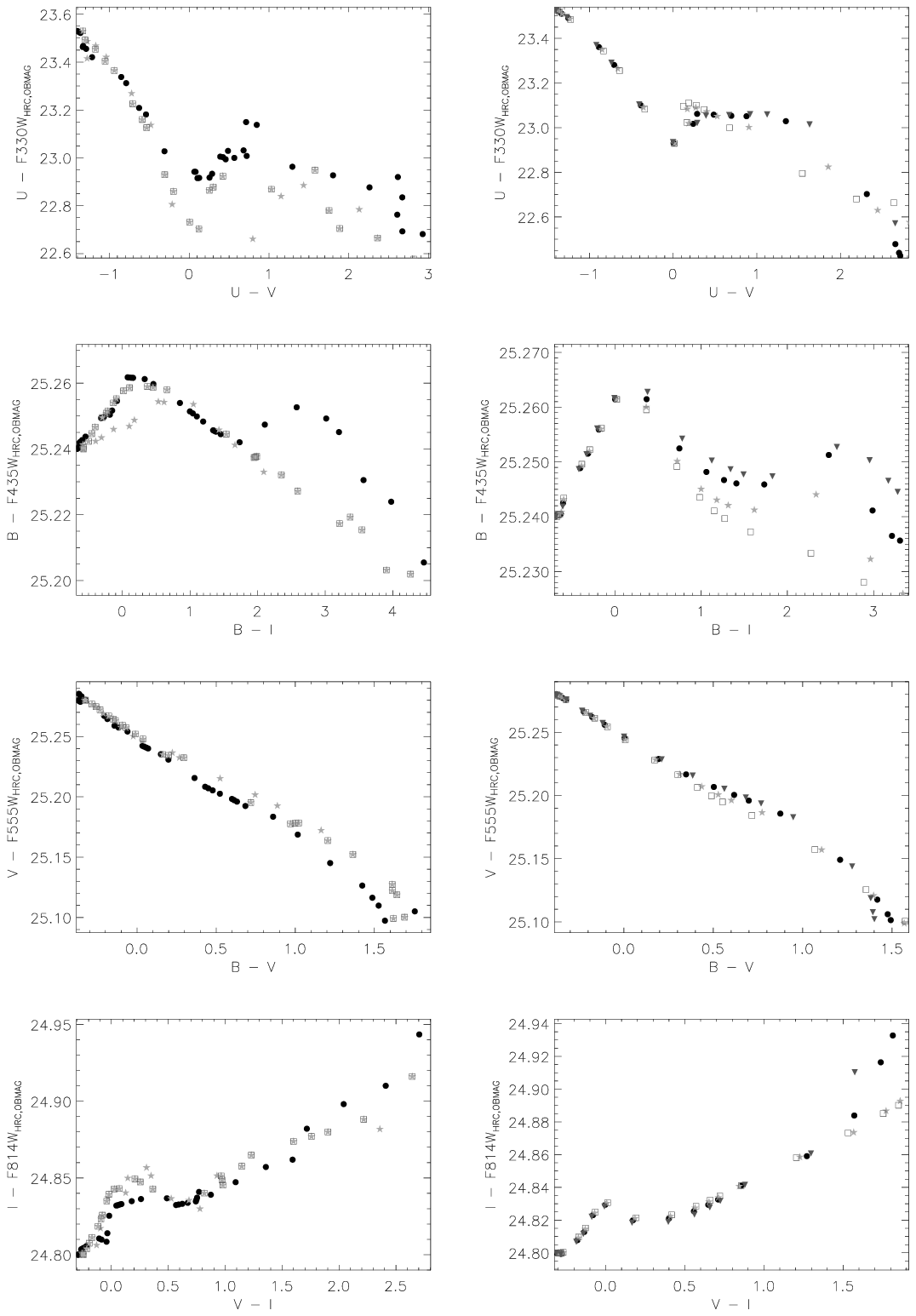}
\end{center}
\caption{Effect of  surface gravity (left) and  metallicity (right) on
  synthetic transformation HRC-to-UBVRI  for a few primary photometric
  filters. Left: black circles are for main-sequence stars, gray stars
  and  boxes are for  giants and  supergiants, respectively  (all from
  BZ77).   Right:   Black  circles  are   for  MS  stars   with  solar
  metallicity, filled triangles  are [Fe/H]~=~0.5, gray-filled stars are
  [Fe/H]~$=-1$ and squares are [Fe/H]~$=-2$ (all from Kurucz).}
\label{tranh2gr_prob}
\end{figure}

\clearpage

\begin{figure}
\centering
\plottwo{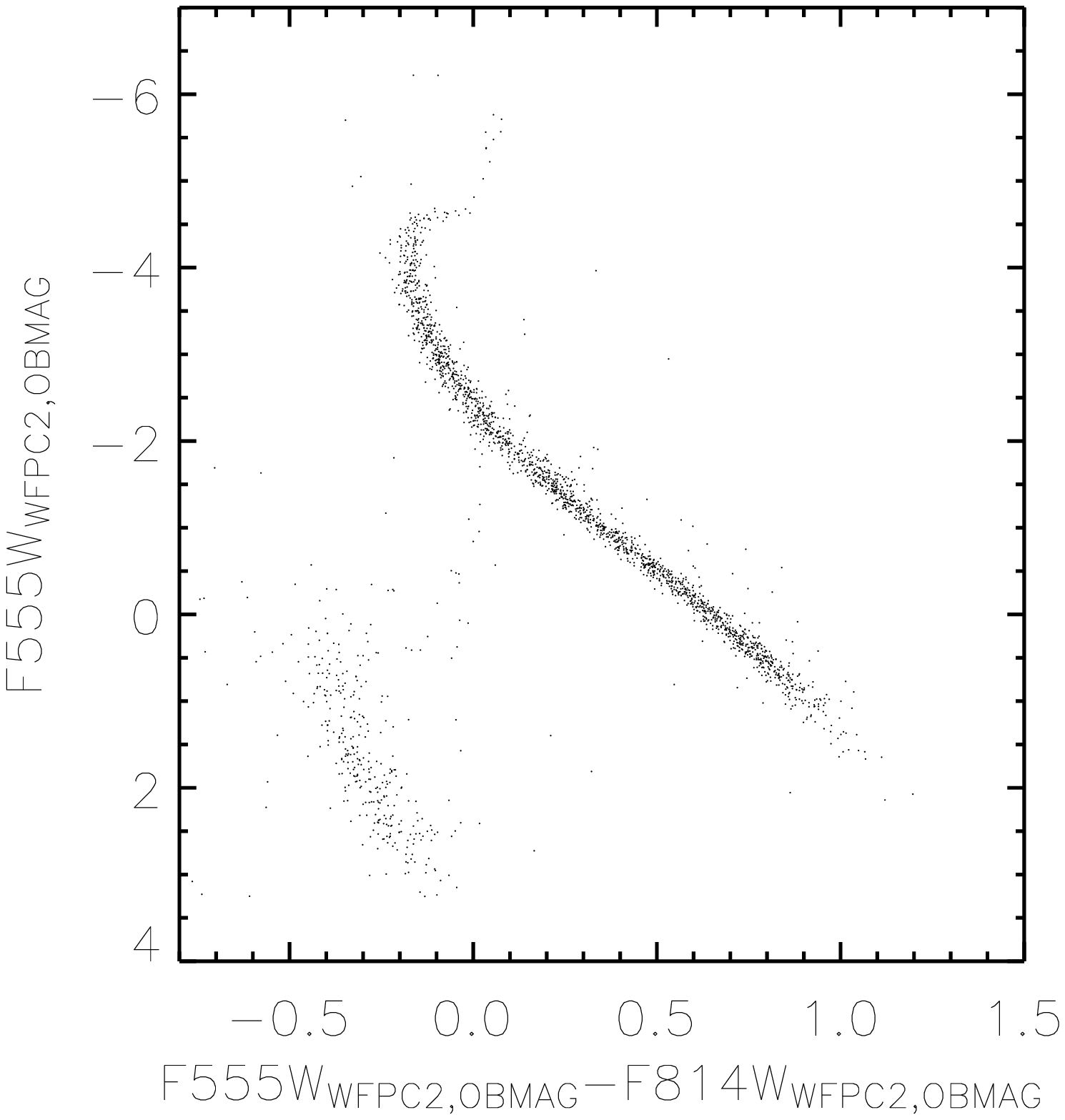}{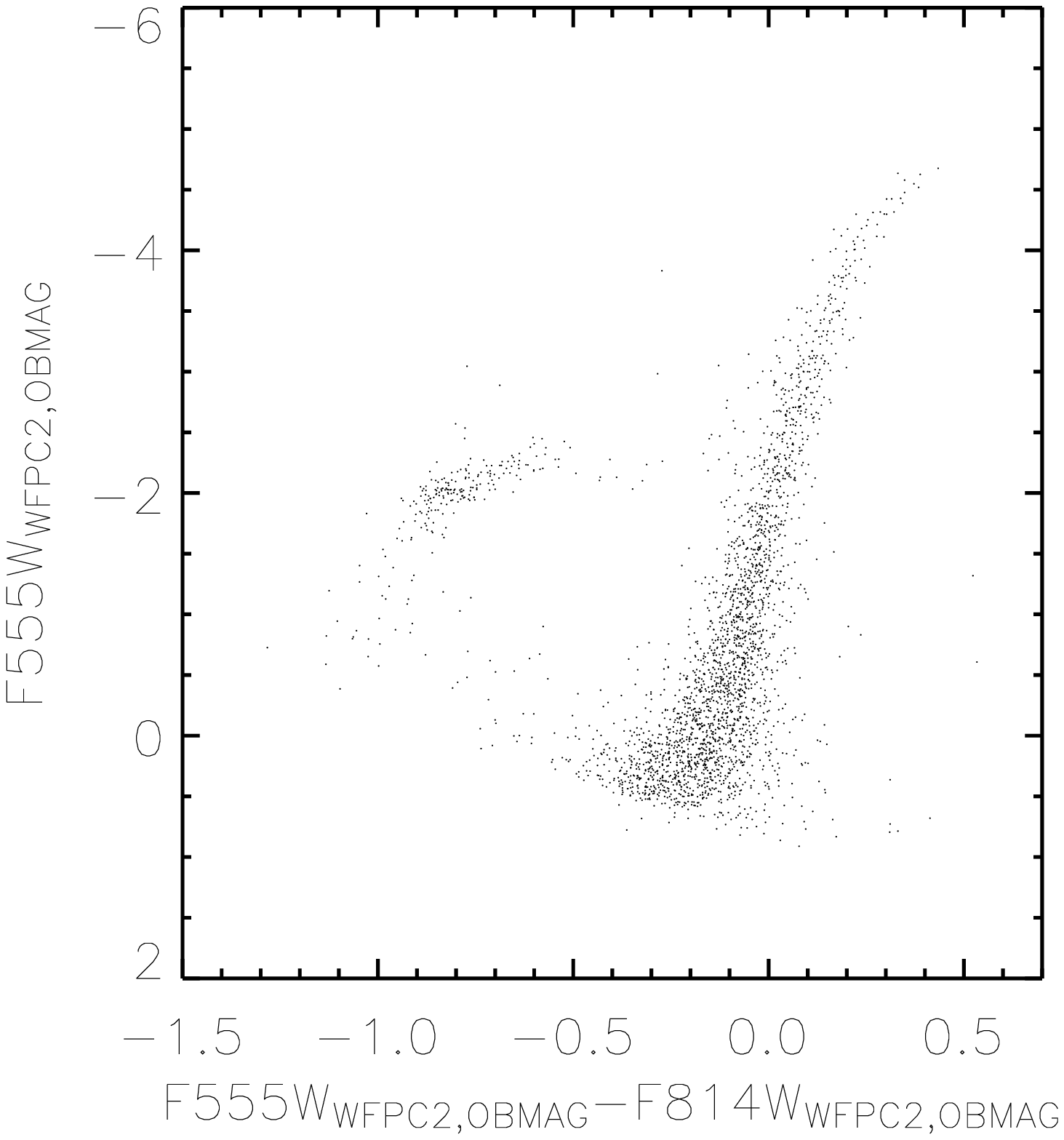}
\caption{Observed  WFPC2 Color Magnitude  Diagrams for  NGC~104 (left)
  and NGC~2419 (right). }
\label{cmds_wfpc2}
\end{figure}

\clearpage
\begin{figure}
\begin{center}
\includegraphics[angle=0,height=7.5in,clip=true]{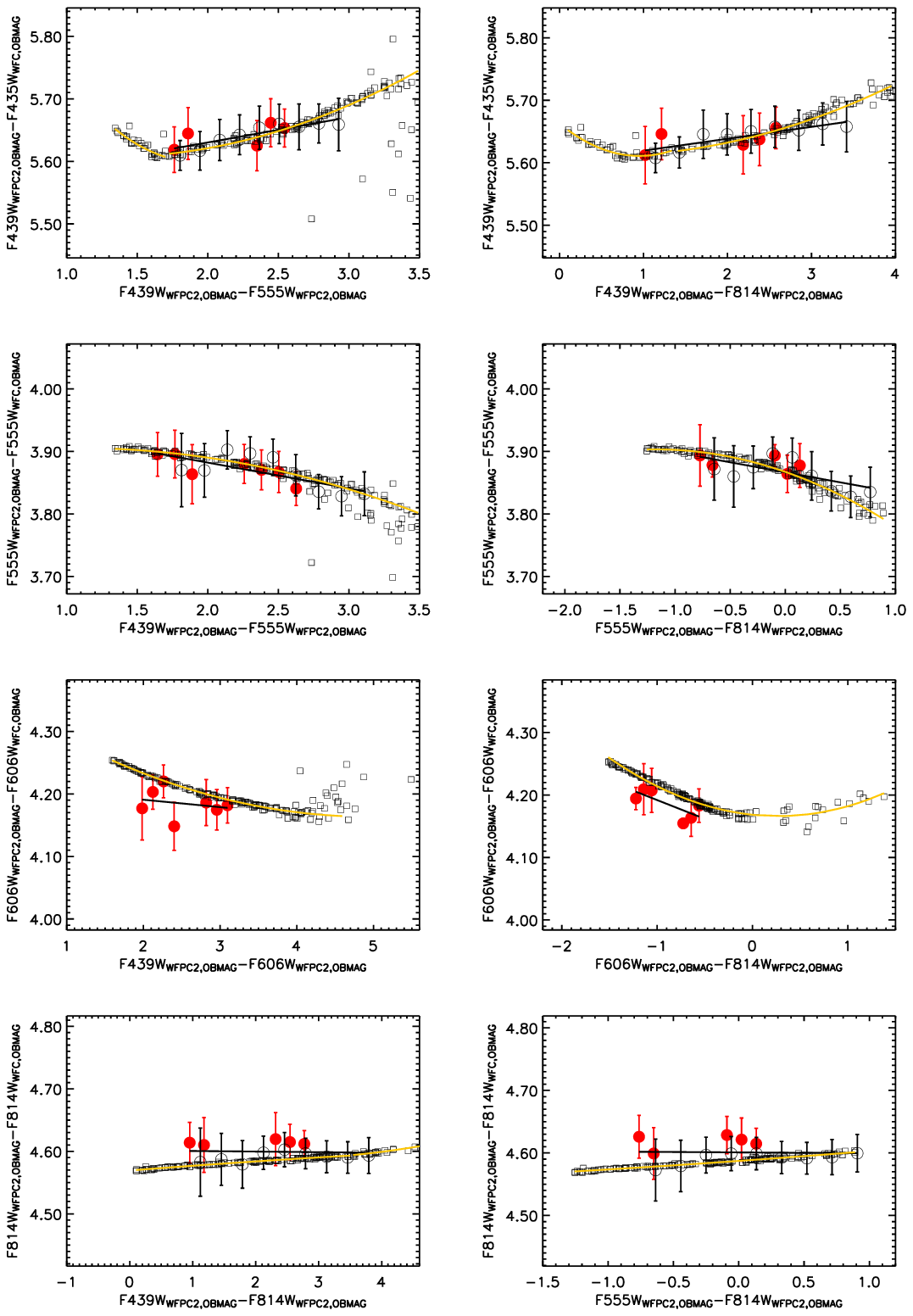}
\end{center}
\caption{Observed  and   synthetic  WFC-to-WFPC2  transformations  for
  primary   photometric   filters.    Squares   show   the   synthetic
  measurements for the BPGS atlas.  The circles are observational data
  (filled for NGC~104,  open for NGC~2419).  The black  line shows the
  linear fit to the observational points, the light curve  
  shows the synthetic transformations.}
\label{tranw2wfpc2}
\end{figure}

\begin{figure}
\begin{center}
\includegraphics[angle=0,height=7.5in,clip=true]{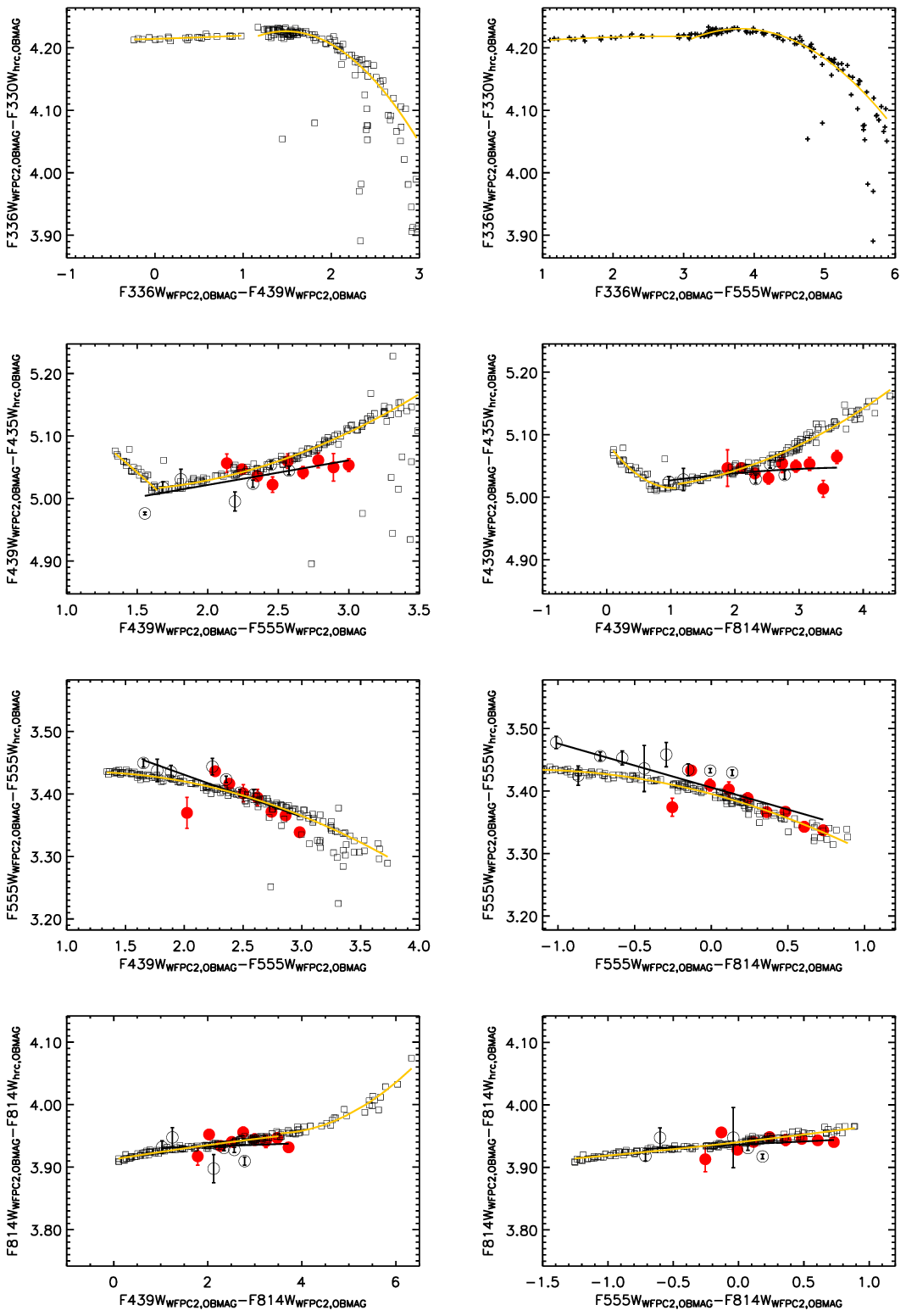}
\end{center}
\caption{Observed  and   synthetic  HRC-to-WFPC2  transformations  for
  primary   photometric   filters.    Squares   show   the   synthetic
  measurements for the BPGS atlas.  The circles are observational data
  (filled for NGC~104,  open for NGC~2419).  The black  line shows the
  linear fit  to the observational  points, the light curve  shows the
  synthetic transformations.}
\label{tranh2wfpc2}
\end{figure}

\clearpage

\begin{figure}
\centering
\epsscale{1}
\plotone{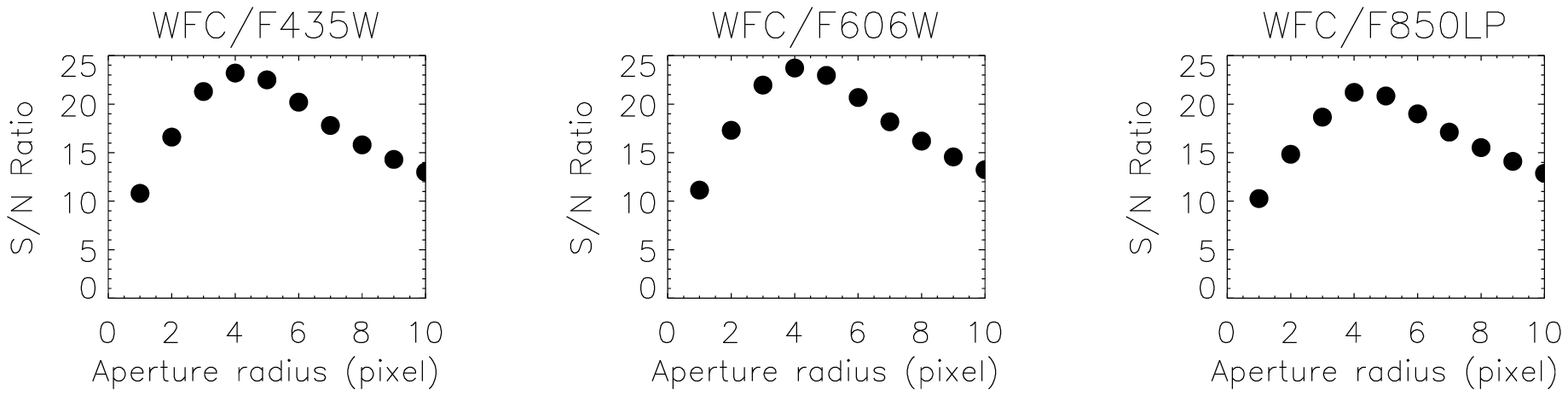}
\caption{S/N  ratio as  a  function  of the  size  of the  photometric
aperture for  three filters  in the WFC.   A star which  produces 1000
electrons  in the  nominal infinite  aperture and  a background  of 10
electrons have been assumed for this test.\label{OAW}}
\end{figure}

\begin{figure}
\centering
\epsscale{1}
\plotone{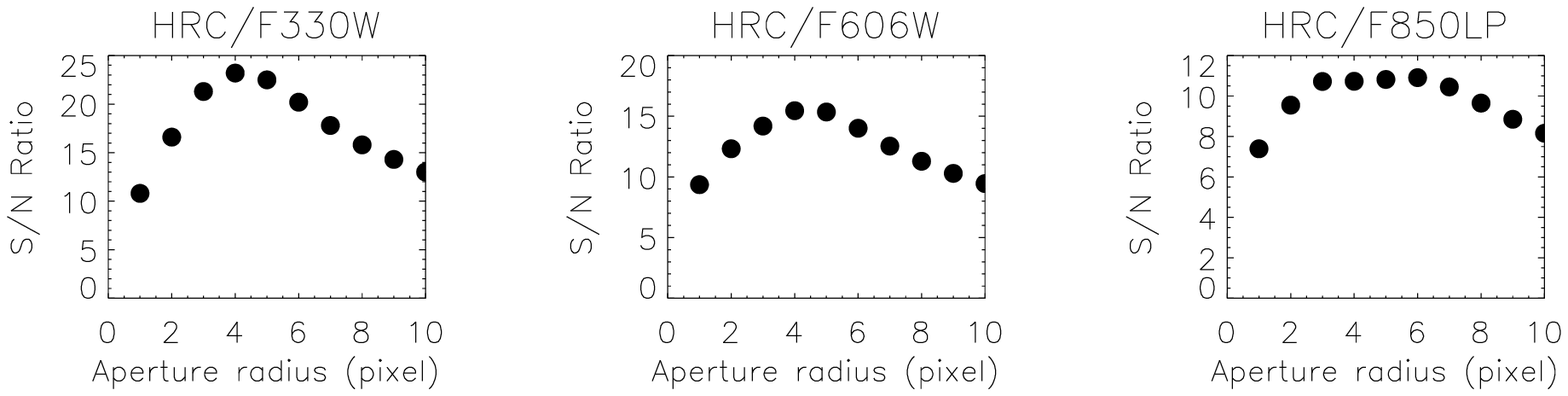}
\caption{Same as Fig.~\ref{OAW} but for the HRC.\label{OAH}}
\end{figure}

\begin{deluxetable}{lcrcl}
\tabletypesize{\small}
\tablecaption{WFC Filters \label{wfilter}}
\tablewidth{0pt}
\tablehead{
\colhead{FILTER} & \colhead{PIVOT $\lambda$} & \colhead{WIDTH} & \colhead{PHOTFLAM\tablenotemark{1}} &\colhead{DESCRIPTION}\\
\colhead{}       & \colhead{(\AA)}             & \colhead{(\AA)}   & \colhead{($erg\,s^{-1}\,cm^{-2}\,$\AA$^{-1}$)}         &\colhead{}
}
\startdata
F435W &	4317.398 &	293.47 &	3.141250E-19 &  Johnson B        \\
F475W &	4744.356 &	420.11 &	1.808802E-19 &	SDSS g'	\\
F502N &	5022.475 &	41.89  &	5.326231E-18 &	[OIII]	\\
F550M &	5581.205 &	163.27 &	3.860100E-19 &	Narrov V	\\
F555W &	5359.547 &	360.02 &	1.955929E-19 &	Johnson V	\\
F606W &	5917.678 &	672.31 &	7.906457E-20 &	Broad V	\\
F625W &	6310.454 &	415.46 &	1.195394E-19 &	SDSS r'	\\
F658N &	6584.015 &	37.15  &	1.999181E-18 &	H$\alpha$	\\
F660N &	6599.374 &	35.53  &	5.347431E-18 &	[NII]	\\
F775W &	7693.026 &	434.60 &	1.006545E-19 &	SDSS i'	\\
F814W &	8059.761 &	654.64 &	7.072360E-20 &	Broad I	\\
F850LP& 9054.768 &	539.43 &	1.507456E-19 &	SDSS z'	\\
F892N &	8914.875 &	72.95  &	1.542548E-18 &	Methane	\\
\enddata
\tablenotetext{1}{Flux density per unit wavelength that generates 1 count/sec}
\end{deluxetable}
\begin{deluxetable}{lcrcl}
\tabletypesize{\small}
\tablecaption{HRC Filters \label{hfilters}}
\tablewidth{0pt}
\tablehead{
\colhead{FILTER} & \colhead{PIVOT $\lambda$} & \colhead{WIDTH} & \colhead{PHOTFLAM\tablenotemark{1}} &\colhead{DESCRIPTION}\\
\colhead{}       & \colhead{(\AA)}             & \colhead{(\AA)}   & \colhead{($erg\,s^{-1}\,cm^{-2}\,$\AA$^{-1}$)}         &\colhead{}
}
\startdata
F220W &	2255.426 &	187.28 & 	8.113312E-18 & Near-UV broadband \\
F250W &	2715.875 &	239.40 &	4.781345E-18 & Near-UV broadband \\
F330W &	3362.675 &	173.82 &	2.236720E-18 & HRC U \\
F344N &	3432.847 &	42.62  &	2.139635E-17 & Ne V \\
F435W &	4310.986 &	309.68 &	5.368556E-19 & Johnson B        \\  
F475W &	4775.740  &	418.81 &	2.937020E-19 & SDSS g'  \\           
F502N &	5021.072 &	83.01  &	8.034045E-18 & [OIII]  \\           
F550M &	5579.729 &	168.05 &	5.985392E-19 & Narrov V        \\   
F555W &	5355.946 &	357.19 &	3.020066E-19 & Johnson V       \\   
F606W &	5887.935 &	664.88 &	1.278104E-19 & Broad V \\           
F625W &	6295.498 &	415.31 &	1.973001E-19 & SDSS r'  \\           
F658N &	6581.980  &	162.69 &	3.357119E-18 & H$\alpha$       \\   
F660N &	6582.439 &	468.30 &	8.976461E-18 & [NII]   \\           
F775W &	7665.083 &	432.03 &	1.949664E-19 & SDSS i'  \\           
F814W &	8115.338 &	703.45 &	1.269102E-19 & Broad I \\           
F850LP& 9145.236  &	538.95 &	2.286987E-19 & SDSS z'  \\           
F892N &	8913.882 &	270.92 &	2.447032E-18 & Methane \\           
\enddata
\tablenotetext{1}{Flux density per unit wavelength that generates 1 count/sec}
\end{deluxetable}

\begin{deluxetable}{rr|ccccccccc}
\tabletypesize{\tiny}
\tablecaption{Encircled Energy -  WFC \label{eeWFC_WD}}
\setlength{\tabcolsep}{0.06in}
\tablewidth{0pt}
\tablehead{
\multicolumn{2}{c}{Aperture radius} & \colhead{F435W} & \colhead{F475W} & \colhead{F555W} & \colhead{F606W} & \colhead{F625W} & \colhead{F775W} & \colhead{F814W} & \colhead{F850LP} & \colhead{F892N} \\
\colhead{pixel} & \colhead{''} & \colhead{ 4323.36\tablenotemark{a}} & \colhead{ 4725.23} & \colhead{ 5342.00} & \colhead{ 5834.71} & \colhead{ 6273.77} & \colhead{ 7662.67} & \colhead{ 7992.32} & \colhead{ 9013.50} & \colhead{ 8914.36} 
}
\startdata
    1&   0.05 &     0.289$\pm$0.011&	 0.318$\pm$0.006&     0.311$\pm$0.005&     0.313$\pm$0.003&	0.312$\pm$0.004&     0.296$\pm$0.001&	  0.293$\pm$0.009&     0.246$\pm$0.004&     0.241$\pm$0.002\\
    2&   0.10 &     0.626$\pm$0.011&	 0.669$\pm$0.006&     0.662$\pm$0.003&     0.656$\pm$0.005&	0.648$\pm$0.004&     0.623$\pm$0.005&	  0.603$\pm$0.006&     0.538$\pm$0.001&     0.546$\pm$0.007\\
    3&   0.15 &     0.775$\pm$0.004&	 0.801$\pm$0.003&     0.795$\pm$0.002&     0.796$\pm$0.003&	0.795$\pm$0.002&     0.780$\pm$0.003&	  0.764$\pm$0.004&     0.689$\pm$0.003&     0.711$\pm$0.007\\
    4&   0.20 &     0.832$\pm$0.002&	 0.850$\pm$0.002&     0.843$\pm$0.001&     0.845$\pm$0.001&	0.844$\pm$0.000&     0.840$\pm$0.001&	  0.833$\pm$0.002&     0.779$\pm$0.002&     0.802$\pm$0.004\\
    5&   0.25 &     0.859$\pm$0.000&	 0.876$\pm$0.002&     0.871$\pm$0.001&     0.871$\pm$0.000&	0.868$\pm$0.001&     0.863$\pm$0.001&	  0.858$\pm$0.001&     0.815$\pm$0.000&     0.835$\pm$0.001\\
    6&   0.30 &     0.875$\pm$0.000&	 0.890$\pm$0.002&     0.891$\pm$0.000&     0.890$\pm$0.001&	0.888$\pm$0.000&     0.882$\pm$0.001&	  0.879$\pm$0.000&     0.837$\pm$0.001&     0.858$\pm$0.000\\
    7&   0.35 &     0.884$\pm$0.001&	 0.900$\pm$0.002&     0.899$\pm$0.001&     0.900$\pm$0.000&	0.899$\pm$0.000&     0.899$\pm$0.001&	  0.895$\pm$0.001&     0.858$\pm$0.001&     0.878$\pm$0.000\\
    8&   0.40 &     0.892$\pm$0.001&	 0.907$\pm$0.002&     0.908$\pm$0.000&     0.908$\pm$0.001&	0.907$\pm$0.000&     0.909$\pm$0.001&	  0.907$\pm$0.001&     0.877$\pm$0.001&     0.895$\pm$0.000\\
\enddata
\tablenotetext{a}{Average effective wavelength (\AA) for the two spectrophotometric standard stars, both white dwarves}  
\end{deluxetable}
\begin{deluxetable}{rr|cccccccccccc}
\tabletypesize{\tiny}
\setlength{\tabcolsep}{0.018in}
\tablecaption{Encircled Energy - HRC \label{eeHRC_WD}}
\rotate
\tablewidth{0pt}
\tablehead{
\multicolumn{2}{c}{Aperture radius} & \colhead{F220W} & \colhead{F250W} & \colhead{F330W} & \colhead{F435W} & \colhead{F475W} & \colhead{F555W} & \colhead{F606W} & \colhead{F625W} & \colhead{F775W} & \colhead{F814W} & \colhead{F850LP} & \colhead{F892N} \\
\colhead{pixel} & \colhead{''} & \colhead{ 2244.64\tablenotemark{a}} & \colhead{ 2702.15} & \colhead{ 3356.23} & \colhead{ 4317.55} & \colhead{ 4756.77} & \colhead{ 5338.72} & \colhead{ 5806.35} & \colhead{ 6258.62} & \colhead{ 7634.95} & \colhead{ 8037.84} & \colhead{ 9104.70} & \colhead{ 8910.36} 
}
\startdata
    1&   0.025 &     0.235$\pm$0.014&     0.245$\pm$0.013&     0.250$\pm$0.011&     0.254$\pm$0.005&     0.230$\pm$0.005&     0.222$\pm$0.002&     0.221$\pm$0.006&     0.208$\pm$0.003&     0.171$\pm$0.004&     0.158$\pm$0.002&     0.109$\pm$0.003&     0.118$\pm$0.001\\
    2&   0.050 &     0.506$\pm$0.010&     0.534$\pm$0.010&     0.542$\pm$0.008&     0.532$\pm$0.003&     0.523$\pm$0.006&     0.514$\pm$0.003&     0.505$\pm$0.004&     0.490$\pm$0.003&     0.433$\pm$0.001&     0.403$\pm$0.008&     0.294$\pm$0.006&     0.324$\pm$0.003\\
    3&   0.075 &     0.623$\pm$0.007&     0.655$\pm$0.006&     0.679$\pm$0.005&     0.687$\pm$0.005&     0.678$\pm$0.002&     0.656$\pm$0.003&     0.640$\pm$0.003&     0.620$\pm$0.003&     0.560$\pm$0.001&     0.534$\pm$0.001&     0.413$\pm$0.008&     0.453$\pm$0.002\\
    4&   0.100 &     0.686$\pm$0.007&     0.720$\pm$0.003&     0.736$\pm$0.004&     0.762$\pm$0.003&     0.764$\pm$0.001&     0.759$\pm$0.002&     0.742$\pm$0.004&     0.723$\pm$0.003&     0.632$\pm$0.002&     0.599$\pm$0.001&     0.465$\pm$0.007&     0.498$\pm$0.005\\
    5&   0.125 &     0.733$\pm$0.001&     0.762$\pm$0.002&     0.775$\pm$0.002&     0.791$\pm$0.001&     0.796$\pm$0.001&     0.803$\pm$0.001&     0.799$\pm$0.002&     0.795$\pm$0.002&     0.723$\pm$0.002&     0.683$\pm$0.001&     0.526$\pm$0.007&     0.561$\pm$0.005\\
    6&   0.150 &     0.758$\pm$0.002&     0.788$\pm$0.001&     0.805$\pm$0.000&     0.813$\pm$0.001&     0.813$\pm$0.002&     0.818$\pm$0.002&     0.818$\pm$0.002&     0.816$\pm$0.003&     0.779$\pm$0.002&     0.742$\pm$0.000&     0.592$\pm$0.008&     0.631$\pm$0.006\\
    7&   0.175 &     0.771$\pm$0.001&     0.803$\pm$0.001&     0.825$\pm$0.002&     0.837$\pm$0.001&     0.833$\pm$0.002&     0.835$\pm$0.001&     0.832$\pm$0.001&     0.829$\pm$0.002&     0.798$\pm$0.001&     0.766$\pm$0.001&     0.634$\pm$0.007&     0.669$\pm$0.006\\
\enddata
\tablenotetext{a}{Average effective wavelength (\AA) for the two spectrophotometric standard stars, both white dwarves}
\end{deluxetable}
 \begin{deluxetable}{lrr|rr}
\tabletypesize{\footnotesize}
\tablecaption{Aperture Correction from $0\farcs5$ Aperture Radius  to Infinite (AC05)\label{ee05}}
\tablewidth{0pt}
\tablehead{
\colhead{}&\multicolumn{2}{c}{HRC}&\multicolumn{2}{c}{WFC}\\
\colhead{FILTER}&\colhead{mag}&\colhead{err}&\colhead{mag}&\colhead{err}
}
\startdata								
 F220W &     0.151 &   0.002 &  	\nodata &	\nodata      \\ 
 F250W &     0.132 &   0.002 &  	\nodata &	\nodata      \\
 F330W &     0.117 &   0.003 &  	\nodata &	\nodata      \\
 F334N &     0.113 &   0.004 &  	\nodata &	\nodata  \\
 F435W &     0.097 &   0.001 &     0.107 &    0.004   \\
 F475W &     0.096 &   0.004 &     0.095 &    0.002   \\
 F502N &     0.093 &   0.001 &     0.094 &    0.001   \\
 F550M &     0.092 &   0.001 &     0.083 &    0.001   \\
 F555W &     0.088 &   0.001 &     0.092 &    0.001   \\
 F606W &     0.090 &   0.002 &     0.088 &    0.000   \\
 F625W &     0.092 &   0.001 &     0.088 &    0.001   \\
 F658N &     0.098 &   0.001 &     0.085 &    0.003   \\
 F660N &     0.110 &   0.001 &     0.090 &    0.003   \\
 F775W &     0.132 &   0.001 &     0.087 &    0.001   \\
 F814W &     0.172 &   0.002 &     0.087 &    0.001   \\
 F850LP&     0.314 &   0.006 &     0.117 &    0.000   \\
 F892N &     0.293 &   0.009 &     0.100 &    0.002   \\
 \tableline
\multicolumn{5}{c}{OBMAG=-2.5 log(count rate $r\, \leq\, 0\farcs5$ $[e^-/s]$) -  AC05}\\

\enddata

\end{deluxetable}

 
\clearpage

\begin{deluxetable}{crrrrrrrrrrrr}
\tabletypesize{\tiny}
\setlength{\tabcolsep}{0.03in}
\tablecaption{WFC - Effective Wavelength  Aperture Correction Relation \label{WFC_acl}}
\tablewidth{0pt}
\tablehead{
\colhead{Effective Wavelength\tablenotemark{a}}&\multicolumn{12}{c}{Aperture radius}\\
\colhead{\AA}  & \colhead{0.10''} & \colhead{0.15''} &  \colhead{0.20'}   
                                     & \colhead{0.25''}   & \colhead{0.30''}& \colhead{0.40''}
                                     & \colhead{0.50''}& \colhead{0.60''} & \colhead{0.80''} & \colhead{1.00''}& \colhead{1.50''}& \colhead{2.00''}
}
\startdata
        3500&   0.651&   0.351&   0.246&   0.208&   0.184&   0.162&   0.140&   0.129&   0.113&   0.096&   0.061&   0.044\\
        3750&   0.610&   0.328&   0.233&   0.195&   0.173&   0.152&   0.130&   0.119&   0.104&   0.088&   0.056&   0.038\\
        4000&   0.566&   0.306&   0.218&   0.182&   0.161&   0.140&   0.120&   0.108&   0.093&   0.080&   0.049&   0.031\\
        4250&   0.521&   0.284&   0.204&   0.169&   0.149&   0.128&   0.110&   0.099&   0.083&   0.071&   0.042&   0.024\\
        4500&   0.481&   0.265&   0.192&   0.157&   0.139&   0.117&   0.101&   0.089&   0.074&   0.063&   0.036&   0.019\\
        4750&   0.454&   0.251&   0.183&   0.151&   0.131&   0.111&   0.095&   0.083&   0.069&   0.057&   0.034&   0.016\\
        5000&   0.443&   0.244&   0.182&   0.147&   0.129&   0.107&   0.093&   0.081&   0.067&   0.056&   0.035&   0.018\\
        5250&   0.446&   0.242&   0.182&   0.145&   0.127&   0.105&   0.091&   0.079&   0.065&   0.055&   0.035&   0.019\\
\enddata

\tablenotetext{a}{Values for $\lambda \, <$ 4300 \AA \,and $\lambda\,>$ 9800 \AA are extrapolated.}

\end{deluxetable}
\begin{deluxetable}{crrrrrrrrrrrrr}
\tabletypesize{\tiny}
\setlength{\tabcolsep}{0.03in}
\tablecaption{HRC - Effective Wavelength  Aperture Correction Relation\label{HRC_acl}}
\tablewidth{0pt}
\tablehead{
\colhead{Effective Wavelength\tablenotemark{a}}&\multicolumn{13}{c}{Aperture radius}\\
\colhead{\AA}  & \colhead{0.075''} & \colhead{0.10''} & \colhead{0.125''} &  \colhead{0.20'}   
                                     & \colhead{0.25''}   & \colhead{0.30''}& \colhead{0.40''}
                                     & \colhead{0.50''}& \colhead{0.60''} & \colhead{0.80''} & \colhead{1.00''}& \colhead{1.50''}& \colhead{2.00''}
}
\startdata

  2000&   0.544&   0.438&   0.359&   0.289&   0.261&   0.231&   0.193&   0.161&   0.134&   0.081&   0.052&   0.031&   0.021\\
        2250&   0.514&   0.409&   0.337&   0.266&   0.240&   0.214&   0.179&   0.151&   0.127&   0.081&   0.053&   0.027&   0.019\\
        2500&   0.489&   0.384&   0.317&   0.244&   0.221&   0.198&   0.165&   0.141&   0.120&   0.082&   0.055&   0.026&   0.016\\
        2750&   0.473&   0.370&   0.304&   0.226&   0.205&   0.187&   0.155&   0.135&   0.117&   0.085&   0.059&   0.027&   0.018\\
        3000&   0.457&   0.358&   0.294&   0.212&   0.192&   0.176&   0.146&   0.128&   0.113&   0.086&   0.063&   0.029&   0.019\\
        3250&   0.451&   0.348&   0.288&   0.198&   0.179&   0.165&   0.139&   0.120&   0.108&   0.085&   0.064&   0.030&   0.018\\
        3500&   0.442&   0.333&   0.279&   0.187&   0.167&   0.155&   0.131&   0.113&   0.102&   0.081&   0.064&   0.030&   0.016\\
        3750&   0.439&   0.323&   0.273&   0.179&   0.159&   0.145&   0.125&   0.107&   0.096&   0.078&   0.061&   0.030&   0.016\\
\enddata

\tablenotetext{a}{Values for $\lambda > $ 9500 \AA\, are extrapolated.}

\end{deluxetable}

\begin{deluxetable}{l|rrr|rrr}
\tabletypesize{\footnotesize}
\tablecaption{Effective Wavelength (\AA) for Stars of  Different Spectral Types in the Near-IR \label{app_efflam}}
\tablewidth{0pt}
\setlength{\tabcolsep}{0.05in}
\tablehead{
\colhead{STAR}&\multicolumn{3}{c}{WFC}&\multicolumn{3}{c}{HRC}\\
\colhead{TYPE}&\colhead{F775W} & \colhead{F814W}  & \colhead{F850LP}  & \colhead{F775W}  & \colhead{F814W}  & \colhead{F850LP}
}
\startdata
O5V & 7659.4 & 7986.0 & 9012.0 & 7631.7 & 8030.7 & 9103.2\\
B0V & 7663.0 & 7995.1 & 9017.6 & 7635.3 & 8041.7 & 9108.5\\
B5V & 7672.4 & 8015.4 & 9028.7 & 7644.6 & 8064.7 & 9119.3\\
A0V & 7678.4 & 8044.3 & 9056.2 & 7650.5 & 8099.1 & 9146.2\\
A5V & 7688.1 & 8061.4 & 9062.7 & 7660.2 & 8118.2 & 9152.2\\
F2V & 7703.7 & 8085.1 & 9064.7 & 7675.8 & 8144.3 & 9154.3\\
F6V & 7707.8 & 8088.0 & 9069.8 & 7679.8 & 8147.5 & 9159.7\\
G0V & 7713.6 & 8097.7 & 9074.9 & 7685.7 & 8158.3 & 9164.9\\
G5V & 7714.9 & 8105.0 & 9082.6 & 7687.0 & 8167.2 & 9172.4\\
K0V & 7717.3 & 8106.5 & 9077.1 & 7689.4 & 8168.4 & 9166.6\\
K5V & 7743.9 & 8156.3 & 9093.1 & 7716.3 & 8223.8 & 9181.8\\
M0V & 7767.2 & 8205.4 & 9120.2 & 7740.0 & 8279.7 & 9207.8\\
M6V & 7837.3 & 8326.2 & 9166.2 & 7812.4 & 8412.7 & 9249.6\\
L3.5& 8003.7 & 8669.6 & 9383.5 & 7980.4 & 8779.7 & 9445.2\\
T6.6& 8133.6 & 8974.3 & 9772.9 & 8118.1 & 9045.6 & 9788.0
 \enddata

\end{deluxetable}

\begin{deluxetable}{ccccc}
\tabletypesize{\footnotesize}
\tablecaption{Gain Correction\label{gainfix}}
\tablewidth{0pt}
\tablehead{
\colhead{Nominal} & \colhead{Amplifier} & \multicolumn{2}{c}{Gain (e$^-$/ADU)} &  \colhead{Correction\,\tablenotemark{a}}\\
\colhead{Gain}    & \colhead{}          & \colhead{old}  & \colhead{new}    & \colhead{$-2.5\, log\,(G_{new}/G_{old} )$}
}
\startdata
\multicolumn{5}{c}{WFC}\\
\tableline
 2 & A & 2.018 & 2.002 & 0.0086 \\
 2 & B & 1.960 & 1.945 & 0.0083 \\
 2 & C & 2.044 & 2.028 & 0.0085 \\
 2 & D & 2.010 & 1.994 & 0.0087 \\
   &   &       &       &        \\
 4 & A & 4.005 & 4.011 & \llap{$-$}0.0016 \\
 4 & B & 3.897 & 3.902 & \llap{$-$}0.0014 \\
 4 & C & 4.068 & 4.074 & \llap{$-$}0.0016 \\
 4 & D & 3.990 & 3.996 & \llap{$-$}0.0016 \\
 &   &       &       &           \\
\multicolumn{5}{c}{HRC}\\
\tableline
 1 & C & 1.185 & 1.163 & 0.0203 \\
&   &       &       &         \\
 4 & C & 4.289 & 4.235 & 0.0138 
\enddata
\tablenotetext{a}{Correction to apply to zeropoints {\bf only} for data retrieved from the HST archive before January 6, 2004.}

\end{deluxetable}
\begin{deluxetable}{lcccc}
\tabletypesize{\footnotesize}
\tablecaption{Synthetic Zeropoints for WFC \label{wfczpt}}
\tablewidth{0pt}
\tablehead{
\colhead{FILTER} & \colhead{VEGAMAG} & \colhead{ABMAG} & \colhead{STMAG}
}
\startdata
F435W  &25.779  &25.673  &25.157  \\
F475W  &26.168  &26.068  &25.757  \\
F502N  &22.352  &22.271  &22.084  \\
F550M  &24.867  &24.892  &24.934  \\
F555W  &25.724  &25.718  &25.672  \\
F606W  &26.398  &26.486  &26.655  \\
F625W  &25.731  &25.898  &26.206  \\
F658N  &22.365  &22.747  &23.148  \\
F660N  &21.389  &21.676  &22.081  \\
F775W  &25.256  &25.654  &26.393  \\
F814W  &25.501  &25.937  &26.776  \\
F850LP &24.326  &24.862  &25.954  \\
F892N  &21.865  &22.371  &23.429  \\
\enddata
\end{deluxetable}

\begin{deluxetable}{lccc}
\tabletypesize{\footnotesize}
\tablecaption{Synthetic Zeropoints for HRC \label{hrczpt}}
\tablewidth{0pt}
\tablehead{
\colhead{FILTER} & \colhead{VEGAMAG} & \colhead{ABMAG} & \colhead{STMAG}
}
\startdata
F220W  &21.871  &23.553  & 21.627 \\
F250W  &22.288  &23.724  & 22.201 \\
F330W  &22.904  &24.085  & 23.026 \\
F344N  &20.439  &21.588  & 20.574 \\
F435W  &25.185  &25.095  & 24.575 \\
F475W  &25.623  &25.527  & 25.230 \\
F502N  &21.906  &21.826  & 21.638 \\
F550M  &24.392  &24.416  & 24.457 \\
F555W  &25.255  &25.248  & 25.200 \\
F606W  &25.893  &25.976  & 26.134 \\
F625W  &25.195  &25.359  & 25.662 \\
F658N  &21.803  &22.185  & 22.585 \\
F660N  &20.827  &21.116  & 21.516 \\
F775W  &24.551  &24.945  & 25.675 \\
F814W  &24.849  &25.287  & 26.141 \\
F850LP &23.850  &24.388  & 25.502 \\
F892N  &21.366  &21.870  & 22.928 \\
\enddata
\end{deluxetable}

\clearpage

\begin{deluxetable}{lclrrrcrrrr}
\tabletypesize{\small}
\setlength{\tabcolsep}{0.04in}
\tablecaption{Coefficients for the Transformation from  WFC to HRC \label{wfc2hrc}}
\tablewidth{0pt}
\rotate
\tablehead{
\colhead{SOURCE}&\multicolumn{2}{c}{TARGET}&\multicolumn{4}{c}{Observed}&\multicolumn{4}{c}{Synthetic}\\
\colhead{SMAG}&\colhead{TMAG}&\colhead{TCOL}&\colhead{c0}&\colhead{c1}&\colhead{c2}& &\colhead{c0}&\colhead{c1}&\colhead{c2}&\colhead{TCOL}\\
\multicolumn{6}{c}{}&\colhead{}&\multicolumn{3}{c}{}&\colhead{range}\tablenotemark{a}
}
\startdata
F435W & F435W & F435W-F475W &    0.580 $\pm$ 0.061 &    0.018 $\pm$ 0.088 &    0.000 $\pm$ 0.000 &  &    0.525 $\pm$ 0.042 &    0.273 $\pm$ 0.227 &   $-$0.266 $\pm$ 0.299 & $< 0.5$ \\
                & & & \multicolumn{4}{c}{  } &    0.592 $\pm$ 0.008 &    0.007 $\pm$ 0.016 &   $-$0.014 $\pm$ 0.008 & $> 0.5$ \\
      & F435W & F435W-F550M & \multicolumn{4}{c}{  } &    0.584 $\pm$ 0.073 &   $-$0.046 $\pm$ 0.154 &   $-$0.043 $\pm$ 0.080 & $<-0.7$ \\
                & & & \multicolumn{4}{c}{  } &    0.589 $\pm$ 0.001 &   $-$0.007 $\pm$ 0.002 &   $-$0.002 $\pm$ 0.002 & $>-0.7$ \\
      & F435W & F435W-F555W &    0.591 $\pm$ 0.031 &    0.009 $\pm$ 0.044 &    0.000 $\pm$ 0.000 &  &    0.592 $\pm$ 0.002 &    0.047 $\pm$ 0.035 &   $-$0.011 $\pm$ 0.159 & $< 0.1$ \\
                & & & \multicolumn{4}{c}{  } &    0.593 $\pm$ 0.003 &   $-$0.001 $\pm$ 0.005 &   $-$0.005 $\pm$ 0.002 & $> 0.1$ \\
      & F435W & F435W-F606W &    0.586 $\pm$ 0.059 &    0.006 $\pm$ 0.040 &    0.000 $\pm$ 0.000 &  &    0.565 $\pm$ 0.028 &    0.044 $\pm$ 0.115 &   $-$0.003 $\pm$ 0.112 & $< 0.7$ \\
                & & & \multicolumn{4}{c}{  } &    0.596 $\pm$ 0.005 &   $-$0.002 $\pm$ 0.005 &   $-$0.001 $\pm$ 0.001 & $> 0.7$ \\
\enddata
\tablenotetext{a}{TCOL range applies only to synthetic transformations.}
 \end{deluxetable}
\begin{deluxetable}{lclrrrcrrrr}
\tabletypesize{\footnotesize}
\setlength{\tabcolsep}{0.04in}
\tablecaption{Coefficients for the Transformation from  HRC to WFC \label{hrc2wfc}}
\tablewidth{0pt}
\rotate
\tablehead{
\colhead{SOURCE}&\multicolumn{2}{c}{TARGET}&\multicolumn{4}{c}{Observed}&\multicolumn{4}{c}{Synthetic}\\
\colhead{SMAG}&\colhead{TMAG}&\colhead{TCOL}&\colhead{c0}&\colhead{c1}&\colhead{c2}& &\colhead{c0}&\colhead{c1}&\colhead{c2}&\colhead{TCOL}\\
\multicolumn{6}{c}{}&\colhead{}&\multicolumn{3}{c}{}&\colhead{range}\tablenotemark{a}
}
\startdata
F435W & F435W & F435W-F475W &   $-$0.573 $\pm$ 0.073 &   $-$0.045 $\pm$ 0.111 &    0.000 $\pm$ 0.000 &  &   $-$0.513 $\pm$ 0.031 &   $-$0.378 $\pm$ 0.178 &    0.439 $\pm$ 0.244 & $< 0.5$ \\
                & & & \multicolumn{4}{c}{   } &   $-$0.592 $\pm$ 0.008 &   $-$0.006 $\pm$ 0.017 &    0.015 $\pm$ 0.008 & $> 0.5$ \\
      & F435W & F435W-F550M & \multicolumn{4}{c}{   } &   $-$0.550 $\pm$ 0.369 &    0.100 $\pm$ 0.629 &    0.060 $\pm$ 0.267 & $<-1.0$ \\
                & & & \multicolumn{4}{c}{   } &   $-$0.588 $\pm$ 0.001 &    0.008 $\pm$ 0.001 &    0.002 $\pm$ 0.002 & $>-1.0$ \\
      & F435W & F435W-F555W &   $-$0.592 $\pm$ 0.022 &   $-$0.016 $\pm$ 0.036 &    0.000 $\pm$ 0.000 &  &   $-$0.594 $\pm$ 0.002 &   $-$0.010 $\pm$ 0.009 &    0.089 $\pm$ 0.040 & $< 0.3$ \\
                & & & \multicolumn{4}{c}{   } &   $-$0.590 $\pm$ 0.004 &   $-$0.002 $\pm$ 0.007 &    0.006 $\pm$ 0.003 & $> 0.3$ \\
      & F435W & F435W-F606W &   $-$0.580 $\pm$ 0.041 &   $-$0.014 $\pm$ 0.028 &    0.000 $\pm$ 0.000 &  &   $-$0.563 $\pm$ 0.008 &   $-$0.076 $\pm$ 0.024 &    0.047 $\pm$ 0.017 & $< 1.2$ \\
                & & & \multicolumn{4}{c}{   } &   $-$0.586 $\pm$ 0.009 &   $-$0.007 $\pm$ 0.008 &    0.004 $\pm$ 0.002 & $> 1.2$ \\
\enddata
\tablenotetext{a}{TCOL range applies only to synthetic transformations.}
\end{deluxetable}

\clearpage

\setlength{\tabcolsep}{0.02in}
\begin{deluxetable}{cccccccccccccc}
\tabletypesize{\scriptsize}
\tablecaption{Extinction Ratios A(P)/E(B-V)  in WFC Filters for Different SEDs\label{est_w}}
\tablewidth{0pt}
\tablehead{
\colhead{SED} & \colhead{F435W} & \colhead{F475W} & \colhead{F502N} & \colhead{F550M} & \colhead{F555W} & \colhead{F606W} & \colhead{F625W} & \colhead{F658N} & \colhead{F660N} & \colhead{F775W} & \colhead{F814W} & \colhead{F850LP} & \colhead{F892N}
}
\startdata
O5 & 4.192 & 3.773 & 3.459 & 3.052 & 3.233 & 2.936 & 2.673 & 2.525 & 2.518 & 2.005 & 1.864 & 1.498 & 1.509 \\
G2 & 4.081 & 3.634 & 3.458 & 3.042 & 3.177 & 2.809 & 2.637 & 2.525 & 2.517 & 1.982 & 1.825 & 1.485 & 1.508 \\
M0 & 3.994 & 3.555 & 3.458 & 3.030 & 3.115 & 2.716 & 2.616 & 2.525 & 2.517 & 1.965 & 1.796 & 1.471 & 1.508 \\
E  & 4.044 & 3.591 & 3.458 & 3.040 & 3.158 & 2.764 & 2.623 & 2.525 & 2.517 & 1.973 & 1.803 & 1.472 & 1.508 \\
Sc & 4.103 & 3.637 & 3.459 & 3.041 & 3.176 & 2.799 & 2.633 & 2.525 & 2.517 & 1.981 & 1.822 & 1.477 & 1.508 \\
Im & 4.127 & 3.659 & 3.460 & 3.044 & 3.200 & 2.830 & 2.637 & 2.525 & 2.517 & 1.979 & 1.818 & 1.476 & 1.508 \\
\enddata
\end{deluxetable}

\setlength{\tabcolsep}{0.01in}
\begin{deluxetable}{cccccccccccccccccc}
\tabletypesize{\tiny}
\tablecaption{Extinction Ratios A(P)/E(B-V)  in HRC Filters for Different SEDs \label{est_h}}
\tablewidth{0pt}
\tablehead{
\colhead{SED} & \colhead{F220W} & \colhead{F250W} & \colhead{F330W} & \colhead{F344N} & \colhead{F435W} & \colhead{F475W} & \colhead{F502N} & \colhead{F550M} & \colhead{F555W} & \colhead{F606W} & \colhead{F625W} & \colhead{F658N} & \colhead{F660N} & \colhead{F775W} & \colhead{F814W} & \colhead{F850LP} & \colhead{F892N}
}
\startdata
O5  & 8.350 & 6.353 & 5.105 & 4.991 & 4.197 & 3.741 & 3.461 & 3.053 & 3.235 & 2.953 & 2.681 & 2.528 & 2.546 & 2.018 & 1.847 & 1.473 & 1.513 \\
G2  & 5.735 & 5.676 & 5.046 & 4.987 & 4.075 & 3.608 & 3.458 & 3.043 & 3.180 & 2.825 & 2.644 & 2.525 & 2.518 & 1.995 & 1.803 & 1.460 & 1.508 \\
M0  & 2.809 & 4.965 & 5.000 & 4.953 & 3.984 & 3.535 & 3.458 & 3.031 & 3.118 & 2.730 & 2.623 & 2.525 & 2.517 & 1.978 & 1.770 & 1.447 & 1.507 \\
E   & 5.479 & 5.654 & 5.043 & 4.980 & 4.036 & 3.569 & 3.458 & 3.041 & 3.160 & 2.780 & 2.630 & 2.525 & 2.518 & 1.985 & 1.779 & 1.447 & 1.508\\
Sc  & 7.773 & 6.133 & 5.066 & 4.987 & 4.100 & 3.611 & 3.459 & 3.042 & 3.179 & 2.816 & 2.640 & 2.525 & 2.518 & 1.994 & 1.800 & 1.452 & 1.508\\
Im  & 7.995 & 6.207 & 5.070 & 4.988 & 4.126 & 3.632 & 3.460 & 3.044 & 3.203 & 2.847 & 2.644 & 2.525 & 2.519 & 1.992 & 1.795 & 1.451 & 1.508\\
\enddata
\end{deluxetable}




\begin{deluxetable}{p{1.9in}|cccc|cccc}
\tabletypesize{\scriptsize}\nonumber
\tablecaption{Aperture Correction - Case 1 \label{halo_ex1}}
\tablewidth{0pt}
\tablehead{
\colhead{Assumption} & \multicolumn{4}{c}{WFC}& \multicolumn{4}{c}{HRC}\\
\colhead{} & \colhead{apt\_c} & \colhead{apt\_c}& \colhead{F850LP}& \colhead{F775W-F850LP}&  \colhead{apt\_c} & \colhead{apt\_c} &\colhead{F850LP}& \colhead{F775W-F850LP}\\
\colhead{} & \colhead{F775W} &\colhead{F850LP}&\colhead{}& \colhead{} &\colhead{F775W} &\colhead{F850LP}&\colhead{}&\colhead{}
}
\startdata
input data & \nodata & \nodata & 26.00 & 1.65 & \nodata & \nodata & 26.00 & 1.71\\
\hline
1- The counts in  0.3$''$ are the total counts & \nodata & \nodata & 26.22 & 1.57 & \nodata & \nodata & 26.43 & 1.51\\
\hline
\multicolumn{9}{l}{2- The counts in 0.3$''$ need to be corrected using the same correction for both filters }\\
   a) F775W apt\_c  from the EE of WDs & 0.14 & 0.14 & 26.08 & 1.57 & 0.18 & 0.18 & 26.25 &  1.51\\
   b) F850LP apt\_c  from the EE of WDs & 0.19 & 0.19 & 26.03 & 1.57 & 0.40 & 0.40 & 26.03 &  1.51\\
\hline
\multicolumn{9}{l}{3- The counts in 0.3$''$ need to be corrected using the different corrections in different  filters}\\
 a) apt\_c from the EE of WDs & 0.14 & 0.19 & 26.03 & 1.62 & 0.18 & 0.40& 26.03 & 1.73\\
 b) apt\_c fromt Tables \ref{WFC_acl} and \ref{HRC_acl} &        0.15 & 0.22 & 26.00 & 1.64 & 0.22 & 0.45 & 26.02 & 1.74\\
 c) using SYNPHOT     &        0.14 & 0.24 & 25.98 & 1.66 & 0.23 & 0.47 & 25.96 & 1.75\\
\enddata
\end{deluxetable}
\begin{deluxetable}{p{1.9in}|cccc|cccc}
\tabletypesize{\scriptsize}\nonumber
\tablecaption{Aperture Correction - Case 2\label{halo_ex2}}
\tablewidth{0pt}
\tablehead{
\colhead{Assumption} & \multicolumn{4}{c}{WFC}& \multicolumn{4}{c}{HRC}\\
\colhead{} & \colhead{apt\_c} & \colhead{apt\_c}& \colhead{F775W}& \colhead{F775W-F850LP}&  \colhead{apt\_c} & \colhead{apt\_c} &\colhead{F775W}& \colhead{F775W-F850LP}\\
\colhead{} & \colhead{F775W} &\colhead{F850LP}&\colhead{}& \colhead{} &\colhead{F775W} &\colhead{F850LP}&\colhead{}&\colhead{}
}
\startdata
input data & \nodata & \nodata & 22.00 & 1.84 & \nodata & \nodata & 22.05 & 1.99\\
\hline
measured data in aperture   & \nodata & \nodata & 22.20 & 1.69 & \nodata & \nodata & 22.55 & 1.54\\
\hline
  apt\_c from the EE of WDs                           & 0.19 & 0.27 & 22.01 & 1.77 & 0.50 & 0.83 & 22.06 & 1.88\\
  apt\_c fromt Tables \ref{WFC_acl} and \ref{HRC_acl} & 0.20 & 0.37 & 22.00 & 1.86 & 0.54 & 0.98 & 22.02 & 1.98\\
  using SYNPHOT                                       & 0.20 & 0.39 & 22.00 & 1.87 & 0.55 & 1.00 & 22.01 & 1.99\\
\enddata
\end{deluxetable}


\clearpage
\begin{deluxetable}{lclrrrcrrrr}
\tabletypesize{\small}
\setlength{\tabcolsep}{0.04in}
\tablecaption{Coefficients for the Transformation from  WFC to BVRI\label{wfc2ground}}
\tablewidth{0pt}
\rotate
\tablehead{
\colhead{SOURCE}&\multicolumn{2}{c}{TARGET}&\multicolumn{4}{c}{Observed}&\multicolumn{4}{c}{Synthetic}\\
\colhead{SMAG}&\colhead{TMAG}&\colhead{TCOL}&\colhead{c0}&\colhead{c1}&\colhead{c2}& &\colhead{c0}&\colhead{c1}&\colhead{c2}&\colhead{TCOL}\\
\multicolumn{6}{c}{}&\colhead{}&\multicolumn{3}{c}{}&\colhead{range}\tablenotemark{a}
}
\startdata
F435W  & B & B-V &   25.842 $\pm$ 0.023 &   $-$0.089 $\pm$ 0.024 &    0.000 $\pm$ 0.000 &  &   25.769 $\pm$ 0.004 &    0.029 $\pm$ 0.012 &   $-$0.156 $\pm$ 0.055 & $< 0.4$ \\
                & & & \multicolumn{4}{c}{ } &   25.709 $\pm$ 0.007 &    0.108 $\pm$ 0.009 &   $-$0.068 $\pm$ 0.002 & $> 0.4$ \\
      & B & B-R & \multicolumn{4}{c}{ } &   25.769 $\pm$ 0.004 &    0.027 $\pm$ 0.008 &   $-$0.073 $\pm$ 0.028 & $< 0.6$ \\
                & & & \multicolumn{4}{c}{ } &   25.743 $\pm$ 0.008 &    0.022 $\pm$ 0.007 &   $-$0.013 $\pm$ 0.001 & $> 0.6$ \\
      & B & B-I &   25.847 $\pm$ 0.021 &   $-$0.045 $\pm$ 0.011 &    0.000 $\pm$ 0.000 &  &   25.768 $\pm$ 0.004 &    0.022 $\pm$ 0.006 &   $-$0.038 $\pm$ 0.011 & $< 1.0$ \\
                & & & \multicolumn{4}{c}{ } &   25.749 $\pm$ 0.009 &    0.008 $\pm$ 0.006 &   $-$0.005 $\pm$ 0.001 & $> 1.0$ \\
F475W & B & B-V & \multicolumn{4}{c}{ } &   26.146 $\pm$ 0.002 &    0.389 $\pm$ 0.004 &    0.032 $\pm$ 0.001 &  .... \\
      & B & B-R & \multicolumn{4}{c}{ } &   26.150 $\pm$ 0.004 &    0.291 $\pm$ 0.008 &   $-$0.110 $\pm$ 0.028 & $< 0.6$ \\
\enddata
\tablenotetext{a}{TCOL range applies only to synthetic transformations.}
\end{deluxetable}
\begin{deluxetable}{lclrrrcrrrr}
\tabletypesize{\scriptsize}
\setlength{\tabcolsep}{0.04in}
\tablecaption{Coefficients for the Transformation from  HRC to UBVRI. \label{hrc2ground}}
\tablewidth{0pt}
\tablehead{
\colhead{SOURCE}&\multicolumn{2}{c}{TARGET}&\multicolumn{4}{c}{Observed}&\multicolumn{4}{c}{Synthetic}\\
\colhead{SMAG}&\colhead{TMAG}&\colhead{TCOL}&\colhead{c0}&\colhead{c1}&\colhead{c2}& &\colhead{c0}&\colhead{c1}&\colhead{c2}&\colhead{TCOL}\\
\multicolumn{6}{c}{}&\colhead{}&\multicolumn{3}{c}{}&\colhead{range}\tablenotemark{a}
}
\startdata
F330W & U & U-B & \multicolumn{4}{c}{ } &   23.020 $\pm$ 0.008 &   $-$0.406 $\pm$ 0.058 &    0.125 $\pm$ 0.079 & $< 0.2$ \\
                & & & \multicolumn{4}{c}{ } &   23.064 $\pm$ 0.029 &   $-$0.104 $\pm$ 0.058 &   $-$0.093 $\pm$ 0.026 & $> 0.2$ \\
      & U & U-V & \multicolumn{4}{c}{ } &   22.995 $\pm$ 0.012 &   $-$0.328 $\pm$ 0.043 &    0.052 $\pm$ 0.040 & $< 0.3$ \\
                & & & \multicolumn{4}{c}{ } &   23.001 $\pm$ 0.017 &    0.065 $\pm$ 0.021 &   $-$0.056 $\pm$ 0.006 & $> 0.3$ \\
      & U & U-R & \multicolumn{4}{c}{ } &   23.012 $\pm$ 0.012 &   $-$0.222 $\pm$ 0.024 &    0.095 $\pm$ 0.024 & $< 0.6$ \\
                & & & \multicolumn{4}{c}{ } &   23.050 $\pm$ 0.023 &    0.017 $\pm$ 0.019 &   $-$0.028 $\pm$ 0.004 & $> 0.6$ \\
     & U & U-I & \multicolumn{4}{c}{ } &   23.014 $\pm$ 0.012 &   $-$0.216 $\pm$ 0.022 &    0.065 $\pm$ 0.020 & $< 0.7$ \\
                & & & \multicolumn{4}{c}{ } &   23.040 $\pm$ 0.024 &    0.015 $\pm$ 0.018 &   $-$0.019 $\pm$ 0.003 & $> 0.7$ \\
\enddata
\tablenotetext{a}{TCOL range applies only to synthetic transformations.}
\end{deluxetable}

\clearpage
\begin{deluxetable}{lclrrrcrrrr}
\tabletypesize{\tiny}
\setlength{\tabcolsep}{0.04in}
\tablecaption{Coefficients for the Transformation from  WFC to WFCP2 \label{wfc2wfpc2}}
\tablewidth{0pt}
\tablehead{
\colhead{SOURCE}&\multicolumn{2}{c}{TARGET}&\multicolumn{4}{c}{Observed}&\multicolumn{4}{c}{Synthetic}\\
\colhead{SMAG}&\colhead{TMAG}&\colhead{TCOL}&\colhead{c0}&\colhead{c1}&\colhead{c2}& &\colhead{c0}&\colhead{c1}&\colhead{c2}&\colhead{TCOL}\\
\multicolumn{6}{c}{}&\colhead{}&\multicolumn{3}{c}{}&\colhead{range}\tablenotemark{a}
}
\startdata
F435W & F439W & F439W-F555W &    5.550 $\pm$ 0.067 &    0.040 $\pm$ 0.029 &    0.000 $\pm$ 0.000 &  &    6.180 $\pm$ 1.210 &   $-$0.605 $\pm$ 1.589 &    0.158 $\pm$ 0.520 & $< 1.7$ \\
                & & & \multicolumn{4}{c}{ } &    5.655 $\pm$ 0.076 &   $-$0.074 $\pm$ 0.060 &    0.029 $\pm$ 0.011 & $> 1.7$ \\
      & F439W & F439W-F606W &    5.783 $\pm$ 2.031 &   $-$0.153 $\pm$ 1.594 &    0.036 $\pm$ 0.304 &  &    5.772 $\pm$ 1.384 &   $-$0.054 $\pm$ 1.537 &   $-$0.014 $\pm$ 0.425 & $< 2.0$ \\
                & & & \multicolumn{4}{c}{ } &    5.638 $\pm$ 0.061 &   $-$0.046 $\pm$ 0.039 &    0.016 $\pm$ 0.006 & $> 2.0$ \\
      & F439W & F439W-F675W &    5.567 $\pm$ 0.176 &    0.030 $\pm$ 0.074 &    0.000 $\pm$ 0.000 &  &    5.694 $\pm$ 0.195 &   $-$0.066 $\pm$ 0.436 &   $-$0.007 $\pm$ 0.237 & $< 1.2$ \\
                & & & \multicolumn{4}{c}{ } &    5.627 $\pm$ 0.031 &   $-$0.030 $\pm$ 0.025 &    0.015 $\pm$ 0.005 & $> 1.2$ \\
      & F439W & F439W-F814W &    5.600 $\pm$ 0.033 &    0.019 $\pm$ 0.015 &    0.000 $\pm$ 0.000 &  &    5.662 $\pm$ 0.016 &   $-$0.108 $\pm$ 0.053 &    0.057 $\pm$ 0.037 & $< 1.4$ \\
                & & & \multicolumn{4}{c}{ } &    5.614 $\pm$ 0.041 &   $-$0.010 $\pm$ 0.031 &    0.009 $\pm$ 0.006 & $> 1.4$ \\
\enddata
\tablenotetext{a}{TCOL range applies only to synthetic transformations.}
\end{deluxetable}
\clearpage
\begin{deluxetable}{lclrrrcrrrr}
\tabletypesize{\tiny}
\setlength{\tabcolsep}{0.04in}
\tablecaption{Coefficients for the Transformation from  HRC to WFPC2 \label{hrc2wfpc2}}
\tablewidth{0pt}
\tablehead{
\colhead{SOURCE}&\multicolumn{2}{c}{TARGET}&\multicolumn{4}{c}{Observed}&\multicolumn{4}{c}{Synthetic}\\
\colhead{SMAG}&\colhead{TMAG}&\colhead{TCOL}&\colhead{c0}&\colhead{c1}&\colhead{c2}& &\colhead{c0}&\colhead{c1}&\colhead{c2}&\colhead{TCOL}\\
\multicolumn{6}{c}{}&\colhead{}&\multicolumn{3}{c}{}&\colhead{range}
}
\startdata
F330W & F336W & F336W-F439W & \multicolumn{4}{c}{   } &    4.214 $\pm$ 0.018 &    0.004 $\pm$ 0.083 &    0.001 $\pm$ 0.104 & $< 1.1$ \\
                & & & \multicolumn{4}{c}{   } &    4.056 $\pm$ 0.097 &    0.229 $\pm$ 0.100 &   $-$0.077 $\pm$ 0.024 & $> 1.1$ \\
      & F336W & F336W-F555W &    5.820 $\pm$ 6.690 &   $-$0.868 $\pm$ 3.329 &    0.118 $\pm$ 0.412 &  &    4.203 $\pm$ 0.142 &    0.011 $\pm$ 0.139 &   $-$0.002 $\pm$ 0.032 & $< 3.1$ \\
                & & & \multicolumn{4}{c}{   } &    3.745 $\pm$ 0.192 &    0.255 $\pm$ 0.089 &   $-$0.034 $\pm$ 0.010 & $> 3.1$ \\
      & F336W & F336W-F606W & \multicolumn{4}{c}{   } &    4.206 $\pm$ 0.160 &    0.006 $\pm$ 0.134 &   $-$0.001 $\pm$ 0.026 & $< 3.6$ \\
                & & & \multicolumn{4}{c}{   } &    3.832 $\pm$ 0.345 &    0.186 $\pm$ 0.148 &   $-$0.022 $\pm$ 0.015 & $> 3.6$ \\
      & F336W & F336W-F675W & \multicolumn{4}{c}{   } &    4.217 $\pm$ 0.148 &   $-$0.002 $\pm$ 0.146 &    0.001 $\pm$ 0.033 & $< 3.1$ \\
                & & & \multicolumn{4}{c}{   } &    3.946 $\pm$ 0.200 &    0.153 $\pm$ 0.093 &   $-$0.021 $\pm$ 0.010 & $> 3.1$ \\
\enddata
\end{deluxetable}

\begin{deluxetable}{lccccccc}
\tabletypesize{\scriptsize}
\tablecaption{AB Colors of the E/S0 Template from Ben\'\i tez et al. (2004) in the WFC Filter System \label{tx_w_el}}
\setlength{\tabcolsep}{0.03in}
\tablewidth{0pt} 
\tablehead{
\colhead{z}&
\colhead{F435W-F606W}&
\colhead{F475W-F606W}&
\colhead{F555W-F606W}&
\colhead{F625W-F606W}&
\colhead{F775W-F606W}&
\colhead{F814W-F606W}&
\colhead{F850LP-F606W}
}
\startdata
0.01 &  1.06 &  0.63 &  0.24 &  $-$0.16 &  $-$0.52 &  $-$0.58 &  $-$0.74 \\
0.06 &  1.26 &  0.72 &  0.25 &  $-$0.16 &  $-$0.56 &  $-$0.63 &  $-$0.79 \\
0.11 &  1.46 &  0.80 &  0.27 &  $-$0.17 &  $-$0.61 &  $-$0.68 &  $-$0.86 \\
0.16 &  1.60 &  0.93 &  0.30 &  $-$0.19 &  $-$0.65 &  $-$0.73 &  $-$0.93 \\
0.21 &  1.76 &  1.06 &  0.34 &  $-$0.23 &  $-$0.71 &  $-$0.80 &  $-$1.01 \\
0.26 &  1.75 &  1.17 &  0.42 &  $-$0.26 &  $-$0.77 &  $-$0.87 &  $-$1.10 \\
0.31 &  1.79 &  1.25 &  0.52 &  $-$0.29 &  $-$0.84 &  $-$0.94 &  $-$1.19 \\
0.36 &  1.83 &  1.34 &  0.62 &  $-$0.32 &  $-$0.91 &  $-$1.01 &  $-$1.28 \\
\enddata
\end{deluxetable}
\begin{deluxetable}{lccccccc}
\tabletypesize{\scriptsize}
\tablecaption{AB Colors of the Sbc Template from
 Ben\'\i tez et al. (2004) in the WFC Filter System\tablenotemark{a}
 \label{tx_w_Sbc}}
\setlength{\tabcolsep}{0.03in}
\tablewidth{0pt} 
\tablehead{
\colhead{z}&
\colhead{F435W-F606W}&
\colhead{F475W-F606W}&
\colhead{F555W-F606W}&
\colhead{F625W-F606W}&
\colhead{F775W-F606W}&
\colhead{F814W-F606W}&
\colhead{F850LP-F606W}
}
\startdata
0.01 &  0.75 &  0.47 &  0.18 &  $-$0.13 &  $-$0.42 &  $-$0.48 &  $-$0.64 \\
0.06 &  0.86 &  0.52 &  0.19 &  $-$0.13 &  $-$0.45 &  $-$0.51 &  $-$0.66 \\
0.11 &  0.98 &  0.57 &  0.21 &  $-$0.14 &  $-$0.48 &  $-$0.54 &  $-$0.70 \\
0.16 &  1.11 &  0.65 &  0.22 &  $-$0.15 &  $-$0.51 &  $-$0.58 &  $-$0.74 \\
0.21 &  1.26 &  0.74 &  0.24 &  $-$0.16 &  $-$0.54 &  $-$0.61 &  $-$0.79 \\
0.26 &  1.38 &  0.82 &  0.27 &  $-$0.18 &  $-$0.58 &  $-$0.66 &  $-$0.85 \\
0.31 &  1.48 &  0.90 &  0.31 &  $-$0.21 &  $-$0.63 &  $-$0.71 &  $-$0.91 \\
0.36 &  1.54 &  1.00 &  0.37 &  $-$0.23 &  $-$0.67 &  $-$0.75 &  $-$0.96 \\
\enddata
\end{deluxetable}
\begin{deluxetable}{lccccccc}
\tabletypesize{\scriptsize}
\tablecaption{AB Colors of the Im Template from
 Ben\'\i tez et al. (2004) in the WFC Filter System \label{tx_w_im}}
\setlength{\tabcolsep}{0.03in}
\tablewidth{0pt} 
\tablehead{
\colhead{z}&
\colhead{F435W-F606W}&
\colhead{F475W-F606W}&
\colhead{F555W-F606W}&
\colhead{F625W-F606W}&
\colhead{F775W-F606W}&
\colhead{F814W-F606W}&
\colhead{F850LP-F606W}
}
\startdata
0.01 &  0.36 &  0.19 &  0.06 &  $-$0.04 &  $-$0.20 &  $-$0.23 &  $-$0.31 \\
0.06 &  0.42 &  0.21 &  0.05 &  $-$0.03 &  $-$0.21 &  $-$0.24 &  $-$0.32 \\
0.11 &  0.50 &  0.28 &  0.05 &  $-$0.04 &  $-$0.22 &  $-$0.25 &  $-$0.33 \\
0.16 &  0.59 &  0.34 &  0.06 &  $-$0.08 &  $-$0.22 &  $-$0.26 &  $-$0.35 \\
0.21 &  0.68 &  0.40 &  0.10 &  $-$0.10 &  $-$0.23 &  $-$0.27 &  $-$0.36 \\
0.26 &  0.76 &  0.47 &  0.15 &  $-$0.11 &  $-$0.23 &  $-$0.27 &  $-$0.38 \\
0.31 &  0.87 &  0.52 &  0.18 &  $-$0.13 &  $-$0.24 &  $-$0.28 &  $-$0.40 \\
0.36 &  0.86 &  0.57 &  0.21 &  $-$0.15 &  $-$0.27 &  $-$0.31 &  $-$0.43 \\
\enddata
\end{deluxetable}

\begin{deluxetable}{lccccccccccc}
\tabletypesize{\scriptsize}
\tablecaption{Multi-instrument Colors of the E/S0 Template from
Ben\'\i tez et al. (2004) in the HRC System (AB mag) and WFC F606W (AB mag)
\label{tx_h_el} }
\setlength{\tabcolsep}{0.03in}
\tablewidth{0pt} 
\tablehead{
\colhead{z}&
\colhead{H$_{F220W}$-}&
\colhead{H$_{F250W}$-}&
\colhead{H$_{F330W}$-}&
\colhead{H$_{F435W}$-}&
\colhead{H$_{F475W}$-}&
\colhead{H$_{F555W}$-}&
\colhead{H$_{F606W}$-}&
\colhead{H$_{F625W}$-}&
\colhead{H$_{F775W}$-}&
\colhead{H$_{F814W}$-}&
\colhead{H$_{F850LP}$-}\\
&
\colhead{W$_{F606W}$}&
\colhead{W$_{F606W}$}&
\colhead{W$_{F606W}$}&
\colhead{W$_{F606W}$}&
\colhead{W$_{F606W}$}&
\colhead{W$_{F606W}$}&
\colhead{W$_{F606W}$}&
\colhead{W$_{F606W}$}&
\colhead{W$_{F606W}$}&
\colhead{W$_{F606W}$}&
\colhead{W$_{F606W}$}
}
\startdata
0.01 &  5.42 &  4.08 &  2.54 &  1.06 &  0.61 &  0.24 &  0.01 &  $-$0.15 &  $-$0.51 &  $-$0.59 &  $-$0.76 \\
0.06 &  5.53 &  4.39 &  2.69 &  1.26 &  0.69 &  0.25 &  0.01 &  $-$0.16 &  $-$0.56 &  $-$0.64 &  $-$0.80 \\
0.11 &  5.61 &  4.64 &  2.88 &  1.45 &  0.77 &  0.27 &  0.01 &  $-$0.16 &  $-$0.60 &  $-$0.68 &  $-$0.87 \\
0.16 &  5.67 &  4.82 &  3.14 &  1.59 &  0.89 &  0.31 &  0.02 &  $-$0.18 &  $-$0.64 &  $-$0.74 &  $-$0.95 \\
0.21 &  5.70 &  4.92 &  3.39 &  1.76 &  1.02 &  0.34 &  0.02 &  $-$0.22 &  $-$0.70 &  $-$0.81 &  $-$1.03 \\
0.26 &  5.69 &  4.98 &  3.66 &  1.76 &  1.14 &  0.43 &  0.02 &  $-$0.26 &  $-$0.76 &  $-$0.88 &  $-$1.12 \\
0.31 &  5.64 &  5.00 &  3.91 &  1.79 &  1.22 &  0.53 &  0.03 &  $-$0.28 &  $-$0.83 &  $-$0.95 &  $-$1.21 \\
0.36 &  5.55 &  5.00 &  4.09 &  1.83 &  1.31 &  0.63 &  0.03 &  $-$0.31 &  $-$0.90 &  $-$1.03 &  $-$1.30 \\
\enddata
\end{deluxetable}
  \begin{deluxetable}{lccccccccccc}
\tabletypesize{\scriptsize}
\tablecaption{Multi-instrument Colors of the Sbc Template from
Ben\'\i tez et al. (2004) in the HRC System (AB mag) and WFC F606W (AB mag)
\label{tx_h_sbc} }
\setlength{\tabcolsep}{0.03in}
\tablewidth{0pt} 
\tablehead{
\colhead{z}&
\colhead{H$_{F220W}$-}&
\colhead{H$_{F250W}$-}&
\colhead{H$_{F330W}$-}&
\colhead{H$_{F435W}$-}&
\colhead{H$_{F475W}$-}&
\colhead{H$_{F555W}$-}&
\colhead{H$_{F606W}$-}&
\colhead{H$_{F625W}$-}&
\colhead{H$_{F775W}$-}&
\colhead{H$_{F814W}$-}&
\colhead{H$_{F850LP}$-}\\
&
\colhead{W$_{F606W}$}&
\colhead{W$_{F606W}$}&
\colhead{W$_{F606W}$}&
\colhead{W$_{F606W}$}&
\colhead{W$_{F606W}$}&
\colhead{W$_{F606W}$}&
\colhead{W$_{F606W}$}&
\colhead{W$_{F606W}$}&
\colhead{W$_{F606W}$}&
\colhead{W$_{F606W}$}&
\colhead{W$_{F606W}$}
}
\startdata
0.01 &  3.12 &  2.78 &  2.05 &  0.75 &  0.46 &  0.19 &  0.01 &  $-$0.12 &  $-$0.42 &  $-$0.48 &  $-$0.65 \\
0.06 &  3.10 &  2.79 &  2.20 &  0.86 &  0.50 &  0.19 &  0.01 &  $-$0.13 &  $-$0.45 &  $-$0.51 &  $-$0.67 \\
0.11 &  3.08 &  2.78 &  2.31 &  0.98 &  0.55 &  0.21 &  0.01 &  $-$0.13 &  $-$0.48 &  $-$0.55 &  $-$0.71 \\
0.16 &  3.06 &  2.76 &  2.35 &  1.11 &  0.62 &  0.22 &  0.01 &  $-$0.14 &  $-$0.51 &  $-$0.58 &  $-$0.75 \\
0.21 &  3.05 &  2.73 &  2.35 &  1.26 &  0.71 &  0.24 &  0.01 &  $-$0.16 &  $-$0.54 &  $-$0.62 &  $-$0.80 \\
0.26 &  3.04 &  2.68 &  2.33 &  1.38 &  0.79 &  0.27 &  0.02 &  $-$0.18 &  $-$0.58 &  $-$0.67 &  $-$0.86 \\
0.31 &  3.03 &  2.63 &  2.28 &  1.47 &  0.87 &  0.32 &  0.02 &  $-$0.20 &  $-$0.63 &  $-$0.72 &  $-$0.93 \\
0.36 &  3.03 &  2.58 &  2.23 &  1.53 &  0.97 &  0.38 &  0.02 &  $-$0.23 &  $-$0.66 &  $-$0.76 &  $-$0.98 \\
\enddata
\end{deluxetable}
 \begin{deluxetable}{lccccccccccc}
\tabletypesize{\scriptsize}
\tablecaption{Multi-instrument Colors of the Im Template from
Ben\'\i tez et al. (2004) in the HRC System (AB mag) and WFC F606W (AB mag)
\label{tx_h_im} }
\setlength{\tabcolsep}{0.03in}
\tablewidth{0pt} 
\tablehead{
\colhead{z}&
\colhead{H$_{F220W}$-}&
\colhead{H$_{F250W}$-}&
\colhead{H$_{F330W}$-}&
\colhead{H$_{F435W}$-}&
\colhead{H$_{F475W}$-}&
\colhead{H$_{F555W}$-}&
\colhead{H$_{F606W}$-}&
\colhead{H$_{F625W}$-}&
\colhead{H$_{F775W}$-}&
\colhead{H$_{F814W}$-}&
\colhead{H$_{F850LP}$-}\\
&
\colhead{W$_{F606W}$}&
\colhead{W$_{F606W}$}&
\colhead{W$_{F606W}$}&
\colhead{W$_{F606W}$}&
\colhead{W$_{F606W}$}&
\colhead{W$_{F606W}$}&
\colhead{W$_{F606W}$}&
\colhead{W$_{F606W}$}&
\colhead{W$_{F606W}$}&
\colhead{W$_{F606W}$}&
\colhead{W$_{F606W}$}
}
\startdata
0.01 &  1.41 &  1.34 &  1.12 &  0.36 &  0.18 &  0.06 &  0.00 &  $-$0.03 &  $-$0.20 &  $-$0.24 &  $-$0.32 \\
0.06 &  1.38 &  1.32 &  1.15 &  0.43 &  0.19 &  0.05 &  0.00 &  $-$0.03 &  $-$0.21 &  $-$0.24 &  $-$0.33 \\
0.11 &  1.36 &  1.30 &  1.17 &  0.50 &  0.27 &  0.05 &  0.00 &  $-$0.04 &  $-$0.21 &  $-$0.25 &  $-$0.34 \\
0.16 &  1.34 &  1.27 &  1.18 &  0.59 &  0.33 &  0.07 &  0.00 &  $-$0.08 &  $-$0.22 &  $-$0.26 &  $-$0.36 \\
0.21 &  1.33 &  1.23 &  1.16 &  0.68 &  0.39 &  0.10 &  0.01 &  $-$0.10 &  $-$0.23 &  $-$0.27 &  $-$0.37 \\
0.26 &  1.33 &  1.20 &  1.13 &  0.76 &  0.45 &  0.15 &  0.01 &  $-$0.11 &  $-$0.23 &  $-$0.28 &  $-$0.39 \\
0.31 &  1.32 &  1.17 &  1.10 &  0.87 &  0.51 &  0.18 &  0.01 &  $-$0.13 &  $-$0.24 &  $-$0.29 &  $-$0.41 \\
0.36 &  1.30 &  1.13 &  1.05 &  0.86 &  0.55 &  0.21 &  0.01 &  $-$0.15 &  $-$0.26 &  $-$0.32 &  $-$0.44 \\
\enddata
\end{deluxetable}

\begin{deluxetable}{lccccc}
\tabletypesize{\footnotesize}
\tablecaption{Multi-instrument Colors of the E/S0 Template from
Ben\'\i tez et al. (2004) in the Landolt System (VEGAMAG) and WFC F606W (VEGAMAG)
\label{tx_lan_el} }
\setlength{\tabcolsep}{0.03in}
\tablewidth{0pt} 
\tablehead{
\colhead{z}&
\colhead{U-F606W}&
\colhead{B-F606W}&
\colhead{V-F606W}&
\colhead{R-F606W}&
\colhead{I-F606W}
}
\startdata
0.01 &  1.57 &  1.14 &  0.27 &  $-$0.33 &  $-$0.89 \\
0.06 &  1.65 &  1.30 &  0.27 &  $-$0.34 &  $-$0.94 \\
0.11 &  1.74 &  1.46 &  0.26 &  $-$0.36 &  $-$1.00 \\
0.16 &  1.86 &  1.61 &  0.27 &  $-$0.38 &  $-$1.04 \\
0.21 &  2.02 &  1.72 &  0.31 &  $-$0.42 &  $-$1.10 \\
0.26 &  2.22 &  1.80 &  0.37 &  $-$0.47 &  $-$1.17 \\
0.31 &  2.43 &  1.86 &  0.46 &  $-$0.51 &  $-$1.23 \\
0.36 &  2.65 &  1.90 &  0.56 &  $-$0.54 &  $-$1.31 \\
0.41 &  2.87 &  1.96 &  0.64 &  $-$0.55 &  $-$1.37 \\
0.46 &  3.03 &  2.01 &  0.68 &  $-$0.57 &  $-$1.44 \\
0.51 &  3.09 &  2.06 &  0.67 &  $-$0.58 &  $-$1.56 \\
0.56 &  3.05 &  2.12 &  0.62 &  $-$0.61 &  $-$1.68 \\
0.61 &  2.96 &  2.19 &  0.57 &  $-$0.64 &  $-$1.81 \\
0.66 &  2.86 &  2.28 &  0.53 &  $-$0.66 &  $-$1.91 \\
0.71 &  2.78 &  2.36 &  0.51 &  $-$0.68 &  $-$1.99 \\
0.76 &  2.70 &  2.41 &  0.49 &  $-$0.72 &  $-$2.07 \\
0.81 &  2.65 &  2.47 &  0.51 &  $-$0.73 &  $-$2.09 \\
0.86 &  2.66 &  2.54 &  0.60 &  $-$0.70 &  $-$2.03 \\
0.91 &  2.65 &  2.57 &  0.69 &  $-$0.69 &  $-$1.97 \\
0.96 &  2.59 &  2.55 &  0.76 &  $-$0.70 &  $-$1.95 \\
1.01 &  2.54 &  2.53 &  0.85 &  $-$0.71 &  $-$1.93 
\enddata
\end{deluxetable}
\begin{deluxetable}{lccccc}
\tabletypesize{\footnotesize}
\tablecaption{Multi-instrument Colors of the Sbc Template from
Ben\'\i tez et al. (2004) in the Landolt System (VEGAMAG) and WFC F606W (VEGAMAG)\tablenotemark{a}
\label{tx_lan_sbc} }
\setlength{\tabcolsep}{0.03in}
\tablewidth{0pt} 
\tablehead{
\colhead{z}&
\colhead{U-F606W}&
\colhead{B-F606W}&
\colhead{V-F606W}&
\colhead{R-F606W}&
\colhead{I-F606W}
}
\startdata
0.01 &  0.96 &  0.88 &  0.22 &  $-$0.29 &  $-$0.79 \\
0.06 &  1.15 &  0.97 &  0.22 &  $-$0.30 &  $-$0.83 \\
0.11 &  1.30 &  1.07 &  0.22 &  $-$0.31 &  $-$0.86 \\
0.16 &  1.42 &  1.19 &  0.23 &  $-$0.32 &  $-$0.89 \\
0.21 &  1.50 &  1.31 &  0.24 &  $-$0.34 &  $-$0.92 \\
0.26 &  1.52 &  1.42 &  0.26 &  $-$0.37 &  $-$0.96 \\
0.31 &  1.49 &  1.52 &  0.29 &  $-$0.40 &  $-$1.02 \\
0.36 &  1.46 &  1.60 &  0.34 &  $-$0.42 &  $-$1.06 \\
0.41 &  1.40 &  1.65 &  0.40 &  $-$0.45 &  $-$1.11 \\
0.46 &  1.34 &  1.66 &  0.45 &  $-$0.48 &  $-$1.17 \\
0.51 &  1.25 &  1.64 &  0.49 &  $-$0.50 &  $-$1.24 \\
0.56 &  1.16 &  1.60 &  0.53 &  $-$0.53 &  $-$1.32 \\
0.61 &  1.06 &  1.53 &  0.55 &  $-$0.55 &  $-$1.41 \\
0.66 &  0.96 &  1.46 &  0.57 &  $-$0.55 &  $-$1.49 \\
0.71 &  0.85 &  1.36 &  0.57 &  $-$0.57 &  $-$1.57 \\
0.76 &  0.74 &  1.26 &  0.55 &  $-$0.58 &  $-$1.64 \\
0.81 &  0.63 &  1.17 &  0.51 &  $-$0.58 &  $-$1.70 \\
0.86 &  0.54 &  1.08 &  0.48 &  $-$0.57 &  $-$1.73 \\
0.91 &  0.47 &  1.00 &  0.43 &  $-$0.55 &  $-$1.74 \\
0.96 &  0.41 &  0.93 &  0.39 &  $-$0.53 &  $-$1.73 \\
1.01 &  0.37 &  0.87 &  0.35 &  $-$0.50 &  $-$1.71 
\enddata
\end{deluxetable}
\begin{deluxetable}{lrcccc}
\tabletypesize{\small}
\tablecaption{Multi-instrument Colors of the Im Template from
Ben\'\i tez et al. (2004) in the Landolt System (VEGAMAG) and WFC F606W (VEGAMAG)\tablenotemark{a}
\label{tx_lan_Im} }
\setlength{\tabcolsep}{0.03in}
\tablewidth{0pt} 
\tablehead{
\colhead{z}&
\colhead{U-F606W}&
\colhead{B-F606W}&
\colhead{V-F606W}&
\colhead{R-F606W}&
\colhead{I-F606W}
}
\startdata
0.01 &  0.18 &  0.52 &  0.13 &  $-$0.18 &  $-$0.56 \\
0.06 &  0.34 &  0.57 &  0.09 &  $-$0.17 &  $-$0.57 \\
0.11 &  0.43 &  0.65 &  0.09 &  $-$0.17 &  $-$0.58 \\
0.16 &  0.46 &  0.73 &  0.12 &  $-$0.22 &  $-$0.58 \\
0.21 &  0.47 &  0.82 &  0.15 &  $-$0.24 &  $-$0.59 \\
0.26 &  0.46 &  0.90 &  0.18 &  $-$0.25 &  $-$0.60 \\
0.31 &  0.44 &  0.96 &  0.21 &  $-$0.25 &  $-$0.61 \\
0.36 &  0.40 &  1.00 &  0.21 &  $-$0.27 &  $-$0.63 \\
0.41 &  0.34 &  1.00 &  0.24 &  $-$0.29 &  $-$0.67 \\
0.46 &  0.24 &  0.94 &  0.25 &  $-$0.34 &  $-$0.79 \\
0.51 &  0.17 &  0.90 &  0.29 &  $-$0.37 &  $-$0.88 \\
0.56 &  0.10 &  0.85 &  0.32 &  $-$0.39 &  $-$0.94 \\
0.61 &  0.03 &  0.79 &  0.34 &  $-$0.38 &  $-$1.00 \\
0.66 &  $-$0.04 &  0.73 &  0.34 &  $-$0.37 &  $-$1.05 \\
0.71 &  $-$0.11 &  0.67 &  0.33 &  $-$0.36 &  $-$1.05 \\
0.76 &  $-$0.17 &  0.61 &  0.31 &  $-$0.35 &  $-$1.04 \\
0.81 &  $-$0.23 &  0.55 &  0.28 &  $-$0.33 &  $-$1.06 \\
0.86 &  $-$0.28 &  0.49 &  0.24 &  $-$0.32 &  $-$1.08 \\
0.91 &  $-$0.34 &  0.43 &  0.19 &  $-$0.32 &  $-$1.10 \\
0.96 &  $-$0.36 &  0.39 &  0.17 &  $-$0.30 &  $-$1.11 \\
1.01 &  $-$0.36 &  0.37 &  0.16 &  $-$0.27 &  $-$1.08 
\enddata
\end{deluxetable}

\begin{deluxetable}{rcccccccccccccc}
\tabletypesize{\footnotesize}
\tablecaption{S/N Calculation for Different Aperture Radii  for WFC/F606W}
\tablewidth{0pt}\label{OAT}
\tablehead{
\multicolumn{2}{c}{Aperture radius}&\colhead{Area}&&\colhead{~EE~}&&\colhead{Signal}&\multicolumn{3}{c}{Main  Noise Sources (ADU)} &&\multicolumn{2}{c}{Other Noise (\%)}&\colhead{Total}&\colhead{S/N}\\ \cline{1-2} \cline{8-10} \cline{12-13}
\colhead{Pix} & \colhead{arcsec} & \colhead{pix} &&\colhead{\%}  &&\colhead{ADU} & \colhead{Source} & 
 \colhead{Read Noise} & \colhead{Background} && \colhead{Breathing} & \colhead{Aperture corr.} &\colhead{Noise} &\colhead{Ratio}
}
\startdata
1  &	0.05&	\phn\phn\phn3.1&&	31.30&&	313.0&		17.7&	\phn4.0&	\phn5.6&&	0.05&	0.07&			28.1&	11.1\\
2  &	0.10&	\phn\phn12.6&&	65.60&&	656.0&		25.6&	\phn7.9&	11.2&&	0.04&	0.03&			37.9&	17.3\\
3  &	0.15&	\phn\phn28.3&&	79.60&&	796.0&		28.2&	11.9&	16.8&&	0.03&	0.02&			36.3&	22.0\\
4  &	0.20&	\phn\phn50.3&&	84.50&&	845.0&		29.1&	15.9&	22.4&&	0.02&	0.02&			35.6&	23.7\\
5  &	0.25&	\phn\phn78.5&&	87.10&&	871.0&		29.5&	19.8&	28.0&&	0.01&	0.01&			37.9&	23.0\\
6  &	0.30&	\phn\phn13.1&&	89.00&&	890.0&		29.8&	23.8&	33.6&&	0.00&	0.01&			43.0&	20.7\\
7  &	0.35&	\phn153.9&&	90.00&&	900.0&		30.0&	27.7&	39.2&&	0.00&	0.01&			49.5&	18.2\\
8  &	0.40&	\phn201.1&&	90.80&&	908.0&		30.1&	31.7&	44.8&&	0.00&	0.01&			56.1&	16.2\\
9  &	0.45&	\phn254.5&&	91.40&&	914.0&		30.2&	35.7&	50.4&&	0.00&	0.01&			62.8&	14.6\\
10 &	0.50&	\phn314.2&&	92.00&&	920.0&		30.3&	39.6&	56.0&&	0.00&	0.01&			69.5&	13.2\\
15 &	0.75&	\phn706.8&&	94.10&&	941.0&		30.7&	59.4&	84.1&&	0.00&	0.01&			\llap{1}03.4&	\phn9.1\\
20 &	1.00&	1256.6&&	95.20&&	952.0&		30.9&	79.3&	\llap{1}12.1&&	0.00&	0.00&			\llap{1}37.6&	\phn6.9\\

\enddata
\tablenotetext{a}{For this simulation we assumed a flux of 1000 electrons in the infinte aperture and a background
level of 10 electrons.}

\end{deluxetable}



\clearpage

\end{document}